\documentclass[11pt,a4paper]{article}

\usepackage[utf8]{inputenc}
\usepackage[T1]{fontenc}
\usepackage[english]{babel}
\usepackage[autolanguage]{numprint}
\usepackage{lmodern}
\usepackage{amsmath}
\usepackage{amssymb}
\usepackage{tensor}
\usepackage{cases}
\usepackage{mathrsfs}
\usepackage{textcomp}
\usepackage{stmaryrd}
\usepackage{inputenc}
\usepackage{moreverb}
\usepackage{listings}
\usepackage{color}
\usepackage[usenames,dvipsnames]{xcolor}
\definecolor{arsenic}{rgb}{0.15, 0.15, 0.15}
\usepackage{fancyvrb}  
\usepackage{esint}
\usepackage{lipsum}
\usepackage{changepage}
\usepackage{pdfpages}
\usepackage[squaren,Gray]{SIunits}
\usepackage{sistyle}
\usepackage[shortlabels]{enumitem}
\usepackage{dsfont}
\usepackage{nicefrac}
\usepackage[vcentermath]{youngtab}
\usepackage{xspace}
\usepackage{pdflscape}

\usepackage[top=2cm,bottom=3cm,inner=2.5cm,outer=2.5cm]{geometry}

\usepackage{graphicx}
\usepackage{tabularx}
\usepackage{multirow}
\usepackage{multicol}
\usepackage[font=small,labelfont=bf,labelsep=space]{caption}
\usepackage{subfig}
\makeatletter

\@addtoreset{figure}{section}
\@addtoreset{equation}{section}
\@addtoreset{table}{section}
\makeatother

\addto\captionsenglish{}
\addto\captionsenglish{}

\usepackage[colorlinks=true,
pdfstartview=FitV,
linkcolor= BrickRed,
citecolor= BrickRed,
urlcolor= BrickRed,
hyperindex=true,
hyperfigures=true]
{hyperref}
\hypersetup{linktoc=page}

\usepackage{doi}
\usepackage{cite}
\def\SO{\ensuremath{\mathrm{SO}}\xspace}
\def\so{\ensuremath{\mathfrak{so}}\xspace}
\def\GL{\ensuremath{\mathrm{GL}}\xspace}
\def\SL{\ensuremath{\mathrm{SL}}\xspace}
\def\SLdeux{\ensuremath{\mathrm{SL}(2,\mathbb{R})}\xspace}
\def\G{\ensuremath{\mathrm{G}}\xspace}
\def\g{\ensuremath{\mathfrak{g}}\xspace}
\def\SU{\ensuremath{\mathrm{SU}}\xspace}
\def\U{\ensuremath{\mathrm{U}}\xspace}
\def\USp{\ensuremath{\mathrm{USp}}\xspace}
\def\Sp{\ensuremath{\mathrm{Sp}}\xspace}
\def\OSphuit{\ensuremath{\mathrm{OSp}(8\vert2,\mathbb{R})}\xspace}
\def\OSpsept{\ensuremath{\mathrm{OSp}(7\vert2,\mathbb{R})}\xspace}
\def\OSpquatre{\ensuremath{\mathrm{OSp}(4^{*}\vert4)}\xspace}
\def\SUquatre{\ensuremath{\SU(4\vert1,1)}\xspace}
\def\F{\ensuremath{\mathrm{F}}\xspace}
\def\GRsym{\ensuremath{\G^{R\text{-sym}}}\xspace}

\newcommand{\bea}{\begin{eqnarray}}
\newcommand{\eea}{\end{eqnarray}}

\makeatletter
\ifcase \@ptsize \relax
  \newcommand{\miniscule}{\@setfontsize\miniscule{4}{5}}
\or
  \newcommand{\miniscule}{\@setfontsize\miniscule{5}{6}}
\or
  \newcommand{\miniscule}{\@setfontsize\miniscule{5}{6}}
\fi
\makeatother

\definecolor{lred}{rgb}{0.3,0,0}
\usepackage[affil-it]{authblk}

\makeatletter
\def\@maketitle{%
  \newpage
  \null
  \vskip 2em%
  \begin{center}%
  \let \footnote \thanks
    {\LARGE\bfseries \@title \par}%
    \vskip 2.5em%
    {\large
      \lineskip .5em%
      \begin{center}
        \begin{minipage}{0.95\textwidth}
            \begin{tabular}[t]{c}%
            \@author
            \end{tabular}
        \end{minipage}    
      \end{center}\par}%
    \vskip 1em%
  \end{center}%
  \par
  \vskip 1.5em}
\makeatother

\title{$\boldsymbol{\mathcal{N}=(8,0)}$ AdS vacua of three-dimensional supergravity}
\author[1]{Nihat Sadik Deger}
\author[2]{Camille Eloy}
\author[2]{Henning Samtleben}
\affil[1]{Department of Mathematics, Bo\v{g}azi\c{c}i University, 34342, Bebek, Istanbul, Turkey}
\affil[2]{Univ Lyon, Ens de Lyon, Univ Claude Bernard, CNRS,
Laboratoire de Physique, F-69342 Lyon, France}

\begin{document}
\maketitle
\thispagestyle{empty}

\begin{abstract}
We give a classification of fully supersymmetric chiral ${\cal N}=(8,0)$ AdS$_3$ vacua
in general three-dimensional half-maximal gauged supergravities coupled to matter.
These theories exhibit a wealth of supersymmetric vacua with background isometries given by the supergroups
OSp$(8|2,\mathbb{R})$, F(4), SU$(4|1,1)$, and OSp$(4^*|4)$, respectively.
We identify the associated embedding tensors and the structure of the associated gauge groups.
We furthermore compute the mass spectra around these vacua.
As an off-spin we include results for a number of ${\cal N}=(7,0)$ vacua
with supergroups OSp$(7|2,\mathbb{R})$ and G$(3)$, respectively.
We also comment on their possible higher-dimensional uplifts.
\end{abstract}

\newpage
\setcounter{page}{1} 
\tableofcontents


\section{Introduction}

Supersymmetric Anti-de Sitter backgrounds of string theory and supergravity are of central importance in the 
holographic AdS/CFT correspondence \cite{Maldacena:1997re,Witten:1998qj,Aharony:1999ti}. Within higher-dimensional supergravity,
these correspond to supersymmetric solutions of the form AdS$_D \times {\cal M}$,
and may give rise after consistent truncation to a $D$-dimensional gauged supergravity with
a stationary point in its scalar potential. The AdS$_D$ solution of the lower-dimensional theory with all
scalars constant located at the stationary point then corresponds to the higher-dimensional 
AdS$_D \times {\cal M}$ solution.

A systematic approach to the classification of such backgrounds may start directly from a classification of
supersymmetric AdS$_D$ backgrounds in $D$-dimensional gauged supergravity.
These supergravities are determined by the choice of a constant embedding tensor which encodes the
gauge structure and couplings of the theories \cite{Nicolai:2000sc,deWit:2002vt,Schon:2006kz}.
Rather than searching AdS vacua within a given theory, one may instead determine the most general embedding tensor
such that the resulting theory admits a supersymmetric AdS$_D$ vacuum,
thereby determining the relevant $D$-dimensional theories together with their solutions.
For half-maximal supergravities in $D\ge4$ dimensions, such an analysis has been performed in Refs~\cite{Louis:2014gxa,Louis:2015mka,Louis:2015dca,Karndumri:2016ruc,Lust:2017aqj},
where the general gauging admitting a fully supersymmetric AdS vacuum has been determined and analyzed.

AdS$_3$ vacua have so far escaped a similar classification. This is mostly due to the fact that
the structure of gauged supergravity theories and their solutions in three space-time dimensions is 
very rich. Already the maximal (${\cal N}=16$) gauged supergravity in three dimensions offers a plethora of
fully supersymmetric AdS$_3$ vacua \cite{Fischbacher:2002fx}. This is in marked contrast
to higher dimensions, where there is a single maximally supersymmetric AdS vacuum in $D=7$ \cite{Pernici:1984xx}
and $D=5$ \cite{Gunaydin:1985cu},
together with a one-parameter family of maximally supersymmetric AdS$_4$ vacua~\cite{deWit:1982bul,DallAgata:2012bb}.
Similarly, many AdS$_3$ vacua have been identified in theories with
${\cal N}=9$ and ${\cal N}=10$ supersymmetry~\cite{Chatrabhuti:2010dh,Chatrabhuti:2010fp}.
The wealth of three-dimensional structures is based on the particular properties of three-dimensional gauge
and gravitational theories. The gravitational (super-)multiplet in three dimensions is non-propagating which allows for
the construction of a gravitational Chern-Simons action for any AdS$_3$ supergroup \cite{Achucarro:1987vz}.
Further coupling to scalar matter offers ample possibilities due to the on-shell duality between scalar and gauge fields
in three dimensions. Again, the possible structures are most conveniently encoded in terms of a properly constrained embedding tensor 
\cite{Nicolai:2000sc,Nicolai:2001sv,deWit:2003ja}.
Finally, the AdS$_3$ isometry group ${\rm SO}(2,2)\sim {\rm SL}(2,\mathbb{R})_L\times {\rm SL}(2,\mathbb{R})_R$ 
is not simple, but a product of two factors. Consequently, the supergroup
of AdS$_3$ background isometries in general factors into a direct product of simple supergroups 
$\mathcal{G}_L\times \mathcal{G}_R$ for which there are various options~\cite{Nahm:1977tg,Gunaydin:1986fe}.
The supercharges accordingly split into ${\cal N}=(p,q)$ charges transforming under 
$\mathcal{G}_R$ and $\mathcal{G}_L$, respectively.
As a result, there is an immense number of half-maximal ${\cal N}=8$ AdS vacua in three dimensions.

\begin{table}
  \centering
  \begin{tabular}{c|cccc}
    Supergroup $\mathcal{G}$ & \OSphuit & {\rm F}(4) & \SUquatre & \OSpquatre \\ \hline
    $\mathrm{Even}\left(\mathcal{G}\right)$ & $\mathrm{SO}(8)\times\mathrm{Sp}(2,\mathbb{R})$ & $\SO(7)\times\SLdeux$ & $\mathrm{U}(4)\times\mathrm{SU}(1,1)$ & $\mathrm{SO}^{*}(4)\times \mathrm{USp}(4)$ \\
    \GRsym & $\SO(8)$ & $\SO(7)$ & $\mathrm{U}(1)\times\SO(6)$ & $\SO(5)\times\SO(3)$ \\
    Supercharges & $\boldsymbol{8_{s}}$ & $\boldsymbol{8}$ & $\boldsymbol{4}^{+1}\boldsymbol{\oplus} \boldsymbol{\bar{4}}{}^{-1}$ & $(\boldsymbol{4,2})$  
  \end{tabular}
  \caption{Supergroups $\mathcal{G}$ with \SLdeux factor and eight supercharges~\cite{Nahm:1977tg,Gunaydin:1986fe}.
  We also list their even part $\mathrm{G_{even}}\simeq\SLdeux\times\GRsym$
  and their $R$-symmetry group $\GRsym$. The supercharges are given in representations of \GRsym.}
  \label{tab:N8supergroups}
\end{table}

\begin{table}
  \footnotesize
  \centering
  \begin{tabular}{ccccc}
  $n\le8$ matter multiplets & Gauge group & $\G_{\mathrm{ext}}$ & free parameter 
  \\\hline
  \multicolumn{4}{c}{\multirow{2}{*}{\normalsize $\boldsymbol{\mathcal{G}_R=\OSphuit}$}} \\
  & & & & \\\hline
  $n_{+}+n_{-}$ & $\SO(8,n_{+})\times\SO(n_{-})$ & $\SO(n_{+})\times\SO(n_{-})$ &  
  \\
  $n_{+}+4$ & $\SO(8,n_{+})\times\SO(4)$ & $\SO(n_{+})\times\SO(4)$ & $\xi$ 
  \\
  $n_{+}+4$ & $\SO(8,n_{+})\times\SO(3)$ & $\SO(n_{+})\times\SO(3)$ &  
  \\
  $n_{+}+6$ & $\SO(8,n_{+})\times\U(3)$ & $\SO(n_{+})\times\U(3)$ &  
  \\
  $n_{+}+7$ & $\SO(8,n_{+})\times\G_{2}$ & $\SO(n_{+})\times\G_{2}$ &  
  \\
  $8$ & $\SO(8)\times\SO(8-p)\times\SO(p)$ & $\SO(8-p)\times\SO(p)$ &  
  \\
  $8$ & $\GL(8)$ & - &  
  \\\hline
\multicolumn{4}{c}{\multirow{2}{*}{\normalsize $\boldsymbol{\mathcal{G}_R=\F(4)}$}} 
\\
  & & & & \\ \hline
  $n_{+}+n_{-}$ & $\SO(7,n_{+})\times\SO(1,n_{-})$ & $\SO(n_{+})\times\SO(n_{-})$ &  
  \\ 
  $n_{+}+3$ & $\SO(7,n_{+})\times\SO(1,3)$ & $\SO(n_{+})\times\SO(3)$ &  $\xi$ 
  \\ 
  $n\geq7$ &  $\GL(7)\times\SO(1,n-7)$ & $\SO(n-7)$ &  
  \\
  $n$ & $\SO(7)\times\SO(n)$ & $\SO(n)$ &  
  \\
  $4$ & $\SO(7)\times\SO(4)$ & $\SO(4)$ &   $\xi$ 
  \\
  $4$ & $\SO(7)\times\SO(3)$ & $\SO(3)$ &  
  \\
  $6$ & $\SO(7)\times\U(3)$ & $\U(3)$ &  
  \\
  $7$ & $\SO(7)\times\G_{2}$ & $\G_{2}$ &  
   \\
  $8$ & $\SO(7)\times\SO(8-p)\times\SO(p)$ & $\SO(8-p)\times\SO(p)$ &  
  \\\hline
\multicolumn{4}{c}{\multirow{2}{*}{\normalsize $\boldsymbol{\mathcal{G}_R=\SU(4|1,1)}$}} 
\\
  & & & & \\ \hline
  $n_{+}+n_{-}$ & $\SO(6,n_{+})\times\SO(2,n_{-})$ & $\SO(n_{+})\times\SO(n_{-})$ &  
  \\
  $n_{+}+2m$ & $\SO(6,n_{+})\times\U(m,1)$ & $\SO(n_{+})\times\U(m)$ &  
  \\
  $n_{+}+2$ & $\SO(6,n_{+})\times\SO(2,1)$ & $\SO(n_{+})$ &  
  \\
  $n_{+}+2$ & $\SO(6,n_{+})\times\SO(2,2)$ & $\SO(n_{+})\times\U(1)$ & $\xi$ 
  \\
  $6\leq n<8$ &  $\GL(6)\times\SO(2,n-6)$ & $\SO(n-6)$ &  
  \\
  $8$ &  $\GL(6)\times\SO(2,1)$ & $\U(1)$ &  
   \\
  $6$ &  $\GL(6)\times\SO(2,2)$ & $\SO(2)$ & $\xi$ 
  \\
  $n$ & $\SO(6)\times\SO(2)\times\SO(n)$ & $\SO(n)$ &  
  \\
  $4$ & $\SO(6)\times\SO(2)\times\SO(4)$ & $\SO(4)$ &   $\xi$
  \\
  $4$ & $\SO(6)\times\SO(2)\times\SO(3)$ & $\SO(3)$ & 
  \\
  $6$ & $\SO(6)\times\SO(2)\times\U(3)$ & $\U(3)$ &  
   \\
  $7$ & $\SO(6)\times\SO(2)\times\G_{2}$ & $\G_{2}$ &  
  \\
  $8$ & $\SO(6)\times\SO(2)\times\SO(8-p)\times\SO(p)$ & $\SO(8-p)\times\SO(p)$ &  
  \\
  $8$ & $\U(4,4)$ & $\U(4)$ &  
  \\
  $8$ & $\SL(2)\times\Sp(4,\mathbb{R})$ & $\U(1)$ &  
  \\
  $2+n_{-}$ & $\U(4,1)\times\SO(n_{-})$ & $\SO(n_{-})\times\U(1)$ &  
   \\\hline
\multicolumn{4}{c}{\multirow{2}{*}{\normalsize $\boldsymbol{\mathcal{G}_R=\OSpquatre}$}}
 \\
  & & & 
  \\ \hline
  $n_{+}+n_{-}$ & $\SO(5,n_{+})\times\SO(3,n_{-})$ & $\SO(n_{+})\times\SO(n_{-})$ &  
  \\
  $n\geq4$ & $\SO(5,n-4)\times\G_{2(2)}$ & $\SO(n-4)\times\SO(3)$ &  
  \\
  $n\geq5$ &  $\GL(5)\times\SO(3,n-5)$ & $\SO(n-5)$&  
   \\
  $n\geq3$ &  $\GL(3)\times\SO(5,n-3)$ & $\SO(n-3)$&  
  \\
  $8$ &  $\GL(5)\times\GL(3)$ & - &  
  \\
  $n$ & $\SO(5)\times\SO(3)\times\SO(n)$ & $\SO(n)$ &  
  \\
  $4$ & $\SO(5)\times\SO(3)\times\SO(4)$ & $\SO(4)$ &  $\xi$ 
  \\
  $4$ & $\SO(5)\times\SO(3)\times\SO(3)$ & $\SO(3)$ &  
  \\
  $6$ & $\SO(5)\times\SO(3)\times\U(3)$ & $\U(3)$ &  
  \\
  $7$ & $\SO(5)\times\SO(3)\times\G_{2}$ & $\G_{2}$ &  
  \\
  $8$ & $\SO(5)\times\SO(3)\times\SO(8-p)\times\SO(p)$ & $\SO(8-p)\times\SO(p)$ &  
  \\
  $8$ & $\Sp(2,2)\times\SO(3)$ & $\USp(4)$ & 
   \\
  \end{tabular}
  \caption{${\cal N}=(8,0)$ AdS$_3$ vacua
  preserving a global ${\rm SL}(2,\mathbb{R})_L \times \mathcal{G}_R \times  \G_{\mathrm{ext}}$ symmetry.
  }
  \label{tab:result01}
\end{table}

In this paper, we take a first step towards their classification, by determining all chiral ${\cal N}=(8,0)$
AdS$_3$ vacua within half-maximal $D=3$ gauged supergravity. 
For these vacua, the background isometries build a supergroup
${\rm SL}(2,\mathbb{R})_L\times \mathcal{G}_R$, where the simple supergroup $\mathcal{G}_R$ is chosen
among the options listed in Tab.~\ref{tab:N8supergroups}.
Some such vacua have recently appeared in Ref.~\cite{Dibitetto:2018ftj} as particular
type IIA AdS$_3$ compactifications on a six sphere $S^6 = {\rm SO}(7)/{\rm SO}(6)$ fibered over an interval,
preserving the exceptional supergroup $F(4)$.
We give a classification of the half-maximal $D=3$ gauged supergravities admitting an 
${\cal N}=(8,0)$ AdS$_3$ vacuum.
These theories couple the non-propagating ${\cal N}=8$ supergravity multiplet to $n$ scalar multiplets,
realizing different gauge groups embedded into the $\SO(8,n)$ isometry group of ungauged supergravity.
We restrict to $n\le8$ matter multiplets\footnote{This choice is discussed in Sec.~\ref{sec:conclusions}.}. 
For each of the supergroups of Tab.~\ref{tab:N8supergroups}, we identify the possible three-dimensional supergravities, 
characterized by their gauge groups which we list in Tab.~\ref{tab:result01},
together with the external global symmetry group ${\rm G}_{\rm ext}$ preserved by the vacuum.
We also indicate in this table, which of these theories 
admit a free parameter, entering in particular the structure constants and the scalar potential.
In the main body of this paper, we perform the explicit analysis of consistency conditions on the embedding
tensor, leading to this classification.
We moreover compute for every vacuum the associated mass spectrum, organized into 
supermultiplets of ${\rm SL}(2,\mathbb{R})_L\times \mathcal{G}_R$.

As a by-product of our constructions, we also identify a number of AdS$_3$ vacua with ${\cal N}=(7,1)$ and
${\cal N}=(7,0)$ supersymmetry, respectively. Some of the latter with ${\rm G}(3)$ superisometry group have also
emerged in the type IIA compactifications of Ref.~\cite{Dibitetto:2018ftj} with fluxes responsible for the breaking of the $R$-symmetry
group down to ${\rm G}_2$\,.
We list our findings in Tab.~\ref{tab:result02}.
Let us recall that there is no three-dimensional supergravity theory with ${\cal N}=7$ local supersymmetries
and non-trivial matter content~\cite{deWit:1992up}. As a consequence, ${\cal N}=(7,0)$ vacua can only be realized
within half-maximal ${\cal N}=8$ theories with 1/8 of supersymmetry spontaneously broken at the vacuum.
Indeed, we find that both our ${\cal N}=(7,0)$ vacua live in half-maximal  theories which also admit a
fully supersymmetric ${\cal N}=(8,0)$ vacuum, \textit{c.f.}~Figs.~\ref{fig:potential_n8} and \ref{fig:potential_n7} below.

\begin{table}
\footnotesize
  \centering
  \begin{tabular}{cccccc}
  $\cal N$ &  $n\le 8$ matter multiplets  & Gauge group & $\G_{\mathrm{ext}}$ & free parameter 
  \\\hline
\multicolumn{5}{c}{\multirow{2}{*}{{\normalsize $\boldsymbol{\mathcal{G}_R=\OSpsept}$}}} 
\\
  & & & 
   \\ \hline
  & & & 
  \\
  $(7,0)$& $8$ &  $\SO(7)\times\SO(8)$ & - & 
  \\
  & & & 
  \\\hline
\multicolumn{5}{c}{\multirow{2}{*}{{\normalsize  $\boldsymbol{\mathcal{G}_R=\G(3)}$}}} 
\\
  & & &
   \\ \hline
  & & & 
  \\
  $(7,1)$& $n$ & $\SO(n,1)\times\G_{2}$ & $\SO(n)$ & 
  \\
  $(7,0)$& $7$ & $\SO(7)\times\G_{2}$ & - &  
\end{tabular}
  \caption{ ${\cal N}=(7,1)$ and ${\cal N}=(7,0)$ AdS$_3$ vacua
  preserving a global ${\rm SL}(2,\mathbb{R})_L \times \mathcal{G}_R \times  \G_{\mathrm{ext}}$ symmetry.}
  \label{tab:result02}
\end{table}

The rest of this paper is organized as follows.
In Sec.~\ref{sec:3D}, we review the structure of half-maximal $D=3$ gauged supergravities,
specifically their embedding tensors and the set of algebraic constraints imposed onto the embedding tensors
in order to ensure consistency of the gauging and the existence of a supersymmetric AdS$_3$ vacuum.
In the following two Secs.~\ref{sec:casei} and~\ref{sec:caseii}, we then turn to the analysis and solution of these constraints.
The computation is organized by choice of supergroup from Tab.~\ref{tab:N8supergroups} 
together with the inequivalent embeddings of the desired AdS$_3$ $R$-symmetry group $\SO(p)\times \SO(q)$ into the 
isometry group $\SO(8,n)$ of ungauged $D=3$ supergravity.
For every solution, we list the explicit form of the embedding tensor, determine the gauge group of the $D=3$ theory,
and compute the mass spectrum of the ${\cal N}=(8,0)$ vacuum.
In Sec.~\ref{sec:N7}, we collect some partial results on AdS$_3$ vacua with ${\cal N}=(7,1)$ and ${\cal N}=(7,0)$ supersymmetry.
In Sec.~\ref{sec:maximal}, we raise and answer the question which of the AdS$_3$ vacua can in fact be
further embedded as vacua in a maximal (${\cal N}=16$) three-dimensional supergravity.
This translates into a couple of additional algebraic constraints to be imposed onto the embedding tensor,
which we check for all our vacua.
Finally, in Sec.~\ref{sec:conclusions} we summarize our findings
in Tabs.~\ref{tab:summary1}--\ref{tab:summary3}, and discuss 
possible generalizations, in particular
the possible higher-dimensional origin of these vacua.


\section{Gauged supergravities in three dimensions}
\label{sec:3D}


In this section, we recall some relevant facts about the three-dimensional half-maximal gauged supergravities. 
Their gauge structure is most conveniently encoded in a constant embedding tensor subject to a set of
algebraic constraints. We spell out the conditions for supersymmetric AdS$_3$ vacua and give the general
formulas for the mass spectra around these vacua.

\subsection{Lagrangian}

Half-maximal gauged supergravities in three dimensions have been 
constructed in Refs.~\cite{Nicolai:2001ac,deWit:2003ja}
by deforming the half-maximal ungauged theory of Ref.~\cite{Marcus:1983hb}. This ungauged theory contains an ${\cal N}=8$ supergravity multiplet composed of a dreibein $e\indices{_{\mu}^{\alpha}}$ and eight Rarita-Schwinger fields $\psi_{\mu}^{A}$, where $\mu$ and $\alpha$ denote respectively the curved and flat spacetime indices, and $A$ is the index of the spinorial representation of the Minkowski $R$-symmetry $\SO(8)$. The matter fields combine into $n$ copies of the ${\cal N}=8$ scalar multiplet, each one composed of eight scalars and eight spin-$\nicefrac{1}{2}$ fermions, transforming in the vectorial and cospinorial representations of $\SO(8)$, respectively. The scalar matter forms an $\SO(8,n)/(\SO(8)\times\SO(n))$ coset space sigma model. In the following, the indices $I,J,\ldots$ and $r,s,\ldots$  denote the vectorial indices of $\SO(8)$ and $\SO(n)$, respectively, which we combine into $\SO(8,n)$ vector indices $\mathcal{M} = \{I,r\}$. The $\SO(8,n)$ invariant tensor is defined as
\begin{equation}
\eta_{\cal MN} = \begin{pmatrix}
                  -\delta_{IJ} & 0 \\
                  0 & \delta_{rs}
                 \end{pmatrix},
\end{equation}
and the generators of $\mathfrak{so}(8,n)$, in the chosen vectorial representation, are given by
\begin{equation}
\big(L^{{\cal MN}}\big){}_{\cal K}{}^{\cal L} = 2\,\delta_{\cal K}{}^{[{\cal M}}\,\eta^{{\cal N}]\cal L}.
\end{equation}
Then, $\{L^{IJ}, L^{rs}\}$ form the generators of $\mathfrak{so}(8)\oplus \mathfrak{so}(n)$ while
the coset is parametrized by the $8n$ scalar fields $\phi_{Ir}$ through the $\SO(8,n)$ matrix
\begin{equation}
\mathcal{V} = e^{\phi_{Ir}L^{Ir}}\;. \label{eq:cosetrepresentative}
\end{equation}
Finally the spin-$\nicefrac{1}{2}$ fermions are denoted by $\chi_{\dot{A}r}$, 
with the index $\dot{A}$ of the cospinorial representation of $\SO(8)$.

The gauging of the theory is described using the embedding tensor formalism~\cite{Nicolai:2000sc,deWit:2002vt}. 
We briefly review its main features in this context. Gauging amounts to promoting a subgroup $\G_{0}\subset\SO(8,n)$ to a local symmetry, in such a way that the local supersymmetry remains preserved. The embedding of $\g_{0}$ in $\so(8,n)$ is given by the embedding tensor $\Theta_{\cal MN|PQ}$, so that the gauge group generators $X_{\cal MN}$ are
\begin{equation}
X_{\cal MN}=\Theta_{\cal MN|PQ}\,L^{\cal PQ}\;.
\label{XTL}
\end{equation}
The embedding tensor
$\Theta_{\cal MN|PQ}$ is antisymmetric in $[\cal M\cal N]$ and $[\cal P\cal Q]$, moreover symmetric under exchange of the two pairs
(in order to allow for an action principle of the gauged theory). 
It is thus contained in the symmetric tensor product of two adjoint representations of $\SO(8,n)$ and 
may accordingly be decomposed into its irreducible parts:
\begin{equation}
\Theta_{\cal MN|PQ} ~\subset~ {\boldsymbol 1} \oplus {\small\yng(2)} \oplus {\small\yng(2,2)} \oplus {\small\yng(1,1,1,1)}\; ,
\label{eq:tt8}
\end{equation}
where each box represents a vector representation $\boldsymbol{8+n}$ of $\SO(8,n)$. With this group-theoretical representation, the constraint on $\Theta_{\cal MN|PQ}$ that ensures supersymmetry of the gauged theory takes a simple form~\cite{deWit:2003ja}:
\begin{equation}
\mathbb{P}_{_{\scalebox{.3}{\yng(2,2)}}}\ \Theta_{\cal MN|PQ} = 0\;, \label{eq:linear}
\end{equation}
\textit{i.e.} one has to project out the ``Weyl-tensor'' type representation.\footnote{In Ref.~\cite{Nicolai:2001ac} a stronger condition has been applied by projecting the embedding tensor on its totally antisymmetric and trace parts. This explains the extra pieces in our fermion mass terms, given in Eq.~\eqref{eq:As} below.} 
This constraint is often called the ``linear constraint''. It can be explicitly solved by parametrizing $\Theta_{\cal MN|PQ}$ as
\begin{equation}
\Theta_{\cal MN|PQ} = \theta_{\cal MNPQ}+2\left(\eta_{{\cal M}[{\cal P}}\,\theta_{{\cal Q}]{\cal N}}-\eta_{{\cal N}[{\cal P}}\,\theta_{{\cal Q}]{\cal M}}\right)+2\,\eta_{{\cal M}[{\cal P}}\,\eta_{{\cal Q}]{\cal N}}\,\theta\;, \label{eq:thetafull}
\end{equation}
where $\theta_{\cal MNPQ}$ is totally antisymmetric and $\theta_{\cal MN}$ is symmetric and traceless.

To ensure that this embedding defines a proper gauge group, the embedding tensor $\Theta_{\cal MN|PQ}$ must be invariant under transformations of $\g_{0}$ itself. Explicitly, this reads
\begin{align}
0 = &\   \left(X_{\cal RS}\right)_{\cal M}{}^{\cal K}\Theta_{\cal KN|PQ} +  \left(X_{\cal RS}\right)_{\cal N}{}^{\cal K}\Theta_{\cal MK|PQ} 
+ \left(X_{\cal RS}\right)_{\cal P}{}^{\cal K}\Theta_{\cal MN|KQ} +  \left(X_{\cal RS}\right)_{\cal Q}{}^{\cal K}\Theta_{\cal MN|PK}\;. \label{eq:quadratic}
\end{align}
Since $X_{\cal MN}$ is defined in terms of the embedding tensor (\ref{XTL}), this condition gives rise to a set of equations
bilinear in $\Theta_{\cal MN|PQ}$, referred to as the ``quadratic constraints''. 
For later use, we rewrite these constraints in the parametrization~\eqref{eq:thetafull}
after contraction with an antisymmetric parameter $\Lambda_{\cal RS}$
\begin{subnumcases}{\label{eq:quadratic_singlets}}
0~=~-\Lambda_{\cal RS}\,\theta^{\cal RSKL}  \theta_{\cal K(M}  \eta_{\cal N)L} +2\, \theta^{\cal RS} \,  \theta_{\cal R(M}\,\Lambda_{\cal N)S}+2\,  \theta\, \theta^{\cal R}{}_{(\cal M}\Lambda_{\cal N)R}\;,\label{eq:QC1} \\
  0 ~=~-\Lambda_{\cal RS}\,\theta^{\cal RSKL} \, \theta_{\cal K[NPQ} \, \eta_{\cal M]L} \, +2\,   \theta^{\cal SK} \, \theta_{\cal K[NPQ}  \, \Lambda_{\cal M]S} \nonumber\\
  \quad\ \ \, ~~~+\,2\,  \Lambda_{\cal RS}\, \theta^{\cal R}{}_{[\cal M} \,  \theta^{\cal S}{}_{\cal NPQ]} \, \,  +2\,\theta\, \theta^{\cal K}{}_{\cal [NPQ} \,  \Lambda_{\cal M]K} \label{eq:QC2}\;.
\end{subnumcases}
In terms of SO$(8,n)$ representations, the quadratic constraints (\ref{eq:QC1}), (\ref{eq:QC2}) can be shown to transform according to
\begin{eqnarray}
{\small\yng(2)} ~\oplus~
{\small\yng(2,1,1)} ~\oplus~
{\small\yng(1,1,1,1)} ~\oplus~
{\small\yng(2,1,1,1,1)} 
\;.
\label{repQCfull}
\end{eqnarray}
Any solution to these constraints defines a viable gauging.

The gauging procedure then follows the standard scheme, introducing covariant derivatives with vectors $A_\mu{}^{\cal MN}$,
and associated field-strengths $F_{\mu\nu}{}^{\cal MN}$.
These fields are not present in the ungauged Lagrangian, but may be defined on-shell upon dualizing the Noether currents
of the global ${\rm SO}(8,n)$ symmetry. In the gauged theory they couple with a Chern-Simons term with the gauge
parameter $g$, and do not carry propagating degrees of freedom.

Their minimal coupling to scalars breaks supersymmetry, 
and necessitates the introduction of new terms to the Lagrangian,
specifically fermionic mass terms, and a scalar potential.
Before presenting the full Lagrangian, it is useful to introduce the so-called $T$-tensor, that encodes these additional terms in the Lagrangian. 
 We define $T_{\cal MN|PQ} = \mathcal{V}\indices{_{\cal M}^{\cal I}}\mathcal{V}\indices{_{\cal N}^{\cal J}}\mathcal{V}\indices{_{\cal P}^{\cal K}}\mathcal{V}\indices{_{\cal Q}^{\cal L}} \Theta_{\cal IJ|KL}$, with the coset representative~\eqref{eq:cosetrepresentative}, or equivalently
\begin{equation}
 \begin{cases}
 T_{\cal MNPQ} = \mathcal{V}\indices{_{\cal M}^{\cal I}}\mathcal{V}\indices{_{\cal N}^{\cal J}}\mathcal{V}\indices{_{\cal P}^{\cal K}}\mathcal{V}\indices{_{\cal Q}^{\cal L}} \theta_{\cal IJKL}\;,\\
 T_{\cal MN} = \mathcal{V}\indices{_{\cal M}^{\cal I}}\mathcal{V}\indices{_{\cal N}^{\cal J}}\theta_{\cal IJ}\;,\\
 T = \theta\;.
\end{cases}
\label{T}
\end{equation}
The full Lagrangian is then
\begin{align}
\mathcal{L} = \ & -\frac{1}{4}\,e\,R + \frac{1}{2}\,\varepsilon^{\mu\nu\rho}\,\bar{\psi}_{\mu}^{A}\,\mathcal{D}_{\nu}\,\psi_{\rho}^{A} + \frac{1}{4}\,e\,\mathcal{P}_{\mu}^{Ir}\mathcal{P}^{\mu\, Ir}- \frac{1}{2}\,ie\,\bar{\chi}^{\dot A r}\gamma^{\mu}\,\mathcal{D}_{\mu}\chi^{\dot A r} 
- \frac{1}{2}\,e\,\mathcal{P}_{\mu}^{Ir}\,\bar{\chi}^{\dot A r}\,\Gamma^{I}_{A\dot A}\,\gamma^{\mu}\gamma^{\nu}\,\psi_{\nu}^{A} 
\nonumber\\
&
 -\frac{1}{4}\,\varepsilon^{\mu\nu\rho}\,\Theta_{\cal MN|PQ}\,A_{\mu}{}^{\cal MN}\left(\partial_{\nu}A_{\rho}{}^{\cal PQ} 
 + \frac{1}{3}\,g\,
 \Theta_{\cal RS|UV}\,f^{\cal PQ,RS}{}_{\cal XY}\, A_{\nu}{}^{\cal UV} A_{\rho}{}^{\cal XY} 
 \right) \nonumber \\
& + \frac{1}{2}\,e\,A_{1}^{AB}\,\bar{\psi}_{\mu}^{A}\,\gamma^{\mu\nu}\,\psi_{\nu}^{B} 
 +i\,e\,A_{2}^{A\dot A r}\,\bar{\chi}^{\dot A r}\,\gamma^{\mu}\psi_{\mu}^{A} + \frac{1}{2}\,,e\,A_{3}^{\dot A r \dot B s}\,\bar{\chi}^{\dot A r}\,\chi^{\dot B s} - e\,V.
 \label{L}
\end{align}
Here, 
$e=\sqrt{|{\rm det}\,g_{\mu\nu}|}$, $R$ denotes the Ricci scalar, and ${f_{\cal MN,PQ}}^{\cal KL}$ describe 
the structure constants of $\mathfrak{so}(8,n)$. 
We refer to App.~\ref{sec:appendix_notations} for further details, particularly for the supersymmetry transformations. 
The last four terms in Eq.~(\ref{L}) are the most relevant for the following as they carry the
fermionic mass matrices and scalar potential characteristic for the given gauging. 
Explicitly, the fermionic mass tensors $A_{1,2,3}$ are given by
\begin{equation}
\begin{cases}
A^{AB}_{1} = {-}\dfrac{1}{48}\,\Gamma^{IJKL}_{AB}\,T_{IJKL}{-}\dfrac{1}{4}\,\delta^{AB}\,(T_{II}-4\,T)\;,\medskip\\
A^{A\dot{A}r}_{2} = -\dfrac{1}{12}\,\Gamma^{IJK}_{A\dot{A}}\,T_{IJKr}
-\dfrac12\,\Gamma^{I}_{A\dot A}\,T_{Ir}\;,
\medskip\\
A^{\dot{A}r\dot{B}s}_{3} = \dfrac{1}{48}\,\delta^{rs}\Gamma^{IJKL}_{\dot{A}\dot{B}}\,T_{IJKL}\!+\!
\dfrac12\,\Gamma^{IJ}_{\dot{A}\dot{B}}\,T_{IJrs}-2\,\delta^{\dot{A}\dot{B}} \delta^{rs}T-2\,\delta^{\dot{A}\dot{B}}T_{rs}+
\dfrac{1}{4}\,\delta^{\dot{A}\dot{B}} \delta^{rs}T_{II},
\end{cases}
\label{eq:As}
\end{equation}
as functions of the $T$-tensor (\ref{T}) and products of the ${\rm SO}(8)$ $\Gamma$-matrices $\Gamma^{I}_{A\dot{A}}$,  
while the scalar potential $V$ is given by
 \begin{align}
 V = &-\frac{1}{4}\,A^{AB}_{1}A^{AB}_{1}+\frac{1}{8}\,A^{A\dot{A}r}_{2} A^{A\dot{A}r}_{2}\nonumber\\
=& - \frac{1}{48} \!\left(T_{IJKL}T_{IJKL}\!+\! \frac{1}{4!}\,\varepsilon^{IJKLMNPQ}\, T_{IJKL}T_{MNPQ}
\!-\! 2\, T_{IJKr}T_{IJKr}\right)
\nonumber\\
&{}
-{\frac18}\,(T_{II}\!-\!4\,T)^2\!+ {\frac14} \,T_{Ir}T_{Ir} \;.
 \label{eq:potential}
 \end{align}

We finally note that the quadratic constraints (\ref{eq:quadratic_singlets}) imply the relation
\begin{eqnarray}
-\frac{1}{4}\,A^{AC}_{1}A^{BC}_{1}+\frac{1}{8}\,A^{A\dot{C}r}_{2} A^{B\dot{C}r}_{2} &=&
\frac18\,\delta^{AB}\,V
\;,
\label{eq:qA1A2}
\end{eqnarray} 
for the fermionic mass tensors, often referred to as a supersymmetric Ward identity.


\subsection[\texorpdfstring{$\mathcal{N}=(p,q)$ supersymmetry}{N=(p,q) supersymetry}]{\texorpdfstring{$\boldsymbol{\mathcal{N}=(p,q)}$ supersymmetry}{N=(p,q) supersymetry}}


In the following, we will classify and analyze supersymmetric AdS vacua of the Lagrangian~(\ref{L}). As a first condition, the existence
of an AdS vacuum necessitates an extremal point of the scalar potential (\ref{eq:potential}), \textit{i.e.}
\begin{align}
0 = \delta_{\phi} V = &-  \frac{1}{48} \, \delta \phi_{Ir}\, \bigg[4\, T_{IJKL}\,T_{JKLr} + \frac{1}{3}\,\varepsilon^{IJKLMNPQ}\, T_{MNPQ} T_{JKLr}+ 12\, T_{IJKs}  T_{JKrs} \nonumber \\
& \qquad\qquad-24\,\Big(T_{JJ} T_{Ir}-T_{Is} T_{sr}-T_{IJ}T_{Jr}-4 \,T\, T_{Ir}\Big)\bigg]\;.
\label{eq:varPot}
\end{align}
The precise amount of preserved supersymmetry can be read off from the eigenvalues of the gravitino mass matrix $A_1^{AB}$.
In units of the AdS length $ \ell^2 =  2/|V_0|$, the condition for 
${\cal N} =(p,q)$ supersymmetry takes the form
\begin{equation}
  \begin{cases}
  A^{ab}_{1} = \delta^{ab}\,{{\ell^{-1}}} \;,\\
  A^{\dot a\dot b}_{1} =- \delta^{\dot a\dot b}\,{{\ell^{-1}}}
  \;,
  \end{cases}
  \label{eq:SUSYconstraints}
\end{equation}
and all other components vanishing, where we have split the index $A$ according to $A=\{a,\dot a, i\}$ with $a\in[\![1,p]\!]$, $\dot a\in[\![p+1,p+q]\!]$ and $i\in[\![p+q+1,8]\!]$. Together with Eq.~\eqref{eq:qA1A2}, this implies
\begin{equation}
    \begin{cases}
    A^{a\dot A r}_{2} = 0\;, \\
    A^{\dot a\dot A r}_{2} = 0\;.
    \end{cases}
  \label{eq:SUSYconstraintsA2}
\end{equation}
For a vacuum with $p+q = 8$, preserving all supercharges, these conditions together with Eq.~\eqref{eq:As} further imply that $T_{IJKr} = T_{Ir} =0$ at the vacuum. From this, it follows that the vacuum condition (\ref{eq:varPot}) is automatically satisfied.
Moreover, for an $\mathcal{N}=(8,0)$ vacuum, 
the tensor $T_{IJKL}$ has to be anti-selfdual at the vacuum.

Around a supersymmetric AdS vacuum,
the matter content of the theory (\ref{L}) organizes into supermultiplets of the associated supergroup that extends the spacetime isometry group. As the $\mathrm{AdS_{3}}$ isometry group $\SO(2,2)\simeq\SLdeux_{L}\times\SLdeux_{R}$ is not simple, the corresponding supergroup in general is a direct product $\mathcal{G}_{L}\times\mathcal{G}_{R}$  of two simple supergroups, whose even parts are isomorphic to the products $\SLdeux_{L,R}\times\GRsym_{L,R}$, of the AdS factor $\SLdeux_{L,R}$ with the respective $R$-symmetry groups $\GRsym_{L,R}$. These supergroups have been classified in Ref.~\cite{Nahm:1977tg}, and further analyzed in Ref.~\cite{Gunaydin:1986fe}. Supersymmetry in three dimensions thus is factorizable and admits the decomposition $\mathcal{N}=(p,q)$, where $p$ and $q$ are the number of fermionic generators of $\mathcal{G}_{R}$ and $\mathcal{G}_{L}$ respectively. In the following, we will mainly focus on chiral $\mathcal{N}=(8,0)$ vacua, for which the even part of the supergroup is of the form 
\bea
{\rm G}_{\rm even} &=& \SLdeux_{L}\times\SLdeux_{R}\times\GRsym_{R}\;. 
\label{eq:Geven}
\eea
The relevant simple supergroups have been given in Tab.~\ref{tab:N8supergroups} above, along with their even subgroups and the representations of supercharges.
In particular, the different $R$-symmetry groups 
are of the form $\GRsym\simeq\SO(p)\times\SO(q)$, with $p+q=8$ and $p\neq4$. 
The supercharges originally transform in the spinor representation of $\SO(8)$ chirally embedded according to
\bea
\SO(8)~\subset~\SO(8)\times\SO(n)~\subset~\SO(8,n)
\;.
\eea
According to Tab.~\ref{tab:N8supergroups}, they remain irreducible $A\longrightarrow A$
(as a real representation) when $\SO(8)$ is broken down to the $R$-symmetry group $\GRsym$. 
For the vector representation of $\SO(8)$, this leaves two options. Depending on performing or not a triality rotation of $\SO(8)$
before embedding the $R$-symmetry group, the vector decomposes as 
\begin{subnumcases} {\label{eq:splitQC1}}
\,\,  {\rm (i)}:   I\longrightarrow I\;, \label{eq:splitQC1i}\\
  {\rm (ii)}:  I\longrightarrow  \{i, {\alpha}\} 
  \;,\quad
  i\in[\![1,p]\!]\;,\quad \alpha\in[\![p+1,8]\!]
  \;, \label{eq:splitQC1ii}
\end{subnumcases}
\textit{i.e.}\  either (i) it remains irreducible, or (ii) it decomposes into the vector of $\SO(p)\times\SO(q)$. 
In the first case, it is the cospinor of  $\SO(8)$ which decomposes
into the vector of $\SO(p)\times\SO(q)$, while it stays irreducible in the second case.
For \OSphuit, \textit{i.e.}\ $p=8$, the two options are equivalent.
The breaking \eqref{eq:splitQC1} determines how the vector of $\SO(8,n)$ transforms under 
the $R$-symmetry group, thus the transformation of the various components of the embedding tensor~\eqref{eq:thetafull}.

\smallskip
For the following analysis of vacua it is convenient to discuss the two cases in \eqref{eq:splitQC1} separately, and we will do so in Secs.~\ref{sec:casei} and \ref{sec:caseii}, respectively.


\subsection{Vacua and spectra}
\label{sec:vacuaspectra}


In this paper, we will classify and analyze AdS$_3$ vacua preserving ${\cal N}=(8,0)$ supersymmetry
for a general choice of the embedding tensor. 
Without loss of generality, we will search for vacua at the scalar origin $\mathcal{V}=\mathds{1}$,
since any extremal point located at a different $\mathcal{V}_0$ can be mapped into an extremal point
at the scalar origin of the theory with embedding tensor rotated by $\mathcal{V}_0^{-1}$~\cite{Dibitetto:2011gm,DallAgata:2011aa}.
We thus simultaneously solve the quadratic constraints 
(\ref{eq:QC1}) and (\ref{eq:QC2}) together with the extremality condition (\ref{eq:varPot}) (evaluated at $T_{\cal MN|PQ}=\Theta_{\cal MN|PQ}$) 
and the supersymmetry condition (\ref{eq:SUSYconstraints}). For every AdS$_3$ vacuum, we then determine the associated gauge group and compute the
mass spectrum of fluctuations.

The analysis is simplified by the symmetries of the desired vacuum. For a given choice of supergroup $\mathcal{G}_{R}$, 
we first parametrize $\Theta_{\cal MN|PQ}$ as a singlet of the $R$-symmetry group $\GRsym_{R}$, \textit{c.f.} Eq.~\eqref{eq:Geven}. 
In general, this group may be embedded in different ways into the compact
$\SO(8)\times\SO(n)\subset\SO(8,n)$, such that the representation of the supercharges branches
into the relevant representation collected in Tab.~\ref{tab:N8supergroups}. All $\GRsym_{R}$ admit a chiral embedding into
$\SO(8)\subset\SO(8)\times\SO(n)$ according to either of the options from \eqref{eq:splitQC1}, 
while for sufficiently large $n$, the group $\GRsym_{R}$ or one of its factors may also admit a
diagonal embedding into $\SO(8)\times\SO(n)$, as we will see in the following.
With the proper parametrization of  $\Theta_{\cal MN|PQ}$, we then solve the Eqs.~\eqref{eq:QC1}, \eqref{eq:QC2}, \eqref{eq:varPot} and \eqref{eq:SUSYconstraints} to identify the vacua.
The gauge group of the theory is deduced from the algebra satisfied by the generators~\eqref{XTL}. At the vacuum, it is spontaneously broken down to its compact subgroup.

For a given vacuum, the spectrum is entirely determined by the embedding tensor $\Theta_{\cal MN|PQ}$. 
In our conventions, the mass matrices for the spin $s=\nicefrac12,\; 1,\; \nicefrac32$ fields are given by~\cite{deWit:2003ja} 
\begin{equation}
\begin{cases}
M^{\dot{A}r\dot{B}s}_{(\nicefrac{1}{2})} = \dfrac12\,A^{\dot{A}r\dot{B}s}_{3}\;, \smallskip\\
M_{(1)\,Ir|Js}= \dfrac12\,\Theta_{Ir|Js}\;, \smallskip\\
M^{AB}_{(\nicefrac{3}{2})}= \dfrac12\,A^{AB}_{1}\;,
\end{cases}
\label{eq:mass1}
\end{equation}
with the tensors $A_1$, $A_3$ from Eq.~\eqref{eq:As}\footnote{We give here the expression of the spin-$\nicefrac{1}{2}$ fermions mass matrix after projecting out the goldstini. This is sufficient for most of the following, since we are mainly dealing with fully supersymmetric vacua. See Ref.~\cite{deWit:2003ja} for the complete expression.}. 
As for the scalars, 
their mass matrix is given by the second order variation of the scalar potential (\ref{eq:potential}) which yields
\begin{align}
  M_{(0)\,Lr,Ms}^{2} ~=  ~            &  \delta_{LM}\left(\frac{1}{3}\,\theta_{IJKr}\theta_{IJKs}-\theta_{IJrp}\theta_{IJsp}+2\,\theta_{II}\theta_{rs}-8\,\theta\,\theta_{rs}-2\,\theta_{Ir}\theta_{Is}\right. -2\,\theta_{pr}\theta_{ps}\bigg) \nonumber \\
&{}+  \delta_{rs}\left(\frac{1}{3}\,\theta_{IJKL}\theta_{IJKM}-\theta_{IJLp}\theta_{IJMp}+\frac{1}{36}\,\theta_{IJKR}\varepsilon^{IJKRMNPQ}\theta_{LNPQ}\right. \nonumber \\
              & \qquad\qquad + 2\,      \theta_{II}\theta_{LM}-2\,\theta_{IL}\theta_{IM} -2\,\theta_{Mp}\theta_{Lp}-8\,\theta\,\theta_{LM}\bigg) \nonumber \\
               & -2\, \theta_{ILMp}\theta_{Irsp}+2\,\theta_{ILsp}\theta_{IMrp}+\frac{1}{12}\,\theta_{IJKN}\varepsilon^{IJKNPQLM}\theta_{PQrs} \nonumber \\
              &-\frac{1}{9}\,\theta_{IJKr}\varepsilon^{IJKLMNPQ}\theta_{NPQs} +4\,\theta_{Lr}\theta_{Ms}-4\,\theta_{Ls}\theta_{Mr}-4\,\theta_{LM}\theta_{rs}\;.
\label{eq:mass2}
              \end{align}
We normalize all masses by the AdS length $\ell={\sqrt{2}}/\sqrt{\vert  V_{0}\vert}$. 
The spectrum is then most conveniently given in terms of the corresponding conformal dimensions $\Delta_{(s)}$,
which allows the identification of the supermultiplets.
In dimension $D=3$, the conformal dimensions 
are related to the normalized masses through~\cite{Klebanov:1999tb,Henningson:1998cd,Muck:1998rr,Mueck:1998iz,lYi:1998trg,Volovich:1998tj}
\begin{equation}
\begin{cases}
\Delta_{(0)}\left(\Delta_{(0)} -2\right) = \left(m_{(0)}\ell\right)^{2}\;,\\
\Delta_{(1)} = 1+\vert m_{(1)}\ell\vert \;,
\end{cases}
{\rm and}\quad
\begin{cases}
\Delta_{(\nicefrac{1}{2})} =1\pm m_{(\nicefrac{1}{2})}\ell\;,\\
\Delta_{(\nicefrac{3}{2})} =1+\vert m_{(\nicefrac{3}{2})}\ell\vert \;.
\end{cases}
\label{eq:confDim}
\end{equation}
Upon projecting out the Goldstone scalars and goldstini, the spectrum organizes into $\mathcal{G}$ supermultiplets.
As the supergroup $\mathcal{G}=\mathcal{G}_{L}\times\mathcal{G}_{R}$ is not simple, each $\Delta$ decomposes itself as $\Delta=\Delta_{L}+\Delta_{R}$, with conformal dimensions $\Delta_{L,R}$ associated to the representations of $\mathcal{G}_{L,R}$. 
The spacetime spin $s$ is identified as $s=\Delta_{R}-\Delta_{L}$.

\section{Solutions with irreducible vector embedding \eqref{eq:splitQC1i}}
\label{sec:casei}

We are now in position to start analyzing the consequences of the algebraic equations (\ref{eq:QC1}) and (\ref{eq:QC2}). 
We will do this analysis separately for all the supergroups given in Tab.~\ref{tab:N8supergroups}, with the two possible embeddings
of the $R$-symmetry groups $\SO(p)\times \SO(q)$ according to Eq.~\eqref{eq:splitQC1}.
In this section, we will consider the case of an irreducible vector embedding \eqref{eq:splitQC1i}.
We restrict to $n\leq8$ matter multiplets.

\subsection{Constraining the embedding tensor}

As explained above, upon implementing Eq.~\eqref{eq:SUSYconstraintsA2} in Eq.~\eqref{eq:As} for ${\cal N}=(8,0)$, the remaining possibly non-vanishing 
components of the embedding tensor 
are \textit{a priori} given by
\begin{equation}
\left\{\theta_{IJ},\; 
\theta_{rs},\; \theta=\kappa,\;\theta_{IJKL}^-,\; \theta_{IJrs},\; \theta_{Ipqr},\; \theta_{pqrs}\right\}, \label{eq:thetafullparam0}
\end{equation}
where $\theta_{IJKL}^-$ is anti-selfdual. 
The fact that the embedding tensor is singlet under the respective $R$-symmetry group,
embedded according to Eq.~\eqref{eq:splitQC1i},
further restricts these components as
\begin{equation}
\left\{\theta_{IJ}=\lambda\,\delta_{IJ},\; 
\theta_{rs},\; \theta=\kappa,\;\theta_{IJKL}^-=\xi^{\dot{A}\dot{B}}\,\Gamma^{IJKL}_{\dot{A}\dot{B}},\; \theta_{IJrs}
,\; \theta_{Ipqr},\; \theta_{pqrs}\right\}\;, \label{eq:thetafullparam}
\end{equation}
with traceless $\xi^{\dot{A}\dot{B}}$ of signature $(p,q)$ (only non-vanishing for $p\not=8$).
The first quadratic constraint \eqref{eq:QC1} with free indices chosen as 
$({\cal M,N})=(I,J)$ is then identically satisfied. Choosing the free indices as
$({\cal M,N})=(I,r)$ gives rise to the equations
\begin{subnumcases}{\label{eq:QC1Ip(i)}}
  \theta_{rp} \,  \theta_{ps}+ \kappa\, \theta_{rs} + \lambda\, (\kappa- \lambda)\, \delta_{rs} = 0\;, \label{eq:QC1Ipa(i)}\\
  \lambda\,\theta_{IJrs} + \theta_{IJrp}\theta_{ps} = 0\;, \label{eq:QC1Ipb(i)} \\
  \lambda\,\theta_{Ipqr} + \theta_{Ipqs}\theta_{sr} = 0\;.\label{eq:QC1Ipc(i)}
  \end{subnumcases}
Eq.~\eqref{eq:QC1Ipa(i)} determines the eigenvalues of the matrix $\theta_{rs}$
to be
\begin{equation}
\theta_{+}= -\lambda\;,\qquad \theta_{-}=\lambda-\kappa\;.
\end{equation}
Accordingly, we choose a basis in which $\theta_{rs}$ is diagonal,
split the indices $r$ into $\{r_+, r_-\}$ and denote by $n_\pm$ the multiplicities of these eigenvalues. 
Tracelessness of $\theta_{{\cal MN}}$ implies that
\begin{equation}
  (8+n_+-n_-)\,\lambda = -n_-\,\kappa\,. \label{eq:tracei}
\end{equation}
If we now set all $\theta_{\cal MNPQ}=0$,
all remaining quadratic constraints are satisfied. 
Up to an arbitrary overall scaling factor, this yields an embedding tensor of the form
\begin{align}
& \theta_{IJ}=n_{-}\,\delta_{IJ}\;,\qquad \theta=-(8+n_+-n_-)\,\;,\nonumber\\
&\theta_{r_{+}s_{+}}=-n_-\,\,\delta_{r_{+}s_{+}}\;,\quad\theta_{r_{-}s_{-}}=(8+n_+)\,\,\delta_{r_{-}s_{-}}\;,
\label{eq:Tklmn=0OSp8}
\end{align}
with all other components vanishing. The gauge group, defined via Eq.~(\ref{XTL}) by this embedding tensor, is
$\mathrm{SO}(8,n_{+})\times\mathrm{SO}(n_{-})$. The vacuum breaks this group down to its compact subgroup 
$\mathrm{SO}(8)\times\mathrm{SO}(n_{+})\times\mathrm{SO}(n_{-})$. 
Together with the preserved ${\cal N}=(8,0)$ supersymmetries and
the AdS symmetries, the isometry group of this vacuum thus is \OSphuit.
Following the discussion in Sec.~\ref{sec:vacuaspectra}
we may compute the spectrum around this vacuum. The result is collected in Tab.~\subref{tab:osp8multchiral},
organized into \OSphuit supermultiplets with the conformal dimensions obtained via Eq.~\eqref{eq:confDim}.
\medskip

It remains to analyze how the solution \eqref{eq:Tklmn=0OSp8} can be extended to 
non-vanishing $\theta_{\cal MNPQ}$.
The first quadratic constraint (\ref{eq:QC1}) with free indices chosen as $({\cal M,N})=(r,s)$ 
together with Eqs.\ \eqref{eq:QC1Ipb(i)}, \eqref{eq:QC1Ipc(i)} gives rise to the equations
\begin{equation}
  \left\{\begin{array}{ccc}
      (2\,\lambda-\kappa)\,\theta_{pqr_{+}s_{-}}  &=& 0\;,\\
    (2\,\lambda-\kappa)\,\theta_{IJrs_{-}}  &=& 0\;,\\
        (2\,\lambda-\kappa)\,\theta_{Ipqr_{-}} &=& 0\;,
        \end{array}\right.
  \label{Mink0}
\end{equation}
respectively.
These equations could be simultaneously solved by choosing $\kappa=2\,\lambda$\,. However, with Eq.~\eqref{eq:tracei} this choice implies that
in fact $\kappa=\lambda=0$, resulting in $\theta=0=\theta_{{\cal MN}}$\,. As a consequence, both tensors $A_1^{AB}$ and $A_2^{A\dot{A}r}$ 
from Eq.~\eqref{eq:As} vanish at the vacuum, inducing a vanishing potential \eqref{eq:potential} and thus a Minkowski vacuum,
which is beyond the scope of the present analysis. In all the following we thus assume that $\kappa\neq2\,\lambda$. Eqs.~\eqref{Mink0} then imply that, 
after complete resolution of the first quadratic constraint \eqref{eq:QC1}, 
the solution \eqref{eq:Tklmn=0OSp8} can be extended to potentially non-vanishing components 
\begin{align}
\Big\{ 
 \theta^-_{IJKL}=\xi^{\dot{A}\dot{B}}\,\Gamma^{IJKL}_{\dot{A}\dot{B}},\ 
 \theta_{IJr_{+}s_{+}},\ \theta_{Ip_{+}q_{+}r_{+}},\ \theta_{p_{+}q_{+}r_{+}s_{+}},\ \theta_{p_{-}q_{-}r_{-}s_{-}}\Big\}\;.
 \label{eq:Theta(i)}
\end{align}
The remaining quadratic constraints restricting these components
follow from evaluating Eq.~\eqref{eq:QC2}. 
For readability, we defer the full set of constraint equations to App.~\ref{app:QC2}, and in the following subsections
treat each of the four possible supergroups separately.


\subsection[\texorpdfstring{\OSphuit}{OSp(8|2,R)}]{\texorpdfstring{$\boldsymbol{\OSphuit}$}{OSp(8|2,R)}}
\label{sec:OSp8(i)}

%

\subsubsection{Chiral embedding}

The $R$-symmetry group in this case is $\SO(8)$. 
Let us first assume that it is chirally embedded into the first factor of ${\rm SO}(8)\times {\rm SO}(n)\subset {\rm SO}(8,n)$. 
Since the embedding tensor must be singlet under this group, its possible non-vanishing components 
within $\theta_{\cal MNPQ}$ further reduce from Eq.~\eqref{eq:Theta(i)} to
\begin{equation}
\left\{ \theta_{p_{+}q_{+}r_{+}s_{+}},\ \theta_{p_{-}q_{-}r_{-}s_{-}}\right\}.
\label{eq:osp82chiral}
\end{equation}
The second quadratic constraint is then reduced to two non-trivial equations, given by Eqs.~\eqref{eq:QC2Ipqrc} and \eqref{eq:QC2pqrsa}. They take the explicit form
\begin{subnumcases}{}
  (2\lambda-\kappa)\,\theta_{pqrs_{+}} = 0\;, \label{eq:TTTOSp8a} \\
  \lambda\,\theta_{l_{-}[q_{-}r_{-}s_{-}} \, \theta_{p_{-}]u_{-}v_{-}l_{-}} = n_-\,(2\,\lambda-\kappa)
  \left(\delta_{u_{-}[p_{-}} \theta_{q_{-}r_{-}s_{-}]v_{-}}-\,\delta_{v_{-}[p_{-}} \theta_{q_{-}r_{-}s_{-}]u_{-}}\right). \label{eq:TTTOSp8} 
\end{subnumcases}
Following the discussion after Eq.~(\ref{Mink0}), we restrict to the case $\kappa\not=2\,\lambda$, after which
the first equation implies that the only non-vanishing components of $\theta_{pqrs}$ is $\theta_{p_-q_-r_-s_-}$. 
Next, we solve the remaining equation~\eqref{eq:TTTOSp8} by considering each value of $n_{-}\ge4$ separately.
Since $\theta_{pqrs}$ does not enter in the mass formulas (\ref{eq:mass1}), (\ref{eq:mass2}), the spectra of all the
resulting theories are still given by Tab.~\subref{tab:osp8multchiral}. But, we will find in the following that non-vanishing $\theta_{pqrs}$
generically reduces the factor $\SO(n_{-})$ of the gauge group to a subgroup ${\rm K}_0$, such that the $\boldsymbol{n_-}$ in 
Tab.~\subref{tab:osp8multchiral} is to be replaced by the corresponding representation of ${\rm K}_0$.


\paragraph*{$\boldsymbol{n_-=4}$} Both sides of Eq.~\eqref{eq:TTTOSp8} identically vanish, such that the general  solution admits a non-vanishing $\theta_{p_{-}q_{-}r_{-}s_{-}}=\xi\, \varepsilon_{p_{-}q_{-}r_{-}s_{-}}$, with a free parameter $\xi$. The full solution then
extends Eq.~\eqref{eq:Tklmn=0OSp8} to
\begin{align}
&\theta_{IJ}=4\,\delta_{IJ}\;,\qquad\theta=-(4+n_+)\;,\qquad \theta_{p_{-}q_{-}r_{-}s_{-}} =(12+n_+)\,\xi\,\varepsilon_{p_{-}q_{-}r_{-}s_{-}}\;,\nonumber\\
&\theta_{r_{+}s_{+}}=-4\,\delta_{r_{+}s_{+}}\;,\quad\theta_{r_{-}s_{-}}=(8+n_+)\,\delta_{r_{-}s_{-}}\;.\label{eq:OSp8SO3}
\end{align}
For $|\xi|\neq 1$ the gauge group is $\mathrm{SO}(8,n_{+})\times\mathrm{SO}(4)$, as in the $\xi=0$ case. 
On the other hand, when $\xi$ takes a critical value $\xi=\pm1$, 
the gauge group reduces to $\mathrm{SO}(8,n_{+})\times\mathrm{SO}(3)_{\pm}$, \textit{i.e.} $\mathrm{SO}(4)=\mathrm{SO}(3)_{+}\times \mathrm{SO}(3)_{-}$ is broken down to one of its chiral factors. 


\paragraph*{$\boldsymbol{n_-=5}$} Setting $\theta_{p_{-}q_{-}r_{-}s_{-}}=\varepsilon_{p_{-}q_{-}r_{-}s_{-}t_{-}}\,\xi^{t_{-}}$, Eq.~\eqref{eq:TTTOSp8} shows that non-vanishing $\xi^{t_{-}}$ implies that $\kappa=2\,\lambda$, thus $\kappa=\lambda=0$ and the vacuum is not AdS.


\paragraph*{$\boldsymbol{n_-=6}$} Setting $\theta_{p_{-}q_{-}r_{-}s_{-}}=\dfrac{1}{2}\,\varepsilon_{p_{-}q_{-}r_{-}s_{-}u_{-}v_{-}}\,\xi^{u_{-}v_{-}}$, Eq.~\eqref{eq:TTTOSp8} leads to
\begin{equation}
(2\lambda-\kappa)\,\xi_{p_{-}q_{-}} = -\frac{1}{8}\,\varepsilon_{p_{-}q_{-}k_{-}l_{-}m_{-}n_{-}}\,\xi^{k_{-}l_{-}}\xi^{m_{-}n_{-}}.
\end{equation}
In particular, this implies that
\begin{equation}
\xi_{p_{-}r_{-}}\xi_{r_{-}q_{-}} = \frac{\delta_{p_{-}q_{-}}}{4!\,2\,(2\lambda-\kappa)} \left(
\varepsilon_{k_{-}l_{-}m_{-}n_{-}r_{-}s_{-}}\,\xi^{k_{-}l_{-}}\xi^{m_{-}n_{-}}\xi^{r_{-}s_{-}}\right)
\;.
\end{equation}
For $\kappa\neq2\lambda$ this implies that there is a basis such that\footnote{The global sign of $\xi_{p_{-}q_{-}}$ is fixed by the choice of convention for the Levi-Civita tensor $\varepsilon_{p_{-}q_{-}k_{-}l_{-}m_{-}n_{-}}$. Here we chose $\varepsilon_{123456}=1$.}
\begin{equation}
\xi_{p_{-}q_{-}}=-(2\lambda-\kappa)\,\begin{pmatrix}
                                      -\sigma & 0 & 0\\
                                      0 &\sigma&  0\\
                                      0 & 0 & \sigma
                                      \end{pmatrix}
                                      \;,
\quad \mathrm{where\ } \sigma = \begin{pmatrix}
                            0 & 1\\
                            -1 & 0
                            \end{pmatrix}.
\label{eq:OSp8n6basis}
\end{equation}
The full solution is then given by
\begin{align}
&\theta_{IJ}=6\,\delta_{IJ}\;,\quad\theta=-(2+n_+)\;,\quad\theta_{r_{+}s_{+}}=-6\,\delta_{r_{+}s_{+}}\;,\quad\theta_{r_{-}s_{-}}=(8+n_+)\,\delta_{r_{-}s_{-}},\nonumber\\
& \theta_{p_{-}q_{-}r_{-}s_{-}} =\frac{1}{2}\,\varepsilon_{p_{-}q_{-}r_{-}s_{-}u_{-}v_{-}}\xi^{u_{-}v_{-}},\quad\xi_{u_{-}v_{-}}=-(14+n_+)\,\begin{pmatrix}
                                    -\sigma & 0 & 0\\
                                      0 &\sigma&  0\\
                                      0 & 0 & \sigma
                                    \end{pmatrix}\;.
\label{eq:OSp8U3}
\end{align}
The gauge group in this case is $\mathrm{SO}(8,n_{+})\times\mathrm{U}(3)$, \textit{i.e.}\
due to the presence of a non-vanishing $\theta_{p_{-}q_{-}r_{-}s_{-}}$, the
$\mathrm{SO}(6)$ factor  is reduced to $\mathrm{U}(3)$ compared to solution \eqref{eq:Tklmn=0OSp8}. 


\paragraph{$\boldsymbol{n_-=7}$}
Setting $\theta_{p_{-}q_{-}r_{-}s_{-}}=\varepsilon_{p_{-}q_{-}r_{-}s_{-}u_{-}v_{-}w_{-}}\,\xi^{u_{-}v_{-}w_{-}}$, Eq.~\eqref{eq:TTTOSp8} leads to
\begin{equation}
\varepsilon_{m_{-}n_{-}p_{-}q_{-}r_{-}s_{-}[u_{-}}\,\xi_{v_{-}w_{-}]q_{-}}\,\xi_{m_{-}n_{-}p_{-}} = (2\lambda-\kappa)\,\left(\delta_{r_{-}[u_{-}}\,\xi_{v_{-}w_{-}]s_{-}}-\delta_{s_{-}[u_{-}}\,\xi_{v_{-}w_{-}]r_{-}}\right)\;.
\end{equation}
For $\kappa\neq2\,\lambda$, this implies
\begin{equation}
\xi^{m_{-}r_{-}s_{-}}\xi^{n_{-}r_{-}s_{-}} = -\frac{\Lambda}{14\,(2\lambda-\kappa)}\,
\delta^{m_{-}n_{-}} 
\;,
\end{equation}
with the constant $\Lambda$ given by
\bea
\Lambda&=& \left(\varepsilon^{p_{-}q_{-}r_{-}s_{-}u_{-}v_{-}w_{-}}\,\xi^{k_{-}p_{-}q_{-}} \xi^{k_{-}r_{-}s_{-}}  \xi^{u_{-}v_{-}w_{-}} \right)
\;.
\eea
This equation is solved by choosing $\xi_{m_{-}r_{-}s_{-}}$ proportional to the 
$\G_{2}\subset\SO(7)$ invariant 3-form 
$\omega_{m_{-}r_{-}s_{-}}$\footnote{We choose the normalisation of $\omega$ so that $\omega_{u_{-}v_{-}w_{-}}\omega_{u_{-}v_{-}w_{-}}=42$.}.
The full embedding tensor is then given by
\begin{align}
&
\theta_{IJ}=7\,\delta_{IJ}\;,\qquad
\theta_{r_{+}s_{+}}=-7\,\delta_{r_{+}s_{+}}\;,\quad\theta_{r_{-}s_{-}}=(8+n_+)\,\delta_{r_{-}s_{-}}\;,\qquad
\theta=-(1+n_+)\;,
\nonumber\\
& \theta_{p_{-}q_{-}r_{-}s_{-}} =\frac{15+n_{+}}{12}\,\varepsilon_{p_{-}q_{-}r_{-}s_{-}u_{-}v_{-}w_{-}}\omega^{u_{-}v_{-}w_{-}}
\;.\label{eq:OSp8G2}
\end{align}
The gauge group is $\mathrm{SO}(8,n_{+})\times\mathrm{G}_{2}$, \textit{i.e.} compared to solution~\eqref{eq:Tklmn=0OSp8} the group
$\mathrm{SO}(7)$ is reduced to $\mathrm{G}_{2}$. 


\paragraph{$\boldsymbol{n_-=8}$}
In this case, Eq.~\eqref{eq:tracei} implies $\kappa=0$ and Eq.~\eqref{eq:TTTOSp8} is solved by a self-dual $\theta_{pqrs}$:
\begin{equation}
\theta_{pqrs} =\theta^+_{pqrs} = \Gamma^{pqrs}_{ab}\,\xi^{ab},
\end{equation}
with ${\rm SO}(8)$ $\Gamma$-matrices $\Gamma^{pqrs}_{ab}$ and traceless $\xi^{ab}$ subject to the equation
\begin{equation}
  \xi^{ac} \xi^{bc} -\frac18\,\delta^{ab}\,\xi^{cd}\xi^{cd} = \frac{\lambda}{2} \;\xi^{ab}.
\end{equation}
This implies that the eigenvalues of $\xi_{ab}$ are
\begin{equation}
\xi^+=\frac{\lambda}{4}\,\frac{8-p}{(4-p)}\;,\qquad \xi^-=-\frac{\lambda}{4}\,\frac{p}{(4-p)}\;,
\end{equation}
with multiplicity $p$ and $8-p$ respectively and $p\neq4$.\footnote{The case $p=4$ implies $\kappa=2\,\lambda$ thus again leading to a Minkowski vacuum.} 
The full embedding tensor then takes the form:
\begin{align}
&\theta_{IJ}=8\,\delta_{IJ}\;,\quad\theta_{rs}=8\,\delta_{rs}\;,\quad
\theta=0\;,\quad \theta_{pqrs} = \Gamma^{pqrs}_{ab}\,\xi^{ab},\nonumber\\
& \mathrm{where}\;\; \xi^{ab}=\frac{2}{4-p}\,{\rm diag}\{\underbrace{8-p, \dots, 8-p}_{p},\underbrace{-p, \dots, -p}_{8-p}\}\;,\quad
\ p\neq4\;. \label{eq:OSp8SOp}
\end{align}
There is an analogous solution for anti-selfdual choice of $\theta_{pqrs}$.
The gauge group is $\mathrm{SO}(8)\times\mathrm{SO}(8-p)\times\mathrm{SO}(p)$, \textit{i.e.} $\mathrm{SO}(8)$ is reduced to $\mathrm{SO}(8-p)\times\mathrm{SO}(p)$ compared to solution~\eqref{eq:Tklmn=0OSp8}.

\begin{table}
  \centering
  \subfloat[$\mathrm{G}=\mathrm{SO}(8,n_{+})\times \mathrm{SO}(n_{-})$]{\begin{tabular}{c|c|c|c|c|c|c}
    $\Delta_{L}$ & $\Delta_{R}$ &  $\Delta$ & $s$ & $\mathrm{SO}(8)$& $\mathrm{SO}(n_{+})$& $\mathrm{SO}(n_{-})$ \\ \hline
    \multirow{2}{*}{$\nicefrac{5}{4}$} & $\nicefrac{3}{4}$ & $2$ & $\nicefrac{-1}{2}$ & $\boldsymbol{8_{c}}$& $\boldsymbol{n_{+}}$ & $\boldsymbol{1}$ \\ 
    & $\nicefrac{1}{4}$ & $\nicefrac{3}{2}$ & $-1$ & $\boldsymbol{8_{v}}$& $\boldsymbol{n_{+}}$ & $\boldsymbol{1}$\\\hline
    \multirow{2}{*}{$\nicefrac{1}{4}$} 
     & $\nicefrac{3}{4}$ & $1$ & $\nicefrac{1}{2}$ & $\boldsymbol{8_{c}}$ & $\boldsymbol{1}$ & $\boldsymbol{n_{-}}$ \\
      & $\nicefrac{1}{4}$ & $\nicefrac{1}{2}$ & $0$ & $\boldsymbol{8_{v}}$& $\boldsymbol{1}$ & $\boldsymbol{n_{-}}$
  \end{tabular} \label{tab:osp8multchiral}
  }
  \qquad
  \subfloat[$\mathrm{G}=\mathrm{GL}(8)$]{
  \begin{tabular}{c|c|c|c|c}
    $\Delta_{L}$ & $\Delta_{R}$ &  $\Delta$ & $s$ & $\mathrm{SO}(8)$ \\ \hline
    \multirow{5}{*}{$\nicefrac{3}{2}$}  & $\nicefrac{5}{2}$ & $4$ & $1$ & $\boldsymbol{1}$ \\
     & $2$ & $\nicefrac{7}{2}$ & $\nicefrac{1}{2}$ & $\boldsymbol{8_{s}}$ \\
     & $\nicefrac{3}{2}$ & $3$ & $0$ & $\boldsymbol{28}$ \\
     & $1$ & $\nicefrac{5}{2}$ & $\nicefrac{-1}{2}$ & $\boldsymbol{56_{s}}$ \\
     & $\nicefrac{1}{2}$ & $2$ & $-1$ & $\boldsymbol{35_{c}}$
  \end{tabular} \label{tab:osp8multdiag}
  }
    \caption{Mass spectra for the ${\rm OSp}(8|2,\mathbb{R})$ solutions with (a) chiral and (b) diagonal embedding
    of the $R$-symmetry group. The spectrum organizes into multiplets of ${\rm OSp}(8|2,\mathbb{R})$, 
    given in Eq.~(5.3) and Tab.~7 of Ref.~\cite{Gunaydin:1986fe}, respectively. For non-vanishing $\theta_{pqrs}$, the factor $\SO(n_{-})$ in (a) reduces to a subgroup ${\rm K}_{0}$, and the representation $\boldsymbol{n_{-}}$ is replaced by the corresponding representation of ${\rm K}_{0}$.}
    \label{tab:osp8mult}
\end{table}


\subsubsection{Diagonal embedding}
For $n=8$ matter multiplets, the $R$-symmetry group ${\rm SO}(8)$ alternatively allows a diagonal embedding as
\bea
\SO(8) ~=~\SO(8)_{\rm diag}~\subset~\SO(8)\times \SO(8)~\subset~\SO(8,8)
\;.
\label{eq:SO8diag}
\eea
Moreover, there are inequivalent diagonal embeddings according to possible triality rotations in the two $\SO(8)$ factors.
In this case the condition of being singlet under the $R$-symmetry group reduces the possible components
of the embedding tensor from Eq.~\eqref{eq:Theta(i)} to
\begin{align}
\Big\{ & \theta_{IJ}= \lambda\,\delta_{IJ},\;
\theta_{r_{+}s_{+}} = - \lambda\,\delta_{r_{+}s_{+}},\;
\theta_{r_{-}s_{-}} = (\lambda-\kappa)\,\delta_{r_{-}s_{-}},\; 
\theta=\kappa,\;
\theta_{IJr_{+}s_{+}}\Big\}\;,
 \end{align}
 with either $(n_{+},n_{-}) = (8,0)$ or $(n_{+},n_{-}) = (0,8)$. For vanishing $\theta_{IJr_{+}s_{+}}$, we are back to solution~\eqref{eq:Tklmn=0OSp8}.
 A non-vanishing $\theta_{IJr_{+}s_{+}}$ on the other hand implies $(n_{+},n_{-}) = (8,0)$ and from
Eq.~\eqref{eq:tracei} we deduce that $\lambda=0$\,. We are then left with the following
surviving components
\begin{equation}
\left\{\theta=\kappa,\; \theta_{IJrs}\right\}\;,
\label{eq:osp82diagonal}
\end{equation}
where we suppress the $+$ subscript.
The remaining equations of the quadratic constraints are given by Eqs.~\eqref{eq:QC2IJKrb}, \eqref{eq:QC2IJrsa}, \eqref{eq:QC2IJrsb} 
and \eqref{eq:QC2Ipqrc}:
\begin{equation}
  \begin{cases}
  \theta_{usM[I}\theta_{JK]rs} + \kappa\,\delta_{M[I}\theta_{JK]ru} = 0\;, \\
  \Lambda_{uv}\,\big[\theta_{uvL[I}\theta_{J]Lrs}+2\,\kappa\,\theta_{IJu[r}\delta_{s]v}\big] = 0\;, \\
  \Lambda_{MN}\,\big[\theta_{MNp[r}\theta_{s]pIJ} +2\,\kappa\,\theta_{rsM[I}\delta_{J]N}\big]=0\;,\\
  \theta_{MLu[p}\theta_{qr]IL} +\kappa\,\delta_{u[p}\theta_{qr]IM} = 0\;.
  \end{cases}
  \label{eq:conqu8}
\end{equation}
An $\SO(8)_{\rm diag}$ singlet in $\theta_{IJrs}$ can be parametrized as
\begin{equation}
\theta_{IJrs}=\rho\, \delta_{r[I}\delta_{J]s}\;.
\end{equation}
Eqs.~\eqref{eq:conqu8} fix $\rho=-2\,\kappa$, thus leading to an embedding tensor
\begin{equation}
  \theta=-\frac{1}{2}\;,\quad \theta_{IJrs}= \delta_{r[I}\delta_{J]s}\;. \label{eq:OSp8GL8}
\end{equation}
The gauge group induced by this tensor is $\mathrm{GL(8)}$. 
The spectrum 
around this vacuum is given in Tab.~\subref{tab:osp8multdiag}.

Upon triality rotation of the second factor in Eq.~\eqref{eq:SO8diag}, the
singlet in $\theta_{IJrs}$ would alternatively be given by
\begin{equation}
\theta_{IJrs}=\rho\, \Gamma^{IJ}_{rs}\;.
\end{equation}
This however does not lead to a non-trivial solution of Eqs.~\eqref{eq:conqu8}.

\subsection[\texorpdfstring{$\F(4)$}{F(4)}]{\texorpdfstring{$\boldsymbol{\F(4)}$}{F(4)}}

\subsubsection{Chiral embedding}

We now consider the supergroup $\F(4)$, with $R$-symmetry group $\SO(7)$. We first assume 
that ${\rm SO}(7)$ is entirely embedded into the first factor of ${\rm SO}(8)\times {\rm SO}(n)\subset {\rm SO}(8,n)$
according to Eq.~\eqref{eq:splitQC1i},
such that the potentially non-vanishing components of the embedding tensor reduce 
from Eq.~\eqref{eq:Theta(i)} to
\begin{align}
\Big\{ & \theta_{IJ}= \lambda\,\delta_{IJ},\;\theta_{r_{+}s_{+}} = - \lambda\,\delta_{r_{+}s_{+}},\;
\theta_{r_{-}s_{-}} = (\lambda-\kappa)\,\delta_{r_{-}s_{-}},\; \theta=\kappa,\;
  \theta_{p_{+}q_{+}r_{+}s_{+}},\ \theta_{p_{-}q_{-}r_{-}s_{-}},\ 
 \nonumber \\
 &
 \;\;
 \theta^-_{IJKL}=
 \Gamma^{IJKL}_{\dot A\dot B}\,\xi^{\dot A \dot B}
  \Big\}\;,
  \quad
  \mbox{with traceless}\;\; \xi^{\dot A\dot B}\propto\,{\rm diag}\{-7,\underbrace{1, \dots, 1}_{7}\}
\;.
\end{align}
Note that a non-vanishing $\xi^{\dot A\dot B}$ is required in order to realize the breaking of $R$-symmetry from ${\rm SO}(8)$ to ${\rm SO}(7)$,
whereas for vanishing $\xi^{\dot A\dot B}$ we are back to the case $\OSphuit$ discussed in Sec.~\ref{sec:OSp8(i)} above.

The remaining quadratic constraints  
for $\theta^-_{IJKL}$ and $\theta_{pqrs}$ are not coupled. For $\theta^-_{IJKL}$, they are given by Eqs.~\eqref{eq:QC2IJKLa} and \eqref{eq:QC2IJKrb}:
\begin{subnumcases}{\label{eq:F4(i)}}
  \Lambda_{MN}\left[\theta^-_{MNP[I}\theta^-_{JKL]P}+2\,(2\,\lambda-\kappa)\,\delta_{M[I}\theta^-_{JKL]N}\right] = 0\;, \label{eq:F4(i)a}\\
  \left(\theta_{rs}+(\kappa-\lambda)\,\delta_{rs}\right)\,\theta^-_{IJKL} = 0\;. \label{eq:F4(i)b}
\end{subnumcases}
The second line shows that non-vanishing $\xi^{\dot A \dot B}$ requires that $n_+=0$, thus
\begin{equation}
  n=n_{-}\;.
  \label{n-}
\end{equation}
We are left with Eq.~\eqref{eq:F4(i)a}, which fixes the proportionality constant in $\xi^{\dot A \dot B}$ to be 
\begin{equation}
\xi^{\dot A\dot B}=\frac{(8+n)\,\lambda}{24\,n}\,{\rm diag}\{-7,
\underbrace{1, \dots, 1}_{7}\}\;.
\end{equation}
Setting $\theta_{pqrs}=0$ solves all remaining equations, in which case the embedding tensor
is given by
\begin{align}
& \theta_{IJ}=n\,\delta_{IJ}\;,\quad \theta_{rs}=8\,\delta_{rs}\;, \quad \theta=-(8-n)\;,\nonumber\\
& \theta^-_{IJKL} = \Gamma^{IJKL}_{\dot A\dot B}\,\xi^{\dot A \dot B}, \quad \mathrm{with}\ \xi^{\dot A\dot B}=\frac{(8+n)}{24}\,{\rm diag}\{-7,\underbrace{1, \dots, 1}_{7}\}\;, 
\label{eq:F4SOn}
\end{align}
up to an arbitrary overall scaling factor.
The associated gauge group is $\SO(7)\times\SO(n)$. 
The spectrum is given in Tab.~\ref{tab:f4multchiral(i)},
organized into supermultiplets of F(4).

\begin{table}
\centering
   \begin{tabular}{c|c|c|c|c|c}
    $\Delta_{L}$ & $\Delta_{R}$ &  $\Delta$ & $s$ & $\mathrm{SO}(7)$ & $\mathrm{G_{ext}}$ \\ \hline
    \multirow{3}{*}{$\nicefrac{5}{6}$} & $\nicefrac{4}{3}$ & $\nicefrac{13}{6}$ & $\nicefrac{1}{2}$ & $\boldsymbol{1}$ & $\boldsymbol{n}$ \\
     & $\nicefrac{5}{6}$ & $\nicefrac{5}{3}$ & $0$ & $\boldsymbol{8}$ & $\boldsymbol{n}$ \\
     & $\nicefrac{1}{3}$ & $\nicefrac{7}{6}$ & $\nicefrac{-1}{2}$ & $\boldsymbol{7}$ & $\boldsymbol{n}$
  \end{tabular}
  \caption{Mass spectrum for ${\rm F(4)}$ solutions with chiral embedding \eqref{eq:splitQC1i}. The gauge groups are of the form $\SO(7)\times\mathrm{G_{ext}}$.}
  \label{tab:f4multchiral(i)}
\end{table}

Finally, we consider the possibility of non-vanishing $\theta_{pqrs}$. 
The equations for $\theta_{pqrs}$ are the same as in Sec.~\ref{sec:OSp8(i)} above, and so are the solutions,
with the only difference that Eq.~(\ref{n-}) restricts to vanishing $n_+$\,.
As in the $\OSphuit$ case, a non-vanishing $\theta_{pqrs}$ does not affect the spectrum of the theories
which is still given by Tab.~\ref{tab:f4multchiral(i)}. Rather, it will restrict the $\SO(n)$ factor of the gauge group
to some subgroup.
Without repeating the details of the derivation, in the rest of this section, we simply list the different solutions for 
non-vanishing $\theta_{pqrs}$,  organized by the different values for $n$.

\paragraph*{$\boldsymbol{n=4}$}
\begin{align}
&\theta_{IJ}=4\,\delta_{IJ}\;,\quad \theta_{rs}=8\,\delta_{rs}\;,\quad\theta=-4\;,\quad \theta_{pqrs} =12\,\xi\,\varepsilon_{pqrs}\;,\nonumber\\
& \theta^-_{IJKL} = \Gamma^{IJKL}_{\dot A\dot B}\,\xi^{\dot A \dot B}, \quad \mathrm{with}\ \xi^{\dot A\dot B}=\frac{1}{2}\,{\rm diag}\{-7,\underbrace{1, \dots, 1}_{7}\}. \label{eq:F4SO3}
\end{align}
If $|\xi|\neq1$, the gauge group is $\mathrm{SO}(7)\times\mathrm{SO}(4)$, as in the case $\xi=0$. Otherwise, the gauge group is $\mathrm{SO}(7)\times\mathrm{SO}(3)_{\pm}$, \textit{i.e.} $\mathrm{SO}(4)=\mathrm{SO}(3)_{+}\times \mathrm{SO}(3)_{-}$ is reduced to one of its chiral factors.

\paragraph*{$\boldsymbol{n=6}$}
\begin{align}
&\theta_{IJ}=6\,\delta_{IJ}\;,\quad\theta_{rs}=8\,\delta_{rs}\;,\quad\theta=-2\;,\quad
\theta_{pqrs} =\frac{1}{2}\,\varepsilon_{pqrsuv}\xi^{uv},\quad\xi_{uv}=-14
                            \begin{pmatrix}
                                                        -\sigma & 0 & 0\\
                                                        0 &\sigma&  0\\
                                                        0 & 0 & \sigma
                                                        \end{pmatrix}, \nonumber \\
& \theta^-_{IJKL} = \Gamma^{IJKL}_{\dot A\dot B}\,\xi^{\dot A \dot B}, \ \mathrm{with}\ \xi^{\dot A\dot B}=\frac{7}{12}\,{\rm diag}\{-7,\underbrace{1, \dots, 1}_{7}\}\;, \label{eq:F4U3}
\end{align}
where $\sigma$ was introduced in Eq.~\eqref{eq:OSp8n6basis}.
The gauge group is $\mathrm{SO}(7)\times\mathrm{U}(3)$. 

\paragraph{$\boldsymbol{n=7}$}
\begin{align}
&
\theta_{IJ}=7\,\delta_{IJ}\;,\quad\theta_{rs}=8\,\delta_{rs}\;,\quad
\theta=-1\;,\quad  \theta_{pqrs} =\frac{5}{4}\,\varepsilon_{pqrsuvw}\,\omega^{uvw}\;,\nonumber\\
& \theta^-_{IJKL} = \Gamma^{IJKL}_{\dot A\dot B}\,\xi^{\dot A \dot B}, \ \mathrm{with}\ \xi^{\dot A\dot B}=\frac{5}{8}\,{\rm diag}\{-7,\underbrace{1, \dots, 1}_{7}\}\;, \label{eq:F4G2}
\end{align}
where the G$_2$ invariant three-form $\omega^{uvw}$ was introduced in Eq.~\eqref{eq:OSp8G2}.
The gauge group is $\mathrm{SO}(7)\times\mathrm{G}_{2}$. 

\paragraph{$\boldsymbol{n=8}$}
\begin{align}
&\theta_{IJ}=8\,\delta_{IJ}\;,\quad\theta_{rs}=8\,\delta_{rs}\;,
\quad\theta=0\;, \quad \theta_{pqrs} = \Gamma^{pqrs}_{ab}\,\tilde\xi^{ab}\;,\nonumber\\
& \mathrm{with}\;\; \tilde\xi^{ab}=\frac{2}{4-p}\,{\rm diag}\{\underbrace{8-p, \dots, 8-p}_{p},
\underbrace{-p, \dots, -p}_{8-p}\}\;,\quad p\neq4, \nonumber \\
& \theta^-_{IJKL} = \Gamma^{IJKL}_{\dot A\dot B}\,{\xi}^{\dot A \dot B}, \ \mathrm{with}\ {\xi}^{\dot A\dot B}=\frac{2}{3}\,{\rm diag}\{-7,\underbrace{1, \dots, 1}_{7}\}\;, \label{eq:F4SOp}
\end{align}
with products of $\SO(8)$ $\Gamma$-matrices $\Gamma^{pqrs}_{ab}$. There is an analogous solution for anti-selfdual choice of $\theta_{pqrs}$.
The gauge group is $\mathrm{SO}(7)\times\mathrm{SO}(8-p)\times\mathrm{SO}(p)$.

\subsubsection{Diagonal embedding}
Similar to Eq.~\eqref{eq:SO8diag},
the $R$-symmetry group $\SO(7)$ could in principle be embedded diagonally into $\SO(8,n)$, but 
unlike for $\OSphuit$
this does not give any new solution.

\subsection[\texorpdfstring{$\SUquatre$}{SU(4|1,1)}]{\texorpdfstring{$\boldsymbol{\SUquatre}$}{SU(4|1,1)}}

\subsubsection{Chiral embedding}

The supergroup \SUquatre has an  $R$-symmetry group $\SO(6)\times\SO(2)$. We first consider its chiral embedding  
 into the first factor of ${\rm SO}(8)\times {\rm SO}(n)\subset {\rm SO}(8,n)$. The components of the embedding tensor are then given by
\begin{align}
\Big\{ & \theta_{IJ}= \lambda\,\delta_{IJ},\
\theta_{r_{+}s_{+}} = - \lambda\,\delta_{r_{+}s_{+}},\
\theta_{r_{-}s_{-}} = (\lambda-\kappa)\,\delta_{r_{-}s_{-}},\
 \theta=\kappa,\
 \theta_{p_{+}q_{+}r_{+}s_{+}},\ \theta_{p_{-}q_{-}r_{-}s_{-}} \nonumber \\
 &\quad\theta^-_{IJKL} = \Gamma^{IJKL}_{\dot A\dot B}\,\xi^{\dot A \dot B}\Big\},
   \quad
  \mbox{with traceless}\;\; \xi^{\dot A\dot B}\propto\,{\rm diag}\{-6,-6,
\underbrace{2, \dots, 2}_{6}\}
\;.
\end{align}
The non-vanishing equations of the second quadratic constraint are Eqs.~\eqref{eq:QC2IJKL}, \eqref{eq:QC2IJKrb}, \eqref{eq:QC2IJrsa}, \eqref{eq:QC2IJrsb}, \eqref{eq:QC2Ipqrc}, \eqref{eq:QC2pqrsa} and \eqref{eq:QC2pqrsc}. Eq.~\eqref{eq:QC2IJKrb} in particular implies
\begin{equation}
  \left(\theta_{rs}+(\kappa-\lambda)\,\delta_{rs}\right)\,\theta^-_{IJKL} = 0.
\end{equation}
As in the $\F(4)$ case, 
$\theta^-_{IJKL}$ is thus non vanishing only if
\begin{equation}
  n=n_{-}\;.
\end{equation}
We are thus left with the same parametrization and set of equations as in the $\F(4)$ case.
The proportionality constant in $\xi^{\dot A \dot B}$ is fixed by the remaining quadratic constraints to be
\begin{equation}
\xi^{\dot A\dot B}=\frac{(8+n)\,\lambda}{16\,n}\,{\rm diag}\{-6,-6,
\underbrace{2, \dots, 2}_{6}\}\;.
\end{equation}
Setting $\theta_{pqrs}=0$ solves all remaining equations, in which case the full embedding tensor
is given by
\begin{align}
& \theta_{IJ}=n\,\delta_{IJ}\;,\quad \theta_{rs}=8\,\delta_{rs}\;, \quad \theta=-(8-n)\;,\nonumber\\
& \theta^-_{IJKL} = \Gamma^{IJKL}_{\dot A\dot B}\,\xi^{\dot A \dot B}, \quad \mathrm{with}\ \xi^{\dot A\dot B}=\frac{(8+n)}{16}\,{\rm diag}\{-6,-6,
\underbrace{2, \dots, 2}_{6}\}\;, \label{eq:SU411SOn}
\end{align}
up to an arbitrary overall scaling factor.
The associated gauge group is $\SO(6)\times\SO(2)\times\SO(n)$. 
The spectrum is given in Tab.~\ref{tab:su411multchiral(i)},
organized into supermultiplets of \SUquatre.

\begin{table}
\centering
   \begin{tabular}{c|c|c|c|c|c}
    $\Delta_{L}$ & $\Delta_{R}$ &  $\Delta$ & $s$ & $\SO(6)\times\U(1)$ & $\mathrm{G_{ext}}$ \\ \hline
    \multirow{3}{*}{$1$} & $\nicefrac{3}{2}$ & $\nicefrac{5}{2}$ & $\nicefrac{1}{2}$ & $\boldsymbol{1}^{+2}\oplus\boldsymbol{1}^{-2}$ & $\boldsymbol{n}$ \\
     & $1$ & $2$ & $0$ & $\boldsymbol{4}^{+1}\oplus\boldsymbol{\bar{4}}^{-1}$ & $\boldsymbol{n}$ \\
     & $\nicefrac{1}{2}$ & $\nicefrac{3}{2}$ & $\nicefrac{-1}{2}$ & $\boldsymbol{6}^{0}$ & $\boldsymbol{n}$
  \end{tabular}
  \caption{Mass spectrum for the \SUquatre solutions with chiral embedding \eqref{eq:splitQC1i}. The gauge groups are of the form $\SO(6)\times\U(1)\times\mathrm{G_{ext}}$.}
  \label{tab:su411multchiral(i)}
\end{table}

In the remainder of this section, in analogy to the $\OSphuit$ case, we 
list the different solutions for non-vanishing $\theta_{pqrs}$, organized by the different values for $n$.
The spectrum of these theories is still given by Tab.~\ref{tab:su411multchiral(i)}.

\paragraph*{$\boldsymbol{n=4}$}
\begin{align}
&\theta_{IJ}=4\,\delta_{IJ}\;,\quad \theta_{rs}=8\,\delta_{rs}\;,\quad\theta=-4\;,\quad \theta_{pqrs} =12\,\xi\,\varepsilon_{pqrs}\;,\nonumber\\
& \theta^-_{IJKL} = \Gamma^{IJKL}_{\dot A\dot B}\,\xi^{\dot A \dot B}, \quad \mathrm{with}\ \xi^{\dot A\dot B}=\frac{3}{4}\,{\rm diag}\{-6,-6,
\underbrace{2, \dots, 2}_{6}\}. \label{eq:SU411SO3}
\end{align}
If $|\xi|\neq 1$ the gauge group is $\SO(6)\times\SO(2)\times\SO(4)$, as in the case $\xi=0$. Otherwise, the gauge group is 
$\SO(6)\times\SO(2)\times\SO(3)_{\pm}$, \textit{i.e.} $\mathrm{SO}(4)=\mathrm{SO}(3)_{+}\times \mathrm{SO}(3)_{-}$ 
is reduced to one of its chiral factors. 

\paragraph*{$\boldsymbol{n=6}$}
\begin{align}
&\theta_{IJ}=6\,\delta_{IJ}\;,\ \theta_{rs}=8\,\delta_{rs}\;,\ \theta=-2\;,\
\theta_{pqrs} =\frac{1}{2}\,\varepsilon_{pqrsuv}\xi^{uv},\quad\xi_{uv}=-14
                            \begin{pmatrix}
                                                        -\sigma & 0 & 0\\
                                                        0 &\sigma&  0\\
                                                        0 & 0 & \sigma
                                                        \end{pmatrix}, \nonumber \\
& \theta^-_{IJKL} = \Gamma^{IJKL}_{\dot A\dot B}\,\xi^{\dot A \dot B}, \ \mathrm{with}\ \xi^{\dot A\dot B}=\frac{7}{8}\,{\rm diag}\{-6,-6,
\underbrace{2, \dots, 2}_{6}\}\;, \label{eq:SU411U3}
\end{align}
where $\sigma$ was introduced in Eq.~\eqref{eq:OSp8n6basis}.
The gauge group is $\SO(6)\times\SO(2)\times\U(3)$.

\paragraph{$\boldsymbol{n=7}$}
\begin{align}
&
\theta_{IJ}=7\,\delta_{IJ}\;,\quad\theta_{rs}=8\,\delta_{rs}\;,\quad
\theta=-1\;,\quad  \theta_{pqrs} =\frac{5}{4}\,\varepsilon_{pqrsuvw}\,\omega^{uvw}\;,\nonumber\\
& \theta^-_{IJKL} = \Gamma^{IJKL}_{\dot A\dot B}\,\xi^{\dot A \dot B}, \ \mathrm{with}\ \xi^{\dot A\dot B}=\frac{15}{16}\,{\rm diag}\{-6,-6,
\underbrace{2, \dots, 2}_{6}\}\;, \label{eq:SU411G2}
\end{align}
where the G$_2$ invariant three-form $\omega^{uvw}$ was introduced in Eq.~\eqref{eq:OSp8G2}.
The gauge group is $\SO(6)\times\SO(2)\times\G_{2}$.

\paragraph{$\boldsymbol{n=8}$}
\begin{align}
&\theta_{IJ}=8\,\delta_{IJ}\;,\quad\theta_{rs}=8\,\delta_{rs}\;,\quad
\theta=0\;,\quad \theta_{pqrs} = \Gamma^{pqrs}_{ab}\,\tilde\xi^{ab},\nonumber\\
& \mathrm{where}\;\; \tilde\xi^{ab}=\frac{2}{4-p}\,{\rm diag}\{\underbrace{8-p, \dots, 8-p}_{p},
\underbrace{-p, \dots, -p}_{8-p}\}\;,\quad p\neq4, \nonumber \\
& \theta^-_{IJKL} = \Gamma^{IJKL}_{\dot A\dot B}\,{\xi}^{\dot A \dot B}, \ \mathrm{with}\ {\xi}^{\dot A\dot B}={\rm diag}\{-6,-6,
\underbrace{2, \dots, 2}_{6}\}\;, \label{eq:SU411SOp}
\end{align}
with products of $\SO(8)$ $\Gamma$-matrices $\Gamma^{pqrs}_{ab}$. There is an analogous solution for anti-selfdual choice of $\theta_{pqrs}$.
The gauge group is $\SO(6)\times\SO(2)\times\SO(8-p)\times\SO(p)$.

\subsubsection{Diagonal embedding}

Alternatively, the $R$-symmetry group $\SO(6)\times\SO(2)$ or one of its factors can be diagonally embedded 
into $\SO(8)\times\SO(n)\subset\SO(8,n)$ for $n\geq2$. 
The non-vanishing components of the embedding tensor are given by
\begin{align}
\Big\{ & \theta_{IJ}= \lambda\,\delta_{IJ},\
\theta_{r_{+}s_{+}} = - \lambda\,\delta_{r_{+}s_{+}},\
\theta_{r_{-}s_{-}} = (\lambda-\kappa)\,\delta_{r_{-}s_{-}},\
 \theta=\kappa, \nonumber \\
 &\theta^-_{IJKL},\ \theta_{IJ r_{+}s_{+}},\  \theta_{p_{+}q_{+}r_{+}s_{+}},\ \theta_{p_{-}q_{-}r_{-}s_{-}}\Big\}.
\end{align}
The set of non-trivial constraints which follow from the second quadratic constraint is the same as the one given for the chiral embedding. 
The new solutions are listed below. They are all defined in terms of the matrices 
\bea
 \xi^{\dot A\dot B}={\rm diag}\{-6,-6,\underbrace{2, \dots, 2}_{6}\}\;, 
\quad 
\tilde{\xi}^{\dot A\dot B}=\begin{pmatrix}
                        \sigma & 0 & 0 & 0\\
                       0 & 0 & 0 & 0\\
                       0 & 0 & 0 & 0 \\
                       0 & 0 & 0 &0  
                       \end{pmatrix}
\;,
\eea
defined in accordance with the embedding \eqref{eq:splitQC1i}.
The $2\times2$ matrix $\sigma$ has been introduced in Eq.~\eqref{eq:OSp8n6basis}.

\paragraph*{$\boldsymbol{n_{+}=8}$ (1)}
The first new solution with $n_{+}=8$ and $\SO(6)\times\SO(2)$ entirely embedded in $\SO(8)\times\SO(8)\subset\SO(8,8)$ has the form
 \begin{align}
&\theta=-16\;,\quad \theta_{pqrs} = \Gamma^{pqrs}_{\dot A\dot B}\,\xi^{\dot A\dot B}\;,\quad
\theta^-_{IJKL} = \Gamma^{IJKL}_{\dot A\dot B}\,\xi^{\dot A \dot B}\;, \quad
\theta_{IJrs} = \pm \,4\,\Gamma^{IJ}_{\dot A\dot B}\tilde{\xi}^{\dot A\dot B}\,\Gamma^{rs}_{\dot C\dot D}\tilde{\xi}^{\dot C\dot D}\;.
\label{eq:SU411U4}
\end{align}
The gauge group is $\U(4,4)$. The spectrum is given in Tab.~\ref{tab:su411multchiral(i)U4}.

\begin{table}
\centering
   \begin{tabular}{c|c|c|c|c|c}
    $\Delta_{L}$ & $\Delta_{R}$ &  $\Delta$ & $s$ & $\SO(6)\times\U(1)$ & $\SO(6)\times\U(1)$ \\ \hline
    \multirow{4}{*}{$\nicefrac{3}{2}$} & $2$ & $\nicefrac{7}{2}$ & $\nicefrac{1}{2}$ & $\boldsymbol{1}^{+4}\oplus\boldsymbol{1}^{-4}$ & $\boldsymbol{4}^{Q}$ \\
     & $\nicefrac{3}{2}$ & $3$ & $0$ & $\boldsymbol{\bar{4}}^{+3}\oplus\boldsymbol{4}^{-3}$ & $\boldsymbol{4}^{Q}$ \\
     & $1$ & $\nicefrac{5}{2}$ & $\nicefrac{-1}{2}$ & $\boldsymbol{6}^{+2}\oplus\boldsymbol{1}^{0}\oplus\boldsymbol{1}^{0}\oplus\boldsymbol{6}^{-2}$ & $\boldsymbol{4}^{Q}$ \\
     & $\nicefrac{1}{2}$ & $2$ & $-1$ & $\boldsymbol{4}^{+1}\oplus\boldsymbol{\bar{4}}^{-1}$ & $\boldsymbol{4}^{Q}$ 
  \end{tabular}
  \caption{Mass spectrum for the \SUquatre solution with diagonal embedding \eqref{eq:splitQC1i} with gauge group $\U(4,4)$.}
  \label{tab:su411multchiral(i)U4}
\end{table}

\paragraph*{$\boldsymbol{n_{+}=8}$ (2)}
The second solution with $n_{+}=8$ and $\SO(6)\times\SO(2)$ entirely embedded in $\SO(8)\times\SO(8)\subset\SO(8,8)$ is
 \begin{align}
&\theta=-16\;,\quad \theta_{pqrs} = \Gamma^{pqrs}_{\dot A\dot B}\,\xi^{\dot A\dot B}\;,\quad
\theta_{IJKL}^- = \Gamma^{IJKL}_{\dot A\dot B}\,\xi^{\dot A \dot B}\;,\nonumber\\
&
\theta_{IJrs} = 32\,\delta_{I[r}\delta_{s]J} - \Gamma^{IJrs}_{\dot A\dot B}\,\xi^{\dot A \dot B}+8\,\Gamma^{IJ}_{\dot A\dot B}\tilde{\xi}^{\dot A\dot B}\,\Gamma^{rs}_{\dot C\dot D}\tilde{\xi}^{\dot C\dot D}
\;.\label{eq:SU411SL2Sp4}
\end{align}
The gauge group is $\SL(2)\times\Sp(4,\mathbb{R})$. The  spectrum is given in Tab.~\ref{tab:su411multchiral(i)SL2Sp4}.

\begin{table}
\centering
   \begin{tabular}{c|c|c|c|c|c}
    $\Delta_{L}$ & $\Delta_{R}$ &  $\Delta$ & $s$ & $\SO(6)^{\U(1)}$ & $\U(1)$ \\ \hline
     \multirow{5}{*}{$2$} & $3$ & $5$ & $1$ & $\boldsymbol{1}^{0}\oplus\boldsymbol{1}^{0}$ & $Q$ \\
     & $\nicefrac{5}{2}$ & $\nicefrac{9}{2}$ & $\nicefrac{1}{2}$ & $\boldsymbol{4}^{+1}\oplus\boldsymbol{\bar{4}}^{-1}\oplus\boldsymbol{4}^{+1}\oplus\boldsymbol{\bar{4}}^{-1}$ & $Q$ \\
     & $2$ & $4$ & $0$ & $\boldsymbol{6}^{+2}\oplus\boldsymbol{15}^{0}\oplus\boldsymbol{15}^{0}\oplus\boldsymbol{6}^{-2}$ & $Q$ \\
     & $\nicefrac{3}{2}$ & $\nicefrac{7}{2}$ & $\nicefrac{-1}{2}$ & $\boldsymbol{\bar{4}}^{+3}\oplus\boldsymbol{20}^{+1}\oplus\boldsymbol{\bar{20}}^{-1}\oplus\boldsymbol{4}^{-3}$ & $Q$ \\
     & $1$ & $3$ & $-1$ & $\boldsymbol{\bar{10}}^{+2}\oplus\boldsymbol{10}^{-2}$ & $Q$
  \end{tabular}
  \caption{Mass spectrum for the \SUquatre solution with diagonal embedding \eqref{eq:splitQC1i} with gauge group $\SL(2)\times\Sp(4,\mathbb{R})$.}
  \label{tab:su411multchiral(i)SL2Sp4}
\end{table}

\paragraph*{$\boldsymbol{n_{+}=2}$}
Finally, there is a solution with $n_{+}=2$ and only the factor $\SO(2)$ embedded in $\SO(8)\times\SO(2+n_{-})\subset\SO(8,2+n_{-})$, given by
\begin{align}
&\theta_{IJ}=\frac{n_{-}}{2}\,\delta_{IJ}\;,\quad\theta_{r_{-}s_{-}}=5\,\delta_{r_{-}s_{-}}\;,\quad\theta_{r_{+}s_{+}}=-\frac{n_{-}}{2}\,\delta_{r_{+}s_{+}}\;,\quad\theta=n_{-}-8\;,\nonumber\\
& \theta^-_{IJKL} = \frac12\,\Gamma^{IJKL}_{\dot A\dot B}\,\xi^{\dot A \dot B}\;, \qquad
\theta_{IJr_{+}s_{+}} = \pm 4\,\Gamma^{IJ}_{\dot A\dot B}\,\tilde{\xi}^{\dot A\dot B}\,\sigma_{r_{+}s_{+}}\;,
\label{eq:SU411SO6U41}
\end{align}
where $\sigma$ was introduced in Eq.~\eqref{eq:OSp8n6basis}.
The gauge group is $\SO(n_{-})\times\U(4,1)$. The spectrum is given in Tab.~\ref{tab:su411multchiral(i)SO6U41}.

\begin{table}[!b]
\centering
   \begin{tabular}{c|c|c|c|c|c}
    $\Delta_{L}$ & $\Delta_{R}$ &  $\Delta$ & $s$ & $\SO(6)^{\U(1)}$ & $\SO(n_{-})^{\U(1)}$ \\ \hline
    \multirow{4}{*}{$\nicefrac{3}{2}$} & $2$ & $\nicefrac{7}{2}$ & $\nicefrac{1}{2}$ & $\boldsymbol{1}^{+4}\oplus\boldsymbol{1}^{-4}$ & $\boldsymbol{1}^{Q}$ \\
     & $\nicefrac{3}{2}$ & $3$ & $0$ & $\boldsymbol{\bar{4}}^{+3}\oplus\boldsymbol{4}^{-3}$ & $\boldsymbol{1}^{Q}$ \\
     & $1$ & $\nicefrac{5}{2}$ & $\nicefrac{-1}{2}$ & $\boldsymbol{6}^{+2}\oplus\boldsymbol{1}^{0}\oplus\boldsymbol{1}^{0}\oplus\boldsymbol{6}^{-2}$ & $\boldsymbol{1}^{Q}$ \\
     & $\nicefrac{1}{2}$ & $2$ & $-1$ & $\boldsymbol{4}^{+1}\oplus\boldsymbol{\bar{4}}^{-1}$ & $\boldsymbol{1}^{Q}$ \\ \hline
     \multirow{3}{*}{$1$} & $\nicefrac{3}{2}$ & $\nicefrac{5}{2}$ & $\nicefrac{1}{2}$ & $\boldsymbol{1}^{+2}\oplus\boldsymbol{1}^{-2}$ & $\boldsymbol{n_{-}}^{0}$ \\
     & $1$ & $2$ & $0$ & $\boldsymbol{4}^{+1}\oplus\boldsymbol{\bar{4}}^{-1}$ & $\boldsymbol{n_{-}}^{0}$ \\
     & $\nicefrac{1}{2}$ & $\nicefrac{3}{2}$ & $\nicefrac{-1}{2}$ & $\boldsymbol{6}^{0}$ & $\boldsymbol{n_{-}}^{0}$
  \end{tabular}
  \caption{Mass spectrum for the \SUquatre solution with diagonal embedding \eqref{eq:splitQC1i} with gauge group $\SO(n_{-})\times\U(4,1)$.}
  \label{tab:su411multchiral(i)SO6U41}
\end{table}

\subsection[\texorpdfstring{$\OSpquatre$}{OSp(4*|4)}]{\texorpdfstring{$\boldsymbol{\OSpquatre}$}{OSp(4*|4){}}}

\subsubsection{Chiral embedding}

We now turn to the last possible supergroup. \OSpquatre has $R$-symmetry group $\SO(5)\times\SO(3)$ and
we first consider its chiral embedding according to Eq.~\eqref{eq:splitQC1i} into the first factor of 
${\rm SO}(8)\times {\rm SO}(n)\subset {\rm SO}(8,n)$. The potentially non-vanishing
components of the embedding tensor are then given by
\begin{align}
\Big\{ & \theta_{IJ}= \lambda\,\delta_{IJ},\
\theta_{r_{+}s_{+}} = - \lambda\,\delta_{r_{+}s_{+}},\
\theta_{r_{-}s_{-}} = (\lambda-\kappa)\,\delta_{r_{-}s_{-}},\
 \theta=\kappa,\
  \theta_{p_{+}q_{+}r_{+}s_{+}},\ \theta_{p_{-}q_{-}r_{-}s_{-}} \;,\nonumber \\
 &\theta_{IJKL}^-= \Gamma^{IJKL}_{\dot A\dot B}\,\xi^{\dot A \dot B} ,\ \Big\}\;,\quad
  \mbox{with traceless}\;\; \xi^{\dot A\dot B}\propto\,{\rm diag}\{-5,-5,-5,
\underbrace{3, \dots, 3}_{5}\}
\;.
\end{align}
Again, the second quadratic constraint \eqref{eq:QC2} implies that $\theta^-_{IJKL}$ is non-vanishing only if
\begin{equation}
  n=n_{-}\;,
\end{equation}
which we will assume in the following.
The only difference with the case $\F(4)$ is the signature of $\xi^{\dot A \dot B}$ in $\theta^-_{IJKL}$, which is fixed by the $\SO(5)\times\SO(3)$ invariance.

Setting $\theta_{pqrs}=0$ solves all remaining equations, in which case the full embedding tensor
is given by
\begin{align}
& \theta_{IJ}=n\,\delta_{IJ}\;,\quad \theta_{rs}=8\,\delta_{rs}\;, \quad \theta=-(8-n)\;,\nonumber\\
& \theta^-_{IJKL} = \Gamma^{IJKL}_{\dot A\dot B}\,\xi^{\dot A \dot B}, \ \mathrm{with}\ \xi^{\dot A\dot B}=\frac{(8+n)}{8}\,{\rm diag}\{-5,-5,-5,
\underbrace{3, \dots, 3}_{5}\}\;, \label{eq:OSp44SOn}
\end{align}
up to an arbitrary overall scaling factor.
The gauge group is $\SO(5)\times\SO(3)\times\SO(n)$. The 
spectrum is given in Tab.~\ref{tab:osp44multchiral(i)} organized into supermultiplets of \OSpquatre.

\begin{table}
\centering
   \begin{tabular}{c|c|c|c|c|c}
    $\Delta_{L}$ & $\Delta_{R}$ &  $\Delta$ & $s$ & $\SO(6)\times\U(1)$ & $\mathrm{G_{ext}}$ \\ \hline
    \multirow{3}{*}{$\nicefrac{3}{2}$} & $2$ & $\nicefrac{7}{2}$ & $\nicefrac{1}{2}$ & $(\boldsymbol{1},\boldsymbol{3})$ & $\boldsymbol{n}$ \\
     & $\nicefrac{3}{2}$ & $3$ & $0$ & $(\boldsymbol{4},\boldsymbol{2})$ & $\boldsymbol{n}$ \\
     & $1$ & $\nicefrac{5}{2}$ & $\nicefrac{-1}{2}$ & $(\boldsymbol{5},\boldsymbol{1})$ & $\boldsymbol{n}$
  \end{tabular}
  \caption{Mass spectrum for the \OSpquatre solution with chiral embedding \eqref{eq:splitQC1i}. The gauge groups are of the form $\SO(5)\times\SO(3)\times\mathrm{G_{ext}}$.}
  \label{tab:osp44multchiral(i)}
\end{table}

In the remainder of this section, in analogy to the $\OSphuit$ case, we 
list the different solutions for non-vanishing $\theta_{pqrs}$, organized by the different values for $n$.
The spectrum of these theories is still given by Tab.~\ref{tab:osp44multchiral(i)},
with non-trivial $\theta_{pqrs}$ only reducing the $\SO(n)$ factor of the gauge group.

\paragraph*{$\boldsymbol{n=4}$}
\begin{align}
&\theta_{IJ}=4\,\delta_{IJ}\;,\quad \theta_{rs}=8\,\delta_{rs}\;,\quad\theta=-4\;,\quad \theta_{pqrs} =12\,\xi\,\varepsilon_{pqrs}\;,\nonumber\\
& \theta^-_{IJKL} = \Gamma^{IJKL}_{\dot A\dot B}\,\xi^{\dot A \dot B}, \ \mathrm{with}\ \xi^{\dot A\dot B}=\frac{3}{2}\,{\rm diag}\{-5,-5,-5,
\underbrace{3, \dots, 3}_{5}\}\;. \label{eq:OSp44SO3}
\end{align}
If $|\xi|\neq 1$ the gauge group is $\SO(5)\times\SO(3)\times\SO(4)$, as in the case $\xi=0$. Otherwise, the gauge group is 
reduced to $\SO(5)\times\SO(3)\times\SO(3)_{\pm}$, 
after factorizing $\mathrm{SO}(4)=\mathrm{SO}(3)_{+}\times \mathrm{SO}(3)_{-}$

\paragraph*{$\boldsymbol{n=6}$}
\begin{align}
&\theta_{IJ}=6\,\delta_{IJ}\;,\quad\theta_{rs}=8\,\delta_{rs}\;,\quad\theta=-2\;,\quad
 \theta_{pqrs} =\frac{1}{2}\varepsilon_{pqrsuv}\xi^{uv},\quad\xi_{uv}=
                            -14\,\begin{pmatrix}
                                                        -\sigma & 0 & 0\\
                                                        0 &\sigma&  0\\
                                                        0 & 0 & \sigma
                                                        \end{pmatrix}, \nonumber \\
& \theta^-_{IJKL} = \Gamma^{IJKL}_{\dot A\dot B}\,\xi^{\dot A \dot B}\;, \quad \mathrm{with}\quad 
\xi^{\dot A\dot B}=\frac{7}{4}\,{\rm diag}\{-5,-5,-5,
\underbrace{3, \dots, 3}_{5}\}\;,
\label{eq:OSp44U3}
\end{align}
where $\sigma$ was introduced in Eq.~\eqref{eq:OSp8n6basis}.
The gauge group is $\SO(5)\times\SO(3)\times\U(3)$.

\paragraph{$\boldsymbol{n=7}$}
\begin{align}
&
\theta_{IJ}=7\,\delta_{IJ}\;,\quad\theta_{rs}=8\,\delta_{rs}\;,\quad
\theta=-1\;,\quad  \theta_{pqrs} =\frac{5}{4}\,\varepsilon_{pqrsuvw}\,\omega^{uvw}\;,\nonumber\\
&\theta^-_{IJKL} = \Gamma^{IJKL}_{\dot A\dot B}\,\xi^{\dot A \dot B}, \ \mathrm{with}\ \xi^{\dot A\dot B}=\frac{15}{8}\,{\rm diag}\{-5,-5,-5,
\underbrace{3, \dots, 3}_{5}\}\;, \label{eq:OSp44G2}
\end{align}
where the G$_2$ invariant three-form $\omega^{uvw}$ was introduced in Eq.~\eqref{eq:OSp8G2}.
The gauge group is $\SO(5)\times\SO(3)\times\G_{2}$.

\paragraph{$\boldsymbol{n=8}$}
\begin{align}
&\theta_{IJ}=8\,\delta_{IJ}\;,\quad\theta_{rs}=8\,\delta_{rs}\;,\quad
\theta=0\;,\quad \theta_{pqrs} = \Gamma^{pqrs}_{ab}\,\tilde\xi^{ab},\nonumber\\
& \mathrm{where}\;\; \tilde\xi^{ab}=\frac{2}{4-p}\,{\rm diag}\{\underbrace{8-p, \dots, 8-p}_{p},
\underbrace{-p, \dots, -p}_{8-p}\},\quad p\neq4\;, \nonumber \\
& \theta^-_{IJKL} = \Gamma^{IJKL}_{\dot A\dot B}\,\xi^{\dot A \dot B}, \ \mathrm{with}\ \xi^{\dot A\dot B}=2\,{\rm diag}\{-5,-5,-5,
\underbrace{3, \dots, 3}_{5}\}\;,  \label{eq:OSp44SOp}
\end{align}
with products of $\SO(8)$ $\Gamma$-matrices $\Gamma^{pqrs}_{ab}$. There is an analogous solution for anti-selfdual choice of $\theta_{pqrs}$.
The gauge group is $\SO(5)\times\SO(3)\times\SO(8-p)\times\SO(p)$.

\subsubsection{Diagonal embedding}

Alternatively, for $n\geq3$ we may consider embedding of the $R$-symmetry group 
$\SO(5)\times\SO(3)$ of \OSpquatre 
or one of its factors diagonally into $\SO(8)\times\SO(n)\subset\SO(8,n)$. 
The potentially non-vanishing components of the embedding tensor are 
\begin{align}
\Big\{ & \theta_{IJ}= \lambda\,\delta_{IJ},\ \theta_{r_{+}s_{+}} = - \lambda\,\delta_{r_{+}s_{+}},\
\theta_{r_{-}s_{-}} = (\lambda-\kappa)\,\delta_{r_{-}s_{-}},\ 
\theta=\kappa, \nonumber \\
 &\theta^-_{IJKL},\ \theta_{IJ r_{+}s_{+}},\theta_{p_{+}q_{+}r_{+}q_{+}}\Big\}.
\end{align}
 The non vanishing equations given by the second quadratic constraint are the same as in the \SUquatre case. There is only one new solution, for $n_{+}=8$ and $\SO(5)\times\SO(3)$ entirely embedded into $\SO(8)\times\SO(8)\subset\SO(8,8)$, given by
 \begin{align}
&\theta=-\frac{1}{2}\;,\;\; 
\theta_{pqrs} = \Gamma^{pqrs}_{\dot A\dot B}\,\xi^{\dot A\dot B}\;,
\;\;
\theta^-_{IJKL} = \Gamma^{IJKL}_{\dot A\dot B}\,\xi^{\dot A \dot B},
\;\; 
\mathrm{with}\ \xi^{\dot A\dot B}=\frac{1}{16}\,{\rm diag}\{-5,-5,-5,
\underbrace{3, \dots, 3}_{5}\}\;, \nonumber \\
& \theta_{IJrs} =  -\frac14\,\Gamma^{IJ}_{\dot\alpha\dot\beta}\Gamma^{rs}_{\dot\alpha\dot\beta} \;,\quad
\dot\alpha,\dot\beta\in\{1, 2, 3\}\;,
\label{eq:OSp44USp44}
\end{align}
where the sum in the last line runs over the $\SO(3)$ vector according to the decomposition of the
cospinor $\dot{A}$ into the vector of $\SO(3)\times\SO(5)$\,.
The gauge group is $\Sp(2,2)\times\SO(3)$. The  spectrum is given in Tab.~\ref{tab:osp44multchiral(i)USp4}.

\begin{table}
\centering
   \begin{tabular}{c|c|c|c|c|c}
    $\Delta_{L}$ & $\Delta_{R}$ &  $\Delta$ & $s$ & $\SO(5)\times\SO(3)$ & $\SO(5)$ \\ \hline
    \multirow{4}{*}{$2$} & $\nicefrac{5}{2}$ & $\nicefrac{9}{2}$ & $\nicefrac{1}{2}$ & $(\boldsymbol{1},\boldsymbol{4})$ & $\boldsymbol{4}$ \\
     & $2$ & $4$ & $0$ & $(\boldsymbol{4},\boldsymbol{3})$ & $\boldsymbol{4}$ \\
     & $\nicefrac{3}{2}$ & $\nicefrac{7}{2}$ & $\nicefrac{-1}{2}$ & $(\boldsymbol{1},\boldsymbol{2})\oplus(\boldsymbol{5},\boldsymbol{2})$ & $\boldsymbol{4}$ \\
     & $1$ & $3$ & $-1$ & $(\boldsymbol{4},\boldsymbol{1})$ & $\boldsymbol{4}$
  \end{tabular}
  \caption{Mass spectrum for the \OSpquatre solution for diagonal embedding \eqref{eq:splitQC1i} with gauge group $\Sp(2,2)\times\SO(3)$.}
  \label{tab:osp44multchiral(i)USp4}
\end{table}



\section{Solutions with reducible vector embedding \eqref{eq:splitQC1ii}}
\label{sec:caseii}


We now consider the second possible embedding \eqref{eq:splitQC1ii} of the $R$-symmetry 
$\SO(p)\times\SO(q)$ into $\SO(8,n)$ according to which the $\SO(8)$ vector breaks into the vector of $\SO(p)\times\SO(q)$. We denote it as
\bea
I\longrightarrow  \{i, {\alpha}\} 
  \;,\quad
  i\in[\![1,p]\!]\;,\quad \alpha\in[\![p+1,8]\!]
\;.
\label{eq:splitQC1ii0}
\eea
The only relevant supergroups are $\F(4)$, \SUquatre and \OSpquatre, as for $\OSphuit$ both embeddings 
\eqref{eq:splitQC1i}, \eqref{eq:splitQC1ii}  are equivalent.
In this case, the potentially non-vanishing components of the embedding tensor are then of the form
\begin{equation}
\left\{\theta_{ij}=\lambda\,\delta_{ij},\; \theta_{\alpha\beta}=\frac{\theta_{rr}-p\,\lambda}{q}\,\delta_{\alpha\beta},\; \theta_{rs},\; 
\theta\equiv\kappa,\;\theta^-_{IJKL},\; \theta_{IJrs},\; \theta_{Ipqr},\; \theta_{pqrs}\right\}\;,
\label{eq:param2}
\end{equation}
with anti-selfdual $\theta^-_{IJKL}$\,.
The fact, that the embedding tensor is singlet under the respective $R$-symmetry group will pose further constraints
on these components that we shall evaluate case by case in the following.

With the parametrization \eqref{eq:param2},
the first quadratic constraint \eqref{eq:QC1}, with free index values $({\cal M,N})=(i,\alpha)$
and depending on the different $\Lambda$ parameters,
gives the following set of independent equations:
\begin{subnumcases}{\label{eq:QC1gen}}
  \big(8\,\lambda-\theta_{rr}\big) \big(\theta_{rr}-(p-q)\,\lambda-q\,\kappa\big) = 0\;,\label{eq:QC1gena}\\
  \left(8\,\lambda-\theta_{rr}\right)\,\theta^-_{ij\alpha\beta} = 0\;, \label{eq:QC1genb}\\
  \left(8\,\lambda-\theta_{rr}\right)\,\theta^-_{\alpha ijk} = 0\;, \label{eq:QC1genc}\\
  \left(8\,\lambda-\theta_{rr}\right)\,\theta^-_{i\alpha\beta\gamma} = 0\;. \label{eq:QC1gend}
\end{subnumcases}
Eq.~\eqref{eq:QC1gena} leaves two options, imposing either of the two factors to vanish. 
Setting $\theta_{rr} = 8\,\lambda$ however implies that $\theta_{IJ}=\lambda\,\delta_{IJ}$,
\textit{i.e.}\ we go back to the case of an irreducible vector embedding \eqref{eq:splitQC1i} 
with the parametrization \eqref{eq:thetafullparam} 
carried out in Sec.~\ref{sec:casei}\footnote{
Strictly speaking, $\theta_{IJ}=\lambda\,\delta_{IJ}$ could also be realized with an embedding \eqref{eq:splitQC1i} 
in which case the breaking~\eqref{eq:splitQC1ii0} of the vector representation would only be visible on the
components of $\theta_{\cal KLMN}$\,.
However, the remaining quadratic constraints rule out this possibility.
}.
We thus set for the rest of this section
\bea
 \theta_{rr} &=& (p-q)\,\lambda+q\,\kappa
 \;,
 \eea
such that Eqs.~\eqref{eq:QC1gen} together with the anti-selfduality of $\theta^-_{IJKL}$ imply that
\bea
 \theta^-_{IJKL}&=&0\;.
\eea
We now focus on the remaining components of the first quadratic constraint \eqref{eq:QC1}. With indices $({\cal M,N})=(i,r)$ and $({\cal M,N})=(\alpha,r)$, 
it gives the independent equations:
\begin{subnumcases}{\label{eq:QC1ip}}
  \theta_{rp} \,  \theta_{ps}+ \kappa\, \theta_{rs} + \lambda\, (\kappa-\lambda)\, \delta_{rs} = 0, \label{eq:QC1ipa}\\
  \lambda\,\theta_{ijrs} + \theta_{ijrp}\theta_{ps} = 0, \label{eq:QC1ipb}\\
  \lambda\,\theta_{ipqr} + \theta_{ipqs}\theta_{sr} = 0.\label{eq:QC1ipc} \\
  (\kappa-\lambda)\,\theta_{\alpha\beta rs} + \theta_{\alpha\beta rp}\theta_{ps} = 0, \label{eq:QC1ipd}\\
  (\kappa-\lambda)\,\theta_{\alpha pqr} + \theta_{\alpha pqs}\theta_{sr} = 0, \label{eq:QC1ipe}\\
  \lambda\,\theta_{\alpha irs} + \theta_{\alpha irp}\theta_{ps} = 0, \label{eq:QC1ipf}\\
  (\kappa-\lambda)\,\theta_{\alpha irs} + \theta_{\alpha irp}\theta_{ps} = 0. \label{eq:QC1ipg}
\end{subnumcases}
As in Eq.~\eqref{eq:QC1Ipa(i)} above, Eq.~\eqref{eq:QC1ipa} implies that we can take $\theta_{rs}$ to be a diagonal matrix with eigenvalues
\begin{equation}
\theta_{+}= -\lambda\;,\qquad \theta_{-}=\lambda-\kappa,
\end{equation}
with multiplicities that, as above, we denote as $n_\pm$. Tracelessness of $\theta_{{\cal MN}}$ then implies that
\begin{equation}
  (p-q+n_+-n_-)\,\lambda = -(n_{-}+q)\,\kappa\,. \label{eq:traceii}
\end{equation}
Finally, the component $({\cal M,N})=(r,s)$ of the first quadratic constraint is unchanged compared to the previous section (see Eqs.~\eqref{Mink0})
and together with Eqs.~\eqref{eq:QC1ip} it implies that the only potentially non-vanishing components of the embedding tensor are
\begin{align}
\bigg\{&\theta_{ij}= \lambda\,\delta_{ij},\;\theta_{\alpha\beta}=-(\lambda-\kappa)\,\delta_{\alpha\beta},\;\theta_{r_{+}s_{+}}=-\lambda\,\delta_{r_{+}s_{+}},\;\theta_{r_{-}s_{-}}=(\lambda-\kappa)\,\delta_{r_{-}s_{-}},\\
&\theta=\kappa,\;\theta_{ijr_{+}s_{+}},\;\theta_{\alpha\beta r_{-}s_{-}},\;\theta_{ip_{+}q_{+}r_{+}},\;\theta_{\alpha p_{-}q_{-}r_{-}},\;\theta_{p_{+}q_{+}r_{+}s_{+}},\;\theta_{p_{-}q_{-}r_{-}s_{-}}\bigg\}.
\end{align}
In the following, we solve the second quadratic constraint~\eqref{eq:QC2} in this parametrization
for the different supergroups with chiral and diagonal embeddings, respectively. Again, for readability we defer the full set of constraint equations to App.~\ref{app:QC2}.

\subsection[F(4)]{\texorpdfstring{$\boldsymbol{\F(4)}$}{F(4)}}
\subsubsection{Chiral embedding}
We first consider the supergroup $\F(4)$, for which $p=7$ and $q=1$, so that the index splitting~\eqref{eq:splitQC1ii0} gives $i\in[\![1,7]\!]$ and $\alpha=8$. We first assume that the $R$-symmetry group $\SO(7)$ is entirely embedded into the first factor of $\SO(8)\times\SO(n)\subset\SO(8,n)$. The potentially non-vanishing components of the embedding tensor are
 \begin{align}
\bigg\{&\theta_{ij}= \lambda\,\delta_{ij},\;\theta_{88}=-(\lambda-\kappa),\;\theta_{r_{+}s_{+}}=-\lambda\,\delta_{r_{+}s_{+}},\;\theta_{r_{-}s_{-}}=(\lambda-\kappa)\,\delta_{r_{-}s_{-}},\\
&\theta=\kappa,\;\theta_{8 p_{-}q_{-}r_{-}},\;\theta_{p_{+}q_{+}r_{+}s_{+}},\;\theta_{p_{-}q_{-}r_{-}s_{-}}\bigg\}.
\end{align}
Setting all $\theta_{\cal MNPQ}$ always gives a solution to the remaining quadratic constraints. This yields an embedding tensor of the form
\begin{align}
&\theta_{ij}=(1+n_-)\,\delta_{ij}\;,\quad\theta_{88} = -(7+n_+)\;,\quad\theta=-\left(6+n_+ -n_-\right)\;,\nonumber\\
&\theta_{r_{+}s_{+}}=-(1+n_{-})\,\delta_{r_{+}s_{+}}\;,\quad\theta_{r_{-}s_{-}}=(7+n_+)\,\delta_{r_{-}s_{-}}\;. \label{eq:Tklmn=0F4}
\end{align}
The gauge group is $\SO(7,n_{+})\times\SO(1,n_{-})$. The spectrum is given in Tab.~\subref{tab:f4multchiral(ii)}.

If $\theta_{\cal MNPQ}$ is not vanishing, the set of equations given by the second quadratic constraint is composed of Eqs.~\eqref{eq:QC2IJrsa}, \eqref{eq:QC2IJrsc}, \eqref{eq:QC2Ipqr}, \eqref{eq:QC2pqrsa} and \eqref{eq:QC2pqrsb}. Eq.~\eqref{eq:QC2Ipqrc} implies
\begin{equation}
    (2\,\lambda-\kappa)\,\theta_{pqrs_{+}}=0\;,
\end{equation}
which implies that $\theta_{pqrs}$ has $\theta_{p_-q_-r_-s_-}$ as only non-vanishing components. We are then left with the following set of independent equations:
 \begin{equation}
 \begin{cases}
 (2\lambda-\kappa)\,\theta_{p_{-}q_{-}r_{-}s_{-}}= -3\,\theta_{8u_{-}[p_{-}q_{-}} \, \theta_{r_{-}s_{-}]u_{-}8} \,\;,\\
3\,\theta_{8s_{-}[p_{-}q_{-}} \, \theta_{r_{-}]u_{-}v_{-}s_{-}}-\theta_{8s_{-}u_{-}v_{-}} \,\theta_{p_{-}q_{-}r_{-}s_{-}}=3\,(2\lambda-\kappa) \left(\theta_{8u_{-}[p_{-}q_{-}}  \, \delta_{r_{-}]v_{-}}-\theta_{8v_{-}[p_{-}q_{-}}  \, \delta_{r_{-}]u_{-}}\right)\;,\\%
\theta_{t_{-}u_{-}v_{-}[p_{-}} \,\theta_{q_{-}r_{-}s_{-}]t_{-}} \,=(2\lambda-\kappa)\,\delta_{u_{-}[p_{-}} \,\theta_{q_{-}r_{-}s_{-}]v_{-}}-(2\lambda-\kappa)\,\delta_{v_{-}[p_{-}} \,\theta_{q_{-}r_{-}s_{-}]u_{-}}
\\
\qquad\qquad\qquad
\qquad\qquad\quad
{}
+\theta_{8u_{-}v_{-}[p_{-}} \,\theta_{q_{-}r_{-}s_{-}]8}\;.\label{eq:QC2F4all_equs}
 \end{cases}
 \end{equation}
 It has a non-vanishing solution for $n_{-}=3$ only, which gives the total embedding tensor
\begin{align}
&\theta_{ij}=4\,\delta_{ij}\;,\quad\theta_{88} = -(7+n_+)\;,\quad\theta_{r_{+}s_{+}}=-4\,\delta_{r_{+}s_{+}}\;,\quad\theta_{r_{-}s_{-}}=(7+n_+)\,\delta_{r_{-}s_{-}},\nonumber\\
&\theta=-\left(3+n_+\right)\;,\quad\theta_{8p_{-}q_{-}r_{-}} = \xi\, \varepsilon_{p_{-}q_{-}r_{-}}\;. \label{eq:F4SO31}
\end{align}
The gauge group is $\SO(7,n_{+})\times\SO(1,3)$, as in the $\xi=0$ case
\eqref{eq:Tklmn=0F4}. The spectrum is also the same as given in Tab.~\subref{tab:f4multchiral(ii)}.

\begin{table}
    \centering
    \subfloat[$\mathrm{SO}(7,n_{+})\times\mathrm{SO}(1,n_{-})$]{\begin{tabular}{c|c|c|c|c|c|c}
    $\Delta_{L}$ & $\Delta_{R}$ &  $\Delta$ & $s$ & $\mathrm{SO}(7)$ & $\mathrm{SO}(n_{+})$ & $\mathrm{SO}(n_{-})$ \\ \hline
    \multirow{3}{*}{$\nicefrac{1}{3}$} & $\nicefrac{4}{3}$ & $\nicefrac{5}{3}$ &$1$ & $\boldsymbol{1}$ & $\boldsymbol{1}$ & $\boldsymbol{n_{-}}$ \\
     & $\nicefrac{5}{6}$ & $\nicefrac{7}{6}$ & $\nicefrac{1}{2}$ & $\boldsymbol{8}$ & $\boldsymbol{1}$ & $\boldsymbol{n_{-}}$ \\
     & $\nicefrac{1}{3}$ & $\nicefrac{2}{3}$ & $0$ & $\boldsymbol{7}$ & $\boldsymbol{1}$ & $\boldsymbol{n_{-}}$ \\ \hline
     \multirow{3}{*}{$\nicefrac{4}{3}$} & $\nicefrac{4}{3}$ & $\nicefrac{8}{3}$ & $0$ & $\boldsymbol{1}$ & $\boldsymbol{n_{+}}$ & $\boldsymbol{1}$ \\
      & $\nicefrac{5}{6}$ & $\nicefrac{13}{6}$ & $\nicefrac{-1}{2}$ & $\boldsymbol{8}$ & $\boldsymbol{n_{+}}$ & $\boldsymbol{1}$  \\
      & $\nicefrac{1}{3}$ & $\nicefrac{5}{3}$ & $-1$ & $\boldsymbol{7}$ & $\boldsymbol{n_{+}}$ & $\boldsymbol{1}$
  \end{tabular}\label{tab:f4multchiral(ii)}
  }
  \ 
  \subfloat[$\mathrm{GL}(7)\times \mathrm{SO}(1,n_-)$]{\begin{tabular}{c|c|c|c|c|c}
    $\Delta_{L}$ & $\Delta_{R}$ &  $\Delta$ & $s$ & $\mathrm{SO}(7)$ & $\mathrm{SO}(n_-)$ \\ \hline
    \multirow{3}{*}{$\nicefrac{1}{3}$} & $\nicefrac{4}{3}$ & $\nicefrac{5}{3}$ & $1$ & $\boldsymbol{1}$ & $\boldsymbol{n_-}$ \\
     & $\nicefrac{5}{6}$ & $\nicefrac{7}{6}$ & $\nicefrac{1}{2}$ & $\boldsymbol{8}$ & $\boldsymbol{n_-}$\\
     & $\nicefrac{1}{3}$ & $\nicefrac{2}{3}$ & $0$ & $\boldsymbol{7}$ & $\boldsymbol{n_-}$\\ \hline
    \multirow{5}{*}{$\nicefrac{5}{3}$}  & $\nicefrac{8}{3}$ & $\nicefrac{13}{3}$ & $1$ & $\boldsymbol{1}$ & $\boldsymbol{1}$\\
     & $\nicefrac{13}{6}$ & $\nicefrac{23}{6}$ & $\nicefrac{1}{2}$ & $\boldsymbol{8}$ & $\boldsymbol{1}$\\
     & $\nicefrac{5}{3}$ & $\nicefrac{10}{3}$ & $0$ & $\boldsymbol{7}\oplus \boldsymbol{21}$ & $\boldsymbol{1}$\\
     & $\nicefrac{7}{6}$ & $\nicefrac{17}{6}$ & $\nicefrac{-1}{2}$ & $\boldsymbol{48}$ & $\boldsymbol{1}$\\
     & $\nicefrac{2}{3}$ & $\nicefrac{7}{3}$ & $-1$ & $\boldsymbol{27}$& $\boldsymbol{1}$
  \end{tabular}\label{tab:f4multdiag(ii)}
  }
  \caption{Mass spectra for the ${\rm F(4)}$ solutions with (a) chiral and (b) diagonal embeddings~\eqref{eq:splitQC1ii}.}
  \label{tab:f4mult(ii)}
\end{table}

\subsubsection{Diagonal embedding}
The $R$-symmetry group could also be embedded diagonally into $\SO(8)\times\SO(n)\subset\SO(8,n)$ for $n\geq7$. There is only one new solution for $n_{+}=7$, given by
\begin{align}
& \theta_{ij}=(n_-+1)\,\delta_{ij}\;,\quad\theta_{88} = -14\;,\quad
\theta_{r_{+}s_{+}}=-(n_-+1)\,\delta_{r_{+}s_{+}}\;,\quad
\theta_{r_{-}s_{-}} = 14\,\delta_{r_{-}s_{-}}\;,\nonumber\\
& \theta=(n_--13)\;,\quad \theta_{ijr_{+}s_{+}}= 2\,(n_-+15)\,\delta_{i[r_{+}}\delta_{s_{+}]j}
\;. \label{eq:F4GL7}
\end{align}
The gauge group is $\mathrm{GL}(7)\times \mathrm{SO}(1,n_-)$ and the spectrum is given in Tab.~\subref{tab:f4multdiag(ii)}.

\subsection[\texorpdfstring{$\SUquatre$}{SU(4|1,1)}]{\texorpdfstring{$\boldsymbol{\SUquatre}$}{SU(4|1,1)}}
\label{sec:SU411(ii)}

\subsubsection{Chiral embedding}
For the supergroup \SUquatre, $p=6$ and $q=2$ and the index splitting~\eqref{eq:splitQC1ii0} gives $i\in[\![1,6]\!]$ and $\alpha\in\{7,8\}$. The $R$-symmetry group $\SO(6)\times\SO(2)$ could first be embedded into the first factor of $\SO(8)\times\SO(n)\subset\SO(8,n)$. The 
potentially non-vanishing components of the embedding tensor are then given by
 \begin{align}
\bigg\{&\theta_{ij}= \lambda\,\delta_{ij},\;\theta_{\alpha\beta}=-(\lambda-\kappa)\,\delta_{\alpha\beta},\;\theta_{r_{+}s_{+}}=-\lambda\,\delta_{r_{+}s_{+}},\;\theta_{r_{-}s_{-}}=(\lambda-\kappa)\,\delta_{r_{-}s_{-}},\\
&\;\;\theta=\kappa,\;\theta_{\alpha\beta r_{-}s_{-}}=\sigma_{\alpha\beta}\,\Omega_{r_{-}s_{-}},\;\theta_{p_{+}q_{+}r_{+}s_{+}},\;\theta_{p_{-}q_{-}r_{-}s_{-}}\bigg\}\;,
\end{align}
where $\Omega_{r_{-}s_{-}}$ is an anti-symmetric matrix and $\sigma$ has been introduced in Eq.~\eqref{eq:OSp8n6basis}. A first solution is given by setting all $\theta_{\cal MNPQ}=0$. The embedding tensor then has the form
\begin{align}
  & \theta_{ij}=(2+n_{-})\,\delta_{ij}\;,\quad \theta_{\alpha\beta}=-(6+n_{+})\,\delta_{\alpha\beta}\;,\quad \theta=-(4+n_+-n_-)\;,\nonumber\\
  &\theta_{r_{+}s_{+}}=-(2+n_{-})\,\delta_{r_{+}s_{+}}\;,\quad\theta_{r_{-}s_{-}}=(6+n_+)\,\delta_{r_{-}s_{-}}  \;. \label{eq:SU411Tklmn=0}
\end{align}
The gauge group is $\SO(6,n_{+})\times\SO(2,n_{-})$. The spectrum is given in Tab.~\subref{tab:SU411multchiral(ii)a}.

\begin{table}
    \centering
    \subfloat[$\mathrm{SO}(6,n_{+})\times\mathrm{SO}(2,n_{-})$]{
    \begin{tabular}{c|c|c|c|c|c|c}
    $\Delta_{L}$ & $\Delta_{R}$ &  $\Delta$ & $s$  & $\mathrm{SO}(6)\times \mathrm{U}(1)$  & $\mathrm{SO}(n_{+})$ & $\mathrm{SO}(n_{-})$ \\ \hline
    \multirow{3}{*}{$\nicefrac{1}{2}$} & $\nicefrac{3}{2}$ & $2$ & $1$ & $\boldsymbol{1}^{+2}\oplus\boldsymbol{1}^{-2}$ & $\boldsymbol{1}$ & $\boldsymbol{n_{-}}$ \\
     & $1$ & $\nicefrac{3}{2}$ & $\nicefrac{1}{2}$ & $\boldsymbol{4}^{+1}\oplus\boldsymbol{\bar4}^{-1}$ & $\boldsymbol{1}$ & $\boldsymbol{n_{-}}$ \\
     & $\nicefrac{1}{2}$ & $1$ & $0$ & $\boldsymbol{6^{0}}$ & $\boldsymbol{1}$ & $\boldsymbol{n_{-}}$ \\ \hline
     \multirow{3}{*}{$\nicefrac{3}{2}$} & $\nicefrac{3}{2}$ & $3$ & $0$ & $\boldsymbol{1}^{+2}\oplus\boldsymbol{1}^{-2}$ & $\boldsymbol{n_{+}}$ & $\boldsymbol{1}$\\
      & $1$ & $\nicefrac{5}{2}$ & $\nicefrac{-1}{2}$ & $\boldsymbol{4}^{+1}\oplus\boldsymbol{\bar4}^{-1}$ & $\boldsymbol{n_{+}}$ & $\boldsymbol{1}$ \\
      & $\nicefrac{1}{2}$ & $2$ & $-1$ & $\boldsymbol{6^{0}}$ & $\boldsymbol{n_{+}}$ & $\boldsymbol{1}$
  \end{tabular}\label{tab:SU411multchiral(ii)a}}
  \qquad
  \subfloat[$\mathrm{SO}(6,n_{+})\times\mathrm{U}(m,1)$]{
  \begin{tabular}{c|c|c|c|c|c|c}
    $\Delta_{L}$ & $\Delta_{R}$&  $\Delta$ & $s$ & $(\mathrm{SO}(6)\times\mathrm{SU}(m))^{\mathrm{U}(1)}$ &$\mathrm{SO}(n_{+})$ & $\mathrm{U_{ext}}(1)$\\ \hline
    \multirow{5}{*}{$1$} & $2$ & $3$ & $1$ & $(\boldsymbol{1},\boldsymbol{m})^{+4}\oplus(\boldsymbol{1},\boldsymbol{\bar m})^{-4}$ & $\boldsymbol{1}$ & $0$ \\
     & $\nicefrac{3}{2}$  & $\nicefrac{5}{2}$ & $\nicefrac{1}{2}$ & $(\boldsymbol{4},\boldsymbol{m})^{+3}\oplus(\boldsymbol{\bar 4},\boldsymbol{\bar m})^{-3}$ & $\boldsymbol{1}$ & $0$\\
     & $1$ & $2$ & $0$ & $(\boldsymbol{6},\boldsymbol{m})^{+2}\oplus(\boldsymbol{6},\boldsymbol{\bar m})^{-2}$ & $\boldsymbol{1}$& $0$\\
     & $\nicefrac{1}{2}$  & $\nicefrac{3}{2}$ & $\nicefrac{-1}{2}$ & $(\boldsymbol{\bar 4},\boldsymbol{m})^{+1}\oplus(\boldsymbol{4},\boldsymbol{\bar m})^{-1}$ & $\boldsymbol{1}$& $0$\\
     & $0$ & $1$ & $-1$ & $(\boldsymbol{1},\boldsymbol{m})^{0}\oplus(\boldsymbol{1},\boldsymbol{\bar m})^{0}$ & $\boldsymbol{1}$ & $0$\\\hline
     \multirow{3}{*}{$\nicefrac{3}{2}$} & $\nicefrac{3}{2}$ & $3$ & $0$ & $(\boldsymbol{1},\boldsymbol{1})^{+2}\oplus(\boldsymbol{1},\boldsymbol{1})^{-2}$ & $\boldsymbol{n_{+}}$& $0$\\
      & $1$ & $\nicefrac{5}{2}$ & $\nicefrac{-1}{2}$ & $(\boldsymbol{4},\boldsymbol{1})^{+1}\oplus(\boldsymbol{\bar 4},\boldsymbol{\bar m})^{-1}$ & $\boldsymbol{n_{+}}$ & $0$\\
      & $\nicefrac{1}{2}$ & $2$ & $-1$ & $(\boldsymbol{6},\boldsymbol{\bar m})^{0}$ & $\boldsymbol{n_{+}}$& $0$
  \end{tabular}\label{tab:SU411multchiral(ii)b}}
  \caption{Mass spectra for ${\rm SU}(4|1,1)$ solutions for chiral embedding \eqref{eq:splitQC1ii} with (a) $\theta_{\cal MNPQ} =0$ and (b) $\theta_{\cal MNPQ} \neq0$. These are the multiplets of ${\rm SU}(|1,1)$ given in Tab.~1 of Ref.~\cite{Gunaydin:1986fe} for $m=n_{-}/2$.}
  \label{tab:SU411multchiral(ii)}
\end{table}

Let us now turn to the second quadratic constraint~\eqref{eq:QC2} with a non-vanishing $\theta_{\cal MNPQ}$. It reduces to Eqs.~\eqref{eq:QC2IJKrb}, \eqref{eq:QC2IJrsa}, \eqref{eq:QC2IJrsb}, \eqref{eq:QC2Ipqrc}, \eqref{eq:QC2pqrsa} and \eqref{eq:QC2pqrsc}. From Eq.~\eqref{eq:QC2Ipqrc}, we find
\begin{equation}
  (2\,\lambda-\kappa)\,\theta_{pqrs_{+}}=0\;,
\end{equation}
which implies once again that the only non-vanishing components of $\theta_{pqrs}$ are $\theta_{p_{-}q_{-}r_{-}s_{-}}$. The other equations are then equivalent to the following set of independent equations:
\begin{subnumcases}{\label{eq:SU411QC2}}
  (2\lambda-\kappa)\,\theta_{p_{-}q_{-}r_{-}s_{-}} = -3\, \Omega_{[p_{-}q_{-}}\Omega_{r_{-}s_{-}]}, \label{eq:SU411QC2a}\\
  \theta_{p_{-}q_{-}r_{-}s_{-}}\Omega_{s_{-}u_{-}} = 3(2\lambda-\kappa)\,\Omega_{[p_{-}q_{-}}\delta_{r_{-}]u_{-}}, \label{eq:SU411QC2b} \\
  \theta_{u_{-}v_{-}l_{-}[p_{-}}\theta_{q_{-}r_{-}s_{-}]l_{-}} = (2\lambda-\kappa)\,\left[\theta_{u_{-}[q_{-}r_{-}s_{-}}\delta_{p_{-}]v_{-}}-\theta_{v_{-}[q_{-}r_{-}s_{-}}\delta_{p_{-}]u_{-}}\right].\label{eq:SU411QC1c}
\end{subnumcases}
As $\Omega_{p_{-}q_{-}}$ is antisymmetric, there exists a basis where it has the form
\begin{equation}
  \begin{pmatrix}
      f_{1}\,\sigma &   &   \\
        &   \ddots   &   \\
    &    &f_{m}\,\sigma 
  \end{pmatrix}\quad \text{if }n_{-}=2m,\ \text{and}\
  \begin{pmatrix}
      f_{1}\,\sigma &   & &  \\
        &   \ddots   & &  \\
    &    &f_{m}\,\sigma &\\
    &&& 0
  \end{pmatrix}\quad 
  \text{if }n_{-}=2m+1\;,
\end{equation}
with scalar functions $f_1, \dots, f_m$, and $\sigma$ from Eq.~\eqref{eq:OSp8n6basis}.
With this parametrization, Eqs.~\eqref{eq:SU411QC2} imply that 
\begin{equation}
  \begin{cases}
    f_{i}f_{j}^{2}= (2\lambda-\kappa)^{2}\,f_{i}, \\
    f_{i}f_{j}f_{k}^{2}= (2\lambda-\kappa)^{2}\,f_{i}f_{j},
  \end{cases}\,i,j,k\in[\![1,m]\!]\,,\,i\neq j\neq k\;,
\end{equation}
if $n_{-}=2m$, whereas they imply that all $f_i$ vanish for $n_{-}=2m+1$.
Thus, for $\kappa\neq2\,\lambda$ there are solutions with non-vanishing $\theta_{\cal MNPQ}$ only if $n_{-}=2m$ is even, and they are given by
\begin{align}
  & \theta_{ij}=(2+n_{-})\,\delta_{ij}\;,\ \theta_{\alpha\beta}=-(6+n_{+})\,\delta_{\alpha\beta}\;,\ \theta_{r_{+}s_{+}}=-(2+n_{-})\,\delta_{r_{+}s_{+}}\;,\ \theta_{r_{-}s_{-}}=(6+n_+)\,\delta_{r_{-}s_{-}}\;, \nonumber \\
  & \theta=-(4+n_+-n_-)\;,\quad \theta_{\alpha\beta p_{-}q_{-}} = \sigma_{\alpha\beta}\,\Omega_{p_{-}q_{-}}\;,\quad
  \theta_{p_{-}q_{-}r_{-}s_{-}} = -\frac{3}{8+n}\,\Omega_{[p_{-}q_{-}}\Omega_{r_{-}s_{-}]}, \nonumber\\
  &\mathrm{with}\   \Omega_{p_{-}q_{-}}=(8+n)\,  
  \begin{pmatrix}
     \sigma &   &   \\
        &   \ddots   &   \\
   &    &\sigma 
  \end{pmatrix}
 \;. \label{eq:SU411Um}
\end{align}
The gauge group is 
$\mathrm{SO}(6,n_{+})\times\SO(2,1)$ if $n_{-}=2$
and
$\SO(6,n_{+})\times\U(n_{-}/2,1)$ otherwise. The spectrum is given in Tab.~\subref{tab:SU411multchiral(ii)b}.

\begin{table}[t]
    \centering
    \subfloat[$1\leq \vert\xi\vert$]{
    \begin{tabular}{c|c|c|c|c|c|c}
    $\Delta_{L}$ & $\Delta_{R}$&  $\Delta$ & $s$ & $\mathrm{SO}(6)^{\mathrm{U}(1)}$ &$\mathrm{SO}(n_{+})$ & $\mathrm{U_{ext}}(1)$\\ \hline
    \multirow{5}{*}{$\frac{1}{2}+\frac{\vert\xi\vert}{2}$} & $\frac{3}{2}+\frac{\vert\xi\vert}{2}$ & $2+\vert\xi\vert$ & $1$ & $\boldsymbol{1}^{+4}\oplus\boldsymbol{1}^{-4}$ & $\boldsymbol{1}$ & $Q$\\
     & $1+\frac{\vert\xi\vert}{2}$ & $\frac{3}{2}+\vert\xi\vert$ & $\nicefrac{1}{2}$ & $\boldsymbol{4}^{+3}\oplus\boldsymbol{\bar4}^{-3}$ & $\boldsymbol{1}$& $Q$\\
     & $\frac{1}{2}+\frac{\vert\xi\vert}{2}$ & $1+\vert\xi\vert$ & $0$ & $\boldsymbol{6}^{+2}\oplus\boldsymbol{6}^{-2}$ & $\boldsymbol{1}$& $Q$\\
     & $\frac{\vert\xi\vert}{2}$ & $\frac{1}{2}+\vert\xi\vert$ & $\nicefrac{-1}{2}$ & $\boldsymbol{\bar4}^{+1}\oplus\boldsymbol{4}^{-1}$ & $\boldsymbol{1}$& $Q$\\
     & $-\frac{1}{2}+\frac{\vert\xi\vert}{2}$ & $\vert\xi\vert$ & $-1$ & $\boldsymbol{1}^{0}\oplus\boldsymbol{1}^{0}$ & $\boldsymbol{1}$& $Q$\\\hline
     \multirow{3}{*}{$\nicefrac{3}{2}$} & $\nicefrac{3}{2}$ & $3$ & 0 & $\boldsymbol{1}^{+2}\oplus\boldsymbol{1}^{-2}$ & $\boldsymbol{n_{+}}$& $0$\\
      & $1$ & $\nicefrac{5}{2}$ & $\nicefrac{-1}{2}$ & $\boldsymbol{4}^{+1}\oplus\boldsymbol{\bar4}^{-1}$ & $\boldsymbol{n_{+}}$ & $0$\\
      & $\nicefrac{1}{2}$ & $2$ & $-1$& $\boldsymbol{6}^{0}$ & $\boldsymbol{n_{+}}$& $0$
  \end{tabular}}

  \subfloat[$\vert\xi\vert< 1$]{
     \begin{tabular}{c|c|c|c|c|c|c}
    $\Delta_{L}$ & $\Delta_{R}$ &  $\Delta$ & $s$ & $\mathrm{SO}(6)^{\mathrm{U}(1)}$ &$\mathrm{SO}(n_{+})$ & $\mathrm{U_{ext}}(1)$\\ \hline
     \multirow{3}{*}{$\frac{1}{2}+\frac{\vert\xi\vert}{2}$}  & $\frac{3}{2}+\frac{\vert\xi\vert}{2}$ & $2+\vert\xi\vert$ & $1$ & $\boldsymbol{1}_{+1}\oplus\boldsymbol{1}^{-2}$ & $\boldsymbol{1}$ & $Q$\\
      & $1+\frac{\vert\xi\vert}{2}$ & $\frac{3}{2}+\vert\xi\vert$ & $\nicefrac{1}{2}$ & $\boldsymbol{4}^{+1}\oplus\boldsymbol{\bar4}^{-1}$ & $\boldsymbol{1}$& $Q$\\
      & $\frac{1}{2}+\frac{\vert\xi\vert}{2}$ & $1+\vert\xi\vert$ & $0$ & $\boldsymbol{6}^{0}$ & $\boldsymbol{1}$ & $Q$\\ \hline
     \multirow{3}{*}{$\frac{1}{2}-\frac{\vert\xi\vert}{2}$} & $\frac{3}{2}-\frac{\vert\xi\vert}{2}$ & $2-\vert\xi\vert$ & $1$ & $\boldsymbol{1}^{+2}\oplus\boldsymbol{1}^{-2}$ & $\boldsymbol{1}$& $Q$\\
     & $1-\frac{\vert\xi\vert}{2}$ & $\frac{3}{2}-\vert\xi\vert$ & $\nicefrac{1}{2}$ & $\boldsymbol{4}^{+1}\oplus\boldsymbol{\bar4}^{-1}$ & $\boldsymbol{1}$& $Q$\\
     & $\frac{1}{2}-\frac{\vert\xi\vert}{2}$ & $1-\vert\xi\vert$ & $0$ & $\boldsymbol{6}^{0}$ & $\boldsymbol{1}$ & $Q$\\ \hline
     \multirow{3}{*}{$\nicefrac{3}{2}$} & $\nicefrac{3}{2}$ & $3$ & $0$ & $\boldsymbol{1}^{+2}\oplus\boldsymbol{1}^{-2}$ & $\boldsymbol{n_{+}}$& $0$\\
     & $1$ & $\nicefrac{5}{2}$ & $\nicefrac{-1}{2}$ & $\boldsymbol{4}^{+1}\oplus\boldsymbol{\bar4}^{-1}$ & $\boldsymbol{n_{+}}$& $0$\\
     & $\nicefrac{1}{2}$ & $2$ & $-1$ & $\boldsymbol{6}^{0}$ & $\boldsymbol{n_{+}}$ & $0$
  \end{tabular}}
  \caption{Mass spectra for the \SUquatre solutions for chiral embedding \eqref{eq:splitQC1ii} with $\theta_{\cal MNPQ} \neq0$ and $n_{-}=2$. The gauge group is 
  $\SO(6,n_{+})\times\SO(2,1)$ for $\vert\xi\vert=1$ and
   $\SO(6,n_{+})\times\SO(2,2)$ otherwise.}
  \label{tab:SU411multchiral(ii)2}
\end{table}

In the case $n_{-}=2$, the previous solution admits a continuous parameter $\xi$, given by
\begin{align}
  & \theta_{ij}=4\,\delta_{ij}\;,\ \theta_{\alpha\beta}=-(6+n_+)\,\delta_{\alpha\beta}\;,\ \theta_{r_{+}s_{+}}=-4\,\delta_{r_{+}s_{+}}\;,\ \theta_{r_{-}s_{-}}=(6+n_+)\,\delta_{r_{-}s_{-}}\;, \nonumber\\
  & \theta=-(2+n_+)\;,\quad\theta_{\alpha\beta p_{-}q_{-}} = (10+n_+)\,\xi\,\sigma_{\alpha\beta}\,\sigma_{p_{-}q_{-}}\;. \label{eq:SU411n2}
\end{align}
For $|\xi|\not=1$, the gauge group enhances from $\SO(6,n_{+})\times\SO(2,1)$ to 
$\SO(6,n_{+})\times\SO(2,2)$. The spectra are given in Tab.~\ref{tab:SU411multchiral(ii)2}.
Although all masses are continuous functions of the parameter $\xi$, the conformal dimensions 
are assigned differently for $|\xi|<1$ and $|\xi|\ge1$, respectively 
(within the values allowed by Eq.~\eqref{eq:confDim}),
in order to combine the different fields into supermultiplets of \SUquatre.

\begin{table}
    \centering
    \subfloat[$1\leq \vert\xi\vert$]{
     \begin{tabular}{c|c|c|c|c|c}
    $\Delta_{L}$ & $\Delta_{R}$&  $\Delta$ & $s$ & $\mathrm{SO}(6)^{\mathrm{U}(1)}$ & $\mathrm{U_{ext}}(1)$ \\ \hline
    \multirow{5}{*}{$\frac{1}{2}+\frac{\vert\xi\vert}{2}$} & $\frac{3}{2}+\frac{\vert\xi\vert}{2}$ & $2+\vert\xi\vert$ & $1$ & $\boldsymbol{1}^{+4}\oplus\boldsymbol{1}^{-4}$ & $Q$ \\
     & $1+\frac{\vert\xi\vert}{2}$ & $\frac{3}{2}+\vert\xi\vert$ & $\nicefrac{1}{2}$ & $\boldsymbol{4}^{+3}\oplus\boldsymbol{\bar4}^{-3}$  & $Q$ \\
     & $\frac{1}{2}+\frac{\vert\xi\vert}{2}$ & $1+\vert\xi\vert$ & $0$ & $\boldsymbol{6}^{+2}\oplus\boldsymbol{6}^{-2}$ & $Q$ \\
     & $\frac{\vert\xi\vert}{2}$ & $\frac{1}{2}+\vert\xi\vert$ & $\nicefrac{-1}{2}$ & $\boldsymbol{\bar4}^{+1}\oplus\boldsymbol{4}^{-1}$ & $Q$ \\
     & $-\frac{1}{2}+\frac{\vert\xi\vert}{2}$ & $\vert\xi\vert$ & $-1$ & $\boldsymbol{1}^{0}\oplus\boldsymbol{1}^{0}$  & $Q$ \\\hline
     \multirow{5}{*}{$2$} & $3$ & $5$ & $1$ & $\boldsymbol{1}^{0}$ & $0$ \\
      & $\nicefrac{5}{2}$ & $\nicefrac{9}{2}$ & $\nicefrac{1}{2}$ & $\boldsymbol{4}^{+1}\oplus\boldsymbol{\bar{4}}^{-1}$ & $0$ \\
      & $2$ & $4$ & $0$ & $\boldsymbol{6}^{+2}\oplus\boldsymbol{15}^{0}\oplus\boldsymbol{6}^{-2}$ & $0$ \\
      & $\nicefrac{3}{2}$ & $\nicefrac{7}{2}$ & $\nicefrac{-1}{2}$ & $\boldsymbol{\bar{20}}^{+1}\oplus\boldsymbol{20}^{-1}$ & $0$ \\
      & $1$ & $3$ & $-1$ & $\boldsymbol{20'}^{0}$ & $0$ 
  \end{tabular}}

  \subfloat[$\vert\xi\vert< 1$]{
     \begin{tabular}{c|c|c|c|c|c}
    $\Delta_{L}$ & $\Delta_{R}$ &  $\Delta$ & $s$ & $\mathrm{SO}(6)^{\mathrm{U}(1)}$ & $\mathrm{U_{ext}}(1)$ \\ \hline
    \multirow{3}{*}{$\frac{1}{2}+\frac{\vert\xi\vert}{2}$}  & $\frac{3}{2}+\frac{\vert\xi\vert}{2}$ & $2+\vert\xi\vert$ & $1$ & $\boldsymbol{1}_{+1}\oplus\boldsymbol{1}^{-2}$ & $Q$\\
      & $1+\frac{\vert\xi\vert}{2}$ & $\frac{3}{2}+\vert\xi\vert$ & $\nicefrac{1}{2}$ & $\boldsymbol{4}^{+1}\oplus\boldsymbol{\bar4}^{-1}$ & $Q$\\
      & $\frac{1}{2}+\frac{\vert\xi\vert}{2}$ & $1+\vert\xi\vert$ & $0$ & $\boldsymbol{6}^{0}$ & $Q$\\ \hline
     \multirow{3}{*}{$\frac{1}{2}-\frac{\vert\xi\vert}{2}$} & $\frac{3}{2}-\frac{\vert\xi\vert}{2}$ & $2-\vert\xi\vert$ & $1$ & $\boldsymbol{1}^{+2}\oplus\boldsymbol{1}^{-2}$ & $Q$\\
     & $1-\frac{\vert\xi\vert}{2}$ & $\frac{3}{2}-\vert\xi\vert$ & $\nicefrac{1}{2}$ & $\boldsymbol{4}^{+1}\oplus\boldsymbol{\bar4}^{-1}$ & $Q$\\
     & $\frac{1}{2}-\frac{\vert\xi\vert}{2}$ & $1-\vert\xi\vert$ & $0$ & $\boldsymbol{6}^{0}$ & $Q$\\ \hline
    \multirow{5}{*}{$2$} & $3$ & $5$ & $1$ & $\boldsymbol{1}^{0}$ & $0$ \\
      & $\nicefrac{5}{2}$ & $\nicefrac{9}{2}$ & $\nicefrac{1}{2}$ & $\boldsymbol{4}^{+1}\oplus\boldsymbol{\bar{4}}^{-1}$ & $0$ \\
      & $2$ & $4$ & $0$ & $\boldsymbol{6}^{+2}\oplus\boldsymbol{15}^{0}\oplus\boldsymbol{6}^{-2}$ & $0$ \\
      & $\nicefrac{3}{2}$ & $\nicefrac{7}{2}$ & $\nicefrac{-1}{2}$ & $\boldsymbol{\bar{20}}^{+1}\oplus\boldsymbol{20}^{-1}$ & $0$\\
      & $1$ & $3$ & $-1$ & $\boldsymbol{20'}^{0}$ & $0$ 
  \end{tabular}}
  \caption{Mass spectra for the \SUquatre solutions with diagonal embedding \eqref{eq:splitQC1ii}. 
  The gauge group is $\GL(6)\times\SO(2,1)$ for $|\xi|=1$  and $\GL(6)\times\SO(2,2)$ otherwise.}
  \label{tab:SU411diag(ii)8}
\end{table}

\subsubsection{Diagonal embedding}
For a diagonal embedding of $\SO(6)\times\SO(2)$ in $\SO(8)\times\SO(n)\subset\SO(8,n)$, there is one new solution for $n_{+}=6$, depending on a free parameter $\xi$:
\begin{align}
  & \theta_{ij}=(n_--+2)\,\delta_{ij}\;,\quad \theta_{\alpha\beta}=-12\,\delta_{\alpha\beta}\;,\quad \theta_{r_{+}s_{+}}=-(n_-+2)\,\delta_{r_{+}s_{+}}\;,\quad \theta_{r_{-}s_{-}}=12\,\delta_{r_{-}s_{-}}\;,\nonumber\\
  & \theta=(n_--10)\;,\quad \theta_{\alpha\beta r_{-}s_{-}} = (14+n_-)\,\xi\,\sigma_{\alpha\beta}\,\sigma_{r_{-}s_{-}}\;,\quad \theta_{ij r_{+}s_{+}} = 2\,(14+n_-)\,\delta_{i[r_{+}}\,\delta_{s_{+}]j}\;. \label{eq:SU411GL6}
\end{align}
In the case $n_-=2$, if $|\xi|\neq1$ the gauge group is  $\GL(6)\times\SO(2,1)$ for $|\xi|=1$ and enhances to 
$\GL(6)\times\SO(2,2)$ otherwise. 
The spectra are collected in Tabs.~\ref{tab:SU411diag(ii)8}.
Again, the conformal dimensions are assigned differently for $|\xi|<1$ and $|\xi|\ge1$, respectively 
(within the values allowed by Eq.~\eqref{eq:confDim}),
in order to combine the different fields into supermultiplets of \SUquatre.
For $0\leq n_-<2$, the gauge group is $\GL(6)\times\SO(2,n_-)$ with the spectrum given in
 Tab.\ \ref{tab:SU411diag(ii)}.

\begin{table}
    \centering
     \begin{tabular}{c|c|c|c|c|c}
    $\Delta_{L}$ & $\Delta_{R}$&  $\Delta$ & $s$ & $\mathrm{SO}(6)^{\mathrm{U}(1)}$ & $\SO(n_-)$ \\ \hline
    \multirow{3}{*}{$\nicefrac{1}{2}$} & $\nicefrac{3}{2}$ & $2$ & $1$ & $\boldsymbol{1}^{+2}\oplus\boldsymbol{1}^{-2}$ & $\boldsymbol{n_-}$ \\
     & $1$ & $\nicefrac{3}{2}$ & $\nicefrac{1}{2}$ & $\boldsymbol{4}^{+1}\oplus\boldsymbol{\bar4}^{-1}$  & $\boldsymbol{n_-}$\\
     & $\nicefrac{1}{2}$ & $1$ & $0$ & $\boldsymbol{6^{0}}$  & $\boldsymbol{n_-}$\\ \hline
     \multirow{5}{*}{$2$} & $3$ & $5$ & $1$ & $\boldsymbol{1}^{0}$ & $\boldsymbol{1}$\\
      & $\nicefrac{5}{2}$ & $\nicefrac{9}{2}$ & $\nicefrac{1}{2}$ & $\boldsymbol{4}^{+1}\oplus\boldsymbol{\bar{4}}^{-1}$ & $\boldsymbol{1}$ \\
      & $2$ & $4$ & $0$ & $\boldsymbol{6}^{+2}\oplus\boldsymbol{15}^{0}\oplus\boldsymbol{6}^{-2}$  & $\boldsymbol{1}$ \\
      & $\nicefrac{3}{2}$ & $\nicefrac{7}{2}$ & $\nicefrac{-1}{2}$ & $\boldsymbol{\bar{20}}^{+1}\oplus\boldsymbol{20}^{-1}$  & $\boldsymbol{1}$ \\
      & $1$ & $3$ & $-1$ & $\boldsymbol{20'}^{0}$ & $\boldsymbol{1}$
  \end{tabular}
  \caption{Mass spectrum for the \SUquatre solution with diagonal embedding \eqref{eq:splitQC1ii} and gauge group $\GL(6)\times\SO(2,n_-)$.}
  \label{tab:SU411diag(ii)}
  \end{table}

\subsection[\texorpdfstring{$\OSpquatre$}{OSp(4*|4)}]{\texorpdfstring{$\boldsymbol{\OSpquatre}$}{OSp(4*|4)}}

\subsubsection{Chiral embedding}
We now consider the last possible supergroup, \OSpquatre. Here $p=5$, and $q=3$, and the index splitting~\eqref{eq:splitQC1ii0} is given by $i\in[\![1,5]\!]$ and $\alpha\in\{6,7,8\}$. Let us first assume that $\mathrm{SO}(5)\times\mathrm{SO}(3)$ is entirely embedded into the first factor of ${\rm SO}(8)\times {\rm SO}(n)\subset {\rm SO}(8,n)$. The potentially non-vanishing components of the embedding tensor are then given by
 \begin{align}
\bigg\{&\theta_{ij}= \lambda\,\delta_{ij},\;\theta_{\alpha\beta}=-(\lambda-\kappa)\,\delta_{\alpha\beta},\;\theta_{r_{+}s_{+}}=-\lambda\,\delta_{r_{+}s_{+}},\\
&\theta_{r_{-}s_{-}}=(\lambda-\kappa)\,\delta_{r_{-}s_{-}},\;\theta=\kappa,\;\theta_{p_{+}q_{+}r_{+}s_{+}},\;\theta_{p_{-}q_{-}r_{-}s_{-}}\bigg\}\;.
\end{align}
The second quadratic constraint~\eqref{eq:QC2} reduces to Eqs.~\eqref{eq:QC2Ipqrc} and \eqref{eq:QC2pqrsa},
of which the former implies
\begin{equation}
  (2\lambda-\kappa)\,\theta_{pqru_{+}} ~=~ 0~=~   (2\lambda-\kappa)\,\theta_{pqru_{-}} \;.
\end{equation}
Thus, all the components of $\theta_{pqrs}$ vanish.
The solution of the embedding tensor then is given by
\begin{align}
  & \theta_{ij}=(3+n_{-})\,\delta_{ij}\;,\quad \theta_{\alpha\beta}=-(5+n_{+})\,\delta_{\alpha\beta}\;,\quad \theta=-(2+n_+-n_-)\;,\nonumber\\
  &\theta_{r_{+}s_{+}}=-(3+n_{-})\,\delta_{r_{+}s_{+}}\;,\quad\theta_{r_{-}s_{-}}=(5+n_+)\,\delta_{r_{-}s_{-}}  \;. \label{eq:OSp44Tklmn=0}
\end{align}
The gauge group is $\SO(5,n_{+})\times\SO(3,n_{-})$ and the spectrum is given in Tab.~\ref{tab:OSp44multchiral(ii)}.

\subsubsection{Diagonal embedding}
Finally, for $n\geq3$, there could be solutions when the entire $R$-symmetry group $\mathrm{SO}(5)\times\mathrm{SO}(3)$ or one of its factors
is embedded diagonally in ${\rm SO}(8)\times {\rm SO}(n)\subset {\rm SO}(8,n)$. The 
 potentially non-vanishing components of the embedding tensor are
 \begin{align}
\bigg\{&\theta_{ij}= \lambda\,\delta_{ij},\;\theta_{\alpha\beta}=-(\lambda-\kappa)\,\delta_{\alpha\beta},\;\theta_{r_{+}s_{+}}=-\lambda\,\delta_{r_{+}s_{+}},\;\theta_{r_{-}s_{-}}=(\lambda-\kappa)\,\delta_{r_{-}s_{-}},\\
&\theta=\kappa,\;\theta_{ijr_{+}s_{+}},\;\theta_{\alpha\beta r_{-}s_{-}},\;\theta_{p_{+}q_{+}r_{+}s_{+}},\;\theta_{p_{-}q_{-}r_{-}s_{-}}\bigg\}.
\end{align}
Evaluating the quadratic constraints with this parametrization gives rise to a number of new solutions for the different embeddings
which we list in the following.

\paragraph*{$\boldsymbol{\SO(5)\times\SO(3)}$}
If $\SO(5)\times\SO(3)$ is embedded diagonally, there is one new solution for $n_{-}=3$ and $n_{+}=5$. It is given by
\begin{align}
  & \theta_{ij}=-6\,\delta_{ij}\;,\quad \theta_{\alpha\beta}=10\,\delta_{\alpha\beta}\;,\quad \theta_{r_{+}s_{+}}=6\,\delta_{r_{+}s_{+}}\;,\quad \theta_{r_{-}s_{-}}=-10\,\delta_{r_{-}s_{-}}\;,\nonumber\\
  & \theta=4\;,\quad \theta_{ijr_{+}s_{+}} = -32\,\delta_{i[r_{+}}\delta_{s_{+}]j}\;,\quad \theta_{\alpha\beta r_{-}s_{-}} = 32\,\delta_{\alpha[r_{-}}\delta_{s_{-}]\beta}\;.\label{eq:OSp44GL5GL3}
\end{align}
The gauge group is $\GL(5)\times\GL(3)$ and the spectrum is given in Tab.~\subref{tab:OSp44multdiag(ii)GL5GL3}.

\begin{table}
    \centering
  \begin{tabular}{c|c|c|c|c|c|c}
    $\Delta_{L}$ & $\Delta_{R}$ &  $\Delta$ & $s$ & $\mathrm{SO}(5)\times\mathrm{SO}(3)$ & $\mathrm{SO}(n_{+})$ & $\mathrm{SO}(n_{-})$ \\ \hline
    \multirow{3}{*}{$1$} & $2$ & $3$ & $1$ & $(\boldsymbol{1},\boldsymbol{3})$ & $\boldsymbol{1}$ & $\boldsymbol{n_{-}}$ \\
     & $\nicefrac{3}{2}$ & $\nicefrac{5}{2}$ & $\nicefrac{1}{2}$ & $(\boldsymbol{4},\boldsymbol{2})$ & $\boldsymbol{1}$ & $\boldsymbol{n_{-}}$ \\
     & $1$ & $2$ & $0$ & $(\boldsymbol{5},\boldsymbol{1})$ & $\boldsymbol{1}$ & $\boldsymbol{n_{-}}$ \\ \hline
     \multirow{3}{*}{$2$} & $2$ & $4$ & $0$ & $(\boldsymbol{1},\boldsymbol{3})$ & $\boldsymbol{n_{+}} $ & $\boldsymbol{1}$  \\
      & $\nicefrac{3}{2}$ & $\nicefrac{7}{2}$ & $\nicefrac{-1}{2}$ & $(\boldsymbol{4},\boldsymbol{2})$  & $\boldsymbol{n_{+}} $ & $\boldsymbol{1}$ \\
      & $1$ & $3$ & $-1$ & $(\boldsymbol{5},\boldsymbol{1})$ & $\boldsymbol{n_{+}} $ & $\boldsymbol{1}$ 
  \end{tabular}
  \caption{Mass spectrum for the \OSpquatre solution for the chiral embedding \eqref{eq:splitQC1ii} with gauge group $\SO(5,n_{+})\times\SO(3,n_{-})$.}
  \label{tab:OSp44multchiral(ii)}
\end{table} 

\paragraph*{$\boldsymbol{\SO(5)}$}
If $\SO(5)$ is embedded diagonally and $\SO(3)$ chirally, there is one new solution, given for $n_{+}=5$ by
\begin{align}
  & \theta_{ij}=(n_-+3)\delta_{ij}\;,\quad \theta_{\alpha\beta}=-10\,\delta_{\alpha\beta}\;,\quad \theta_{r_{+}s_{+}}=-(n_-+3)\,\delta_{r_{+}s_{+}}\;,\quad \theta_{r_{-}s_{-}}=10\,\delta_{r_{-}s_{-}}\;, \nonumber \\ 
  & \theta=(n_--7)\;,\quad\quad\theta_{ij r_{+}s_{+}} = 2\,(n_-+13)\,\delta_{i[r_{+}}\delta_{s_{+}]j}\;. \label{eq:OSp44GL5}
\end{align}
The gauge group is $\GL(5)\times\SO(3,n_-)$ and the spectrum is given in Tab.~\subref{tab:OSp44multdiag(ii)GL5}.

\begin{table}
    \centering
    \subfloat[$\mathrm{GL}(5)\times\mathrm{GL}(3)$]{
     \begin{tabular}{c|c|c|c|c}
     $\Delta_{L}$ & $\Delta_{R}$ &  $\Delta$ & $s$ & $\mathrm{SO}(5)\times\mathrm{SO}(3)$ \\ \hline
    \multirow{5}{*}{$2$} & $3$ & $5$ & $1$ & $(\boldsymbol{1},\boldsymbol{5})$ \\
     & $\nicefrac{5}{2}$ & $\nicefrac{9}{2}$ & $\nicefrac{1}{2}$ & $(\boldsymbol{4},\boldsymbol{4})$\\
     & $2$ &$4$ & $0$ & $(\boldsymbol{1},\boldsymbol{3})\oplus(\boldsymbol{5},\boldsymbol{3})$\\
     & $\nicefrac{3}{2}$ & $\nicefrac{7}{2}$ & $\nicefrac{-1}{2}$ & $(\boldsymbol{4},\boldsymbol{2})$\\
     & $1$ & $3$ & $-1$ & $(\boldsymbol{1},\boldsymbol{1})$ \\\hline
    \multirow{5}{*}{$3$} & $4$ & $7$ & $1$ & $(\boldsymbol{1},\boldsymbol{1})$\\
     & $\nicefrac{7}{2}$ & $\nicefrac{13}{2}$ & $\nicefrac{1}{2}$ & $(\boldsymbol{4},\boldsymbol{2})$ \\
     & $3$ &$6$ & $0$  & $(\boldsymbol{10},\boldsymbol{1})\oplus(\boldsymbol{5},\boldsymbol{3})$\\
     & $\nicefrac{5}{2}$ & $\nicefrac{11}{2}$ & $\nicefrac{-1}{2}$ & $(\boldsymbol{16},\boldsymbol{2})$ \\
     & $2$ & $5$ & $-1$ & $(\boldsymbol{14},\boldsymbol{1})$ 
  \end{tabular}\label{tab:OSp44multdiag(ii)GL5GL3}}

  \subfloat[$\mathrm{GL}(5)\times\mathrm{SO}(3,n_-)$]{
    \begin{tabular}{c|c|c|c|c|c}
    $\Delta_{L}$ & $\Delta_{R}$ &  $\Delta$ & $s$ & $\mathrm{SO}(5)\times\mathrm{SO}(3)$ & $\mathrm{SO}(n_-)$ \\ \hline
    \multirow{3}{*}{$1$} & $2$ & $3$ & $1$ & $(\boldsymbol{1},\boldsymbol{3})$ & $\boldsymbol{n_-}$ \\
      & $\nicefrac{3}{2}$ & $\nicefrac{5}{2}$ & $\nicefrac{1}{2}$& $(\boldsymbol{4},\boldsymbol{2})$ & $\boldsymbol{n_-}$\\
      & $1$ & $2$ & $0$ & $(\boldsymbol{5},\boldsymbol{1})$ & $\boldsymbol{n_-}$\\\hline
    \multirow{5}{*}{$3$} & $4$ & $7$ & $1$ & $(\boldsymbol{1},\boldsymbol{1})$ & $\boldsymbol{1}$ \\
     & $\nicefrac{7}{2}$ & $\nicefrac{13}{2}$ & $\nicefrac{1}{2}$ & $(\boldsymbol{4},\boldsymbol{2})$ & $\boldsymbol{1}$\\
     & $3$ &$6$ & $0$  & $(\boldsymbol{10},\boldsymbol{1})\oplus(\boldsymbol{5},\boldsymbol{3})$& $\boldsymbol{1}$\\
     & $\nicefrac{5}{2}$ & $\nicefrac{11}{2}$ & $\nicefrac{-1}{2}$ & $(\boldsymbol{16},\boldsymbol{2})$ & $\boldsymbol{1}$\\
     & $2$ & $5$ & $-1$ & $(\boldsymbol{14},\boldsymbol{1})$ & $\boldsymbol{1}$ 
  \end{tabular}\label{tab:OSp44multdiag(ii)GL5}}

      \subfloat[$\mathrm{GL}(3)\times\mathrm{SO}(5,n_+)$]{
    \begin{tabular}{c|c|c|c|c|c}
     $\Delta_{L}$ & $\Delta_{R}$ &  $\Delta$ & $s$ & $\mathrm{SO}(5)\times\mathrm{SO}(3)$ & $\mathrm{SO}(n_+)$ \\ \hline
    \multirow{5}{*}{$2$} & $3$ & $5$ & $1$ & $(\boldsymbol{1},\boldsymbol{5})$ & $\boldsymbol{1}$  \\
     & $\nicefrac{5}{2}$ & $\nicefrac{9}{2}$ & $\nicefrac{1}{2}$ & $(\boldsymbol{4},\boldsymbol{4})$& $\boldsymbol{1}$\\
     & $2$ &$4$ & $0$ & $(\boldsymbol{1},\boldsymbol{3})\oplus(\boldsymbol{5},\boldsymbol{3})$& $\boldsymbol{1}$\\
     & $\nicefrac{3}{2}$ & $\nicefrac{7}{2}$ & $\nicefrac{-1}{2}$ & $(\boldsymbol{4},\boldsymbol{2})$& $\boldsymbol{1}$\\
     & $1$ & $3$ & $-1$ & $(\boldsymbol{1},\boldsymbol{1})$ & $\boldsymbol{1}$\\ \hline
     \multirow{3}{*}{$2$} & $2$ & $4$ & $0$ & $(\boldsymbol{1},\boldsymbol{3})$ & $\boldsymbol{n_+}$ \\
      & $\nicefrac{3}{2}$ & $\nicefrac{7}{2}$ & $\nicefrac{-1}{2}$ & $(\boldsymbol{4},\boldsymbol{2})$ & $\boldsymbol{n_+}$ \\
      & $1$ & $3$ & $-1$ & $(\boldsymbol{5},\boldsymbol{1})$ & $\boldsymbol{n_+}$ 
  \end{tabular}\label{tab:OSp44multdiag(ii)GL3}}

  \subfloat[$\mathrm{SO}(5,n_+)\times\mathrm{G}_{2(2)}$]{
     \begin{tabular}{c|c|c|c|c|c|c}
     $\Delta_{L}$ & $\Delta_{R}$ &  $\Delta$ & $s$ & $\mathrm{SO}(5)\times\mathrm{SO}(3)$ & $\mathrm{SO}(3)$ & $\mathrm{SO}(n_+)$ \\ \hline
    \multirow{4}{*}{$\nicefrac{3}{2}$} & $\nicefrac{5}{2}$ & $4$ & $1$ & $(\boldsymbol{1},\boldsymbol{4})$ & $\boldsymbol{2}$ & $\boldsymbol{1}$  \\
     & $2$ & $\nicefrac{7}{2}$ & $\nicefrac{1}{2}$ & $(\boldsymbol{4},\boldsymbol{3})$& $\boldsymbol{2}$& $\boldsymbol{1}$ \\
     & $\nicefrac{3}{2}$ &$3$ & $0$ & $(\boldsymbol{5},\boldsymbol{2})\oplus(\boldsymbol{1},\boldsymbol{4})$& $\boldsymbol{2}$& $\boldsymbol{1}$ \\
     & $1$ & $\nicefrac{5}{2}$ & $\nicefrac{-1}{2}$ & $(\boldsymbol{4},\boldsymbol{1})$ & $\boldsymbol{2}$& $\boldsymbol{1}$  \\ \hline
     \multirow{3}{*}{$2$} & $2$ & $4$ & $0$ & $(\boldsymbol{1},\boldsymbol{3})$ & $\boldsymbol{1}$ & $\boldsymbol{n_+}$ \\
      & $\nicefrac{3}{2}$ & $\nicefrac{7}{2}$ & $\nicefrac{-1}{2}$ & $(\boldsymbol{4},\boldsymbol{2})$ & $\boldsymbol{1}$ & $\boldsymbol{n_+}$  \\
      & $1$ & $3$ & $-1$ & $(\boldsymbol{5},\boldsymbol{1})$ & $\boldsymbol{1}$ & $\boldsymbol{n_+}$ 
  \end{tabular}\label{tab:OSp44multdiag(ii)G22}}
  \caption{Mass spectra for the \OSpquatre solutions with diagonal embedding \eqref{eq:splitQC1ii}.}
  \label{tab:OSp44multdiag(ii)}
\end{table}

\paragraph*{$\boldsymbol{\SO(3)\ (1)}$}
Finally, if $\SO(3)$ is embedded diagonally and $\SO(5)$ chirally, there are two new solutions. The first one requires $n_-=3$ and is given by
\begin{align}
  & \theta_{ij}=-6\,\delta_{ij}\;,\quad \theta_{\alpha\beta}=(n_++5)\,\delta_{\alpha\beta}\;,\quad \theta_{r_{+}s_{+}}=6\,\delta_{r_{+}s_{+}}\;,\quad \theta_{r_{-}s_{-}}=-(n_++5)\,\delta_{r_{-}s_{-}}\;,\nonumber\\
  & \theta=(n_+-1)\;,\quad\theta_{\alpha\beta r_{-}s_{-}} = 2\,(n_++11)\,\delta_{\alpha[r_{-}}\delta_{s_{-}]j}\;.\label{eq:OSp44GL3}
\end{align}
The gauge group is $\GL(3)\times\SO(5,n_+)$ and the spectrum is given in Tab.~\subref{tab:OSp44multdiag(ii)GL3}.

\paragraph*{$\boldsymbol{\SO(3)\ (2)}$}
The second solution for diagonal embedding of $\SO(3)$ requires $n_-=4$ and is given by
\begin{align}
  & \theta_{ij}=-14\,\delta_{ij}\;,\quad \theta_{\alpha\beta}=2\,(n_++5)\,\delta_{\alpha\beta}\;,\quad \theta_{r_{+}s_{+}}=14\,\delta_{r_{+}s_{+}}\;,
  \quad
  \theta_{r_{-}s_{-}}=-2\,(n_++5)\,\delta_{r_{-}s_{-}}\;,\nonumber\\
  & \theta=2\,(n_+-2)\;,\quad\theta_{\alpha\beta \bar r_{-}\bar s_{-}} = 2\,(n_++12)\,\delta_{\alpha[\bar r_{-}}\delta_{\bar s_{-}]\beta}\;,\nonumber \\
  &\theta_{\alpha\beta \bar0\bar r_{-}} = -(n_++12)\,\varepsilon_{\alpha\beta \bar r_{-}}\;,\quad \theta_{\bar0\bar p_{-}\bar q_{-}\bar r_{-}} = -(n_++12)\,\varepsilon_{\bar p_{-}\bar q_{-}\bar r_{-}}\;,\label{eq:OSp44G22}
\end{align}
where the index $r_-$ has been split into $r_-\rightarrow\{\bar0, \bar{r}_-\}$, $\bar{r}_-\in\{1,2,3\}$\,.
The gauge group is $\SO(5,n_+)\times\G_{2(2)}$ and the spectrum is given in Tab.~\subref{tab:OSp44multdiag(ii)G22}.


\section{Some \texorpdfstring{$\boldsymbol{\mathcal{N}=(7,1)}$}{N=(7,1)} and \texorpdfstring{$\boldsymbol{\mathcal{N}=(7,0)}$}{N=(7,0)} vacua}
\label{sec:N7}


Having given an exhaustive classification of ${\cal N}=(8,0)$ vacua for $n\leq8$, 
in this section we present some partial analysis of vacua with ${\cal N}=(7,1)$ and ${\cal N}=(7,0)$ supersymmetries, 
respectively.
The relevant supergroups with ${\cal N}=7$ supercharges are \OSpsept and G(3),
whose $R$-symmetry subgroups are ${\rm SO}(7)$ and G$_2$, respectively.
The supergroup with one supercharge is ${\rm OSp}(1|2,\mathbb{R})$.

In these cases, the constraints~\eqref{eq:SUSYconstraints} and \eqref{eq:SUSYconstraintsA2} get replaced by
\begin{equation}
{\cal N}=(7,1): \quad \begin{cases}
                        A_{1}^{ab} = \ell^{-1}\,\delta^{ab}\;, \\
                        A_{1}^{88} = -\ell^{-1}\;,\\
                       A_{2}^{a\dot Ar} = 0= A_{2}^{8\dot Ar}\;,
                      \end{cases} \qquad\quad
{\cal N}=(7,0): \quad \begin{cases}
                        A_{1}^{ab} = \ell^{-1}\,\delta^{ab}\;, \\
                        A_{1}^{88} \neq \pm \ell^{-1}\;, \\
                        A_{2}^{a\dot Ar} = 0\;,
                      \end{cases}
\end{equation}
respectively, where the index $A$ splits according to $A=\{a,8\}$ with $a\in[\![1,7]\!]$. 
For ${\cal N}=(7,1)$ the potentially non-vanishing components of the embedding tensor are thus given by
\begin{equation}
\left\{\theta_{IJ},\; \theta_{rs},\; \theta,\;\theta_{IJKL},\; \theta_{IJrs},\; \theta_{Ipqr},\; \theta_{pqrs}\right\}, \label{eq:thetafullN71}
\end{equation}
where $\theta_{IJKL}$ now also has a selfdual contribution, unlike the ${\cal N}=(8,0)$ case of Eq.~\eqref{eq:thetafullparam}. 
For ${\cal N}=(7,0)$ vacua, the potentially non-vanishing  components of the embedding tensor are
\begin{equation}
\left\{\theta_{IJ},\; \theta_{rs},\; \theta_{Ir},\; \theta,\;\theta_{IJKL},\;\theta_{IJKr},\; \theta_{IJrs},\; \theta_{Ipqr},\; \theta_{pqrs}\right\}\;,
 \label{eq:thetafullN70}
\end{equation}
again with a $\theta_{IJKL}$ which is not restricted to anti-selfdual tensors.
In the rest of this section, we present our findings for such vacua.


\subsection[\texorpdfstring{\OSpsept}{OSp(7|2,R)}]{\texorpdfstring{$\boldsymbol{\OSpsept}$}{OSp(7|2,R)}}


\begin{table}
    \centering
    \begin{tabular}{c|c|c|c|c}
     $\Delta_{L}$ & $\Delta_{R}$ &  $\Delta$ & $s$ & $\mathrm{SO}(7)$ \\ \hline
    \multirow{6}{*}{$\nicefrac{3}{2}$}& $3$ & $\nicefrac{9}{2}$ & $\nicefrac{3}{2}$ & $\boldsymbol{1}$\\
     & $\nicefrac{5}{2}$ & $4$ & $1$ & $\boldsymbol{7}$  \\
     & $2$ & $\nicefrac{7}{2}$ & $\nicefrac{1}{2}$ & $\boldsymbol{21}$\\
     & $\nicefrac{3}{2}$ &$3$ & $0$ & $\boldsymbol{1}\oplus\boldsymbol{35}$\\
     & $1$ & $\nicefrac{5}{2}$ & $\nicefrac{-1}{2}$ & $\boldsymbol{7}\oplus\boldsymbol{35}$\\
     & $\nicefrac{1}{2}$ & $2$ & $-1$ & $\boldsymbol{21}$
  \end{tabular}
  \caption{Mass spectrum for the \OSpsept  solution with ${\cal N}=(7,0)$ supersymmetries. 
  The gauge group is $\mathrm{SO}(8)\times\mathrm{SO}(7)$.}
  \label{tab:OSp7mult}
\end{table}

For \OSpsept, a pair of ${\cal N}=(7,0)$ vacua  is given by the embedding tensor
\begin{align}
&\theta_{IJ}=56\,\delta_{IJ}\;,\quad \theta_{rs}=56\,\delta_{rs}\;,\quad\theta_{Ir}=\pm\,32\,\sqrt{3}\,\delta_{Ir}\;,\quad\theta=0\;,\quad \theta_{IJKL} =-6\,\Gamma^{IJKL}_{AB}\xi^{AB}\;,\nonumber \\
&\theta_{IJKr} =\mp\,4\,\sqrt{3}\,\Gamma^{IJKr}_{AB}\xi^{AB}\;,\quad\Theta_{IJrs} =-8\,\Gamma^{IJrs}_{AB}\xi^{AB}\;,\quad\theta_{Ipqr} =\mp\frac{16}{\sqrt{3}}\,\Gamma^{Ipqr}_{AB}\xi^{AB}\;,\nonumber\\
&\theta_{pqrs} =-\frac{32}{3}\,\Gamma^{pqrs}_{AB}\xi^{AB}\;,\quad\mathrm{with}\ \xi^{AB} = {\rm diag}\Big\{-7,\underbrace{1, \dots, 1}_{7} \Big\}\;,\label{eq:OSp7SO8SO7}
\end{align}
with $n=8$\,.
The gauge group is $\mathrm{SO}(8)\times\mathrm{SO}(7)$,
which at the vacuum is spontaneously broken down to a diagonal $\mathrm{SO}(7)$ subgroup. 
The spectrum is given in Tab.~\ref{tab:OSp7mult}.

Closer inspection shows that these embedding tensors may be related to the embedding tensor of Eq.~\eqref{eq:OSp8SOp}
(with selfdual $\theta_{pqrs}$) by an $\SO(8,8)$ rotation of the form
\begin{equation}
{\cal V}\indices{_{\cal M}^{\cal I}}(\phi)=\begin{pmatrix}
                                \cosh\left(\phi\right)\,\mathds{1}_{8} & \sinh\left(\phi\right)\,\mathds{1}_{8} \\
                                \sinh\left(\phi\right)\,\mathds{1}_{8}  & \cosh\left(\phi\right)\,\mathds{1}_{8} 
                              \end{pmatrix},
                              \label{eq:1so7}
\end{equation}
with 
\bea
\phi=\phi_{\pm}=\ln\left(7\pm4\,\sqrt{3}\right)/2
\;.
\label{eq:posphi}
\eea
In view of our discussion in Sec.~\ref{sec:vacuaspectra},
we have thus identified three vacua which all belong to the same three-dimensional theory. 
To illustrate this structure, we evaluate the scalar potential~\eqref{eq:potential} on the 1-scalar truncation (\ref{eq:1so7})
to ${\rm SO}(7)$ singlets, which takes the form
\begin{figure}[b!]
\centering
\includegraphics[scale=1]{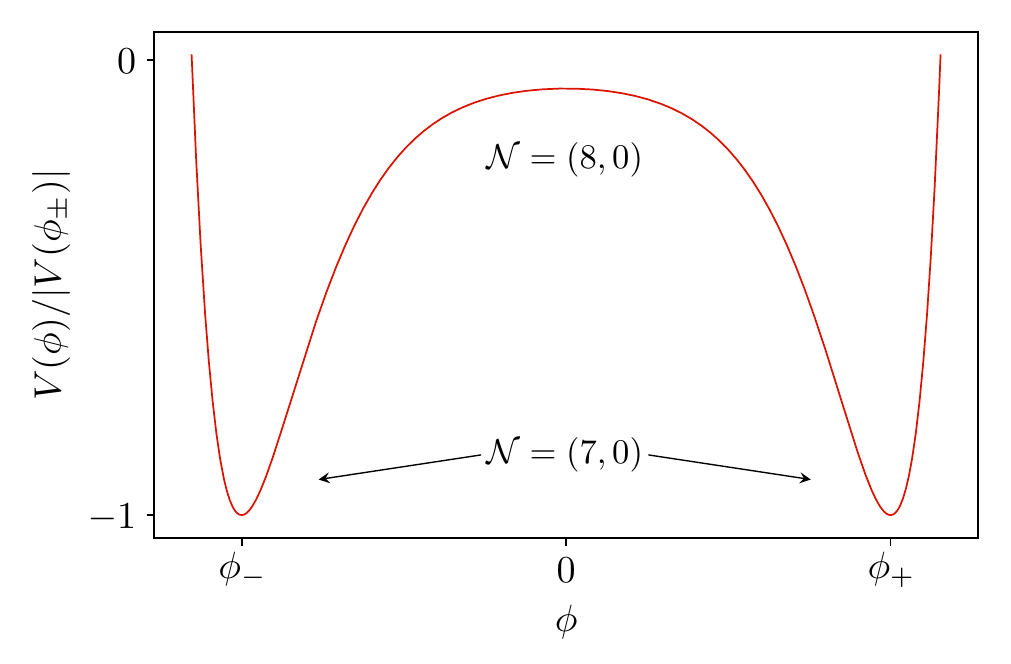}
\caption{Potential for the 1-scalar truncation (\ref{eq:1so7}) of the theory with gauge group $\SO(8)\times\SO(7)$. The 
vacuum at the scalar origin $\phi=0$ preserves full $\mathcal{N} = (8,0)$ supersymmetry,
while the two vacua at $\phi_{\pm} =\ln\left(7\pm4\,\sqrt{3}\right)/2$ spontaneously break supersymmetry down to $\mathcal{N} = (7,0)$.}
\label{fig:potential_n8}
\end{figure}
\begin{equation}
V(\phi) = \frac{16}{9} \,  \big(-250+105 \cosh (2 \,\phi )-150\cosh (4 \,\phi )+7 \cosh (6\, \phi )\big)\;, \label{eq:potential_n8}
\end{equation}
that is sketched in Fig.~\ref{fig:potential_n8}. 
It exhibits the fully symmetric ${\cal N}=(8,0)$ vacuum at the scalar origin $\phi=0$,
together with the two ${\cal N}=(7,0)$ vacua at $\phi=\phi_\pm$ from Eq.~\eqref{eq:posphi}.
This potential may be cast into the following form
\begin{equation}
V(\phi) = -2\,W(\phi)^{2}+\frac{1}{8}\,\left(\frac{{\rm d}W}{{\rm d}\phi}(\phi)\right)^{2},
\end{equation}
in terms of a real superpotential 
\begin{equation}
W(\phi) = \frac{2}{3}\, \big(3-28\,\cosh(2\,\phi)+\cosh(4\,\phi)\big) \;, 
\label{eq:superpotential_n8}
\end{equation}
which shares its stationary points with $V(\phi)$.


\subsection[G(3)]{\texorpdfstring{$\boldsymbol{\G(3)}$}{G(3)}}


For the supergroup $\G(3)$, we present vacua with both, ${\cal N}=(7,1)$ and ${\cal N}=(7,0)$ supersymmetry, respectively.

\paragraph*{$\boldsymbol{{\cal N}=(7,1)}$}
The embedding tensor
\begin{align}
&\theta_{ij}=(n+1)\,\delta_{ij}\;,\quad\theta_{88}=-7\;,\quad\theta_{rs}=7\,\delta_{rs}\;,\quad \theta=(n-6)\;,\nonumber\\
&\theta_{ijkl} =-\frac{(8+n)}{12}\,\varepsilon_{ijklxyz}\,\omega^{xyz}\;,\label{eq:G3SOn1G2}
\end{align}
with $i\in[\![1,7]\!]$ 
and the ${\rm G}_{2}$-invariant 3-form $\omega^{xyz}$ from Eq.~\eqref{eq:OSp8G2},
describes an ${\cal N}=(7,1)$ vacuum within an $\SO(n,1)\times\G_{2}$
gauged theory whose gauge group at the vacuum is broken down to its compact $\SO(n)\times\G_{2}$ subgroup. Its spectrum is given in Tab.~\ref{tab:G3mult71}.

\begin{table}
   \centering
   \begin{tabular}{c|c|c|c|c|c}
    $\Delta_{L}$ & $\Delta_{R}$ &  $\Delta$ & $s$ &$\mathrm{G}_{2}$ & $\mathrm{SO}(n)$  \\ \hline
    $\nicefrac{5}{4}$ & $\nicefrac{3}{4}$ & $2$ & $\nicefrac{-1}{2}$ & $\boldsymbol{1}$ & $\boldsymbol{n}$ \\
    $\nicefrac{5}{4}$ & $\nicefrac{1}{4}$ & $\nicefrac{3}{2}$ & $-1$ & $\boldsymbol{1}$ & $\boldsymbol{n}$ \\
    $\nicefrac{3}{4}$ & $\nicefrac{3}{4}$ & $\nicefrac{3}{2}$ & $0$ & $\boldsymbol{7}$ & $\boldsymbol{n}$ \\
    $\nicefrac{3}{4}$ & $\nicefrac{1}{4}$ & $1$ & $\nicefrac{-1}{2}$ & $\boldsymbol{7}$ & $\boldsymbol{n}$
  \end{tabular}
  \caption{Mass spectrum for the $\G(3)$ solution with $\mathcal{N}=(7,1)$ supersymmetry. The gauge group is $\SO(n,1)\times\G_{2}$.}
  \label{tab:G3mult71}
\end{table}

\paragraph*{$\boldsymbol{{\cal N}=(7,0)}$}
For $n=7$, a pair of $\G(3)$ solutions with ${\cal N}=(7,0)$ supersymmetry is given by the embedding tensors
\begin{align}
&\theta_{ij}=127\,\delta_{ij}\;,\ \theta_{88}=7\;,\ \theta_{rs}=128\,\delta_{rs}\;,\ \theta_{ir}=\pm\,90\,\sqrt{2}\,\delta_{ir}\;,\ \theta_{8r}=0\;,\ \theta=-1\;,\nonumber \\
&\theta_{ijkl} =75\,\tilde\omega_{ijkl}\;,\ \theta_{8ijk} = -135\,\omega_{ijk}\;,\ \theta_{ijkr} =\pm\,90\,\sqrt{2}\,\tilde\omega_{ijkr}\;,\ \theta_{8ijr} = \mp\,90\,\sqrt{2}\,\omega_{ijr}\;,\nonumber\\
&\theta_{ijrs} =180\,\tilde\omega_{ijrs}\;,\ \theta_{8irs} = -120\,\omega_{irs}\;,\ \theta_{ipqr} =\pm 660\,\sqrt{2}\,\tilde\omega_{ipqr}\;,\ \theta_{8pqr} = \mp\,80\,\sqrt{2}\,\omega_{pqr}\;,\nonumber\\
&\theta_{pqrs} =\frac{575}{2}\,\tilde\omega_{pqrs} \;,
 \label{eq:G3SO7G2}
\end{align}
with the index $i\in[\![1,7]\!]$, the G$_2$ invariant three-form $\omega^{kmn}$ from Eq.~\eqref{eq:OSp8G2}
and its dual defined by $\tilde\omega_{ijkl}\equiv \dfrac16\,\varepsilon_{ijklmnp}\,\omega^{mnp}$\,.
The gauge group is $\mathrm{SO}(7)\times\mathrm{G}_{2}$,
broken down at the vacuum to a diagonal $\mathrm{G}_{2}$ subgroup. The spectrum is given in Tab.~\ref{tab:G3mult70}.

\begin{table}
    \centering
    \begin{tabular}{c|c|c|c|c}
     $\Delta_{L}$ & $\Delta_{R}$ &  $\Delta$ & $s$ & $\mathrm{G}_{2}$ \\ \hline
    \multirow{6}{*}{$\nicefrac{7}{4}$}& $\nicefrac{13}{4}$ & $5$ & $\nicefrac{3}{2}$ & $\boldsymbol{1}$\\
     & $\nicefrac{11}{4}$ & $\nicefrac{9}{2}$ & $1$ & $\boldsymbol{7}$  \\
     & $\nicefrac{9}{4}$ & $4$ & $\nicefrac{1}{2}$ & $\boldsymbol{7}\oplus\boldsymbol{14}$\\
     & $\nicefrac{7}{4}$ & $\nicefrac{7}{2}$ & $0$ & $\boldsymbol{1}\oplus\boldsymbol{7}\oplus\boldsymbol{27}$\\
     & $\nicefrac{5}{4}$ & $3$ & $\nicefrac{-1}{2}$ & $\boldsymbol{7}\oplus\boldsymbol{27}$\\
     & $\nicefrac{3}{4}$ & $\nicefrac{5}{2}$ & $-1$ & $\boldsymbol{14}$
  \end{tabular}
  \caption{Mass spectrum for the $\G(3)$ solution with ${\cal N}=(7,0)$ supersymmetry. The gauge group is $\mathrm{SO}(7)\times\mathrm{G}_{2}$.}
  \label{tab:G3mult70}
\end{table}

\begin{figure}[b!] 
\centering
\includegraphics[scale=1]{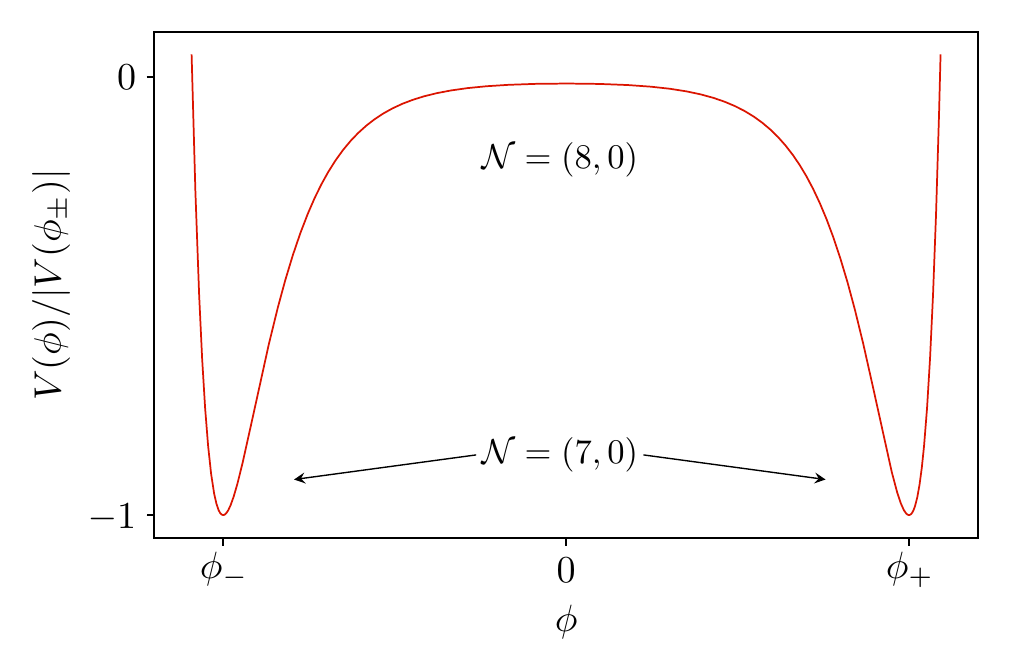}
\caption{Potential for the 1-scalar truncation (\ref{eq:1g2}) of the theory with gauge group $\SO(7)\times\G_{2}$. The 
vacuum at the scalar origin $\phi=0$ preserves full $\mathcal{N} = (8,0)$ supersymmetry,
while the two vacua at $\phi_{\pm} =\ln\left(3\pm2\,\sqrt{2}\right)$ spontaneously break supersymmetry down to $\mathcal{N} = (7,0)$.}
\label{fig:potential_n7}
\end{figure}

The embedding tensors \eqref{eq:G3SO7G2} turn out to be related to the previously found ${\cal N}=(8,0)$ solution~\eqref{eq:F4G2} 
by an $\SO(8,7)$ rotation of the form
\begin{equation}
{\cal V}\indices{_{\cal M}^{\cal I}}(\phi)=\begin{pmatrix}
                                \cosh\left(\phi\right)\,\mathds{1}_{7} & 0 & \sinh\left(\phi\right)\,\mathds{1}_{7} \\
                                0 & 1 & 0 \\
                                \sinh\left(\phi\right)\,\mathds{1}_{7} & 0 & \cosh\left(\phi\right)\,\mathds{1}_{7} 
                              \end{pmatrix}\;.
                              \label{eq:1g2}
\end{equation}
Starting from Eq.~\eqref{eq:F4G2} (up to a change of basis) at the origin $\phi=0$, the tensors \eqref{eq:G3SO7G2} are obtained
at $\phi_{\pm}=\ln\left(3\pm2\,\sqrt{2}\right)$. 
In view of our discussion in Sec.~\ref{sec:vacuaspectra},
we have thus identified three vacua which all belong to the same three-dimensional theory. 
To illustrate this structure, we evaluate the scalar potential~\eqref{eq:potential} on the 1-scalar truncation (\ref{eq:1g2})
to ${\rm G}_2$ singlets, which takes the form
\bea
V(\phi) &=& \frac{25}{1024} \,\big(-15163+3416\cosh(\phi)-3640\cosh(2\,\phi)+2856\cosh(3\,\phi)-6860\cosh (4 \,\phi ) \nonumber \\
&&{}
\qquad \quad+952\cosh(5\,\phi)+56\cosh(6\,\phi)-56\cosh(7\,\phi)+7\cosh(8\,\phi)\big)\;, \label{eq:potential_n7}
\eea
that is sketched in Fig.~\ref{fig:potential_n7}. 
It exhibits the fully symmetric ${\cal N}=(8,0)$ vacuum at the scalar origin $\phi=0$,
together with the two ${\cal N}=(7,0)$ vacua at $\phi_{\pm}=\ln\left(3\pm2\,\sqrt{2}\right)$.

Let us finally note, that the scalar potential~\eqref{eq:potential_n7} admits a 
``fake'' superpotential
\begin{equation}
W(\phi) = - \frac{5}{32} \,\big(33+84\cosh(\phi)-56\cosh(2\,\phi)+28\cosh(3\,\phi)+7\cosh (4 \,\phi)\big])
\;, 
\label{eq:superpotential_n7}
\end{equation}
in terms of which it may be written as
\begin{equation}
V(\phi) = -2\,W(\phi)^{2}+\frac{1}{7}\,\left(\frac{{\rm d}W}{{\rm d}\phi}(\phi)\right)^{2}
\;.
\end{equation}
However, the $\mathcal{N} = (7,0)$ vacua are not stationary points of $W(\phi)$.


\section{Embedding into the maximal theory}
\label{sec:maximal}


For a given supersymmetric vacuum, identified as a solution of Eqs.~\eqref{eq:QC1}, \eqref{eq:QC2}, \eqref{eq:varPot} and \eqref{eq:SUSYconstraints}, 
the embedding tensor defines the gauge group generators according to Eq.~\eqref{XTL} from which we have determined the 
specific half-maximal theory that fulfils all the requirements.
It is an interesting question to ask, which of the vacua identified in this analysis can actually be
embedded into a maximally supersymmetric (${\cal N}=16$) three-dimensional supergravity,
thus spontaneously breaking half of the supersymmetries of the theory.
For $D=4$ supergravities, the analogous question has been addressed in Ref.~\cite{Dibitetto:2011eu}.

To answer this question, we first recall some of the basic structures of maximal $D=3$ supergravities
 \cite{Nicolai:2000sc,Nicolai:2001sv}. In this case, the scalar sector describes an ${\rm E}_{8(8)}/{\rm SO}(16)$
 coset space sigma model and the embedding tensor which defines the gauge group generators within
 ${\rm E}_{8(8)}$ in analogy to Eq.~\eqref{XTL} transforms in the ${\bf 1}\oplus{\bf 3875}$ representation of  ${\rm E}_{8(8)}$.
The maximal theory can be truncated to a half-maximal subsector upon truncating the coset space
\bea
{\rm E}_{8(8)}/{\rm SO}(16) &\longrightarrow& 
\SO(8,8)/(\SO(8)\times\SO(8))
\;.
\label{eq:E8SO88}
\eea
Under $\SO(8,8)$, the embedding tensor of the maximal theory decomposes as
\bea
{\bf 1}\oplus{\bf 3875} &\longrightarrow& 
{\bf 1}\oplus {\bf 135}\oplus {\bf 1820}\oplus {\bf 1920}_c
\;,
\label{eq:maxhalf}
\eea
of which the first three parts reproduce the embedding tensor (\ref{eq:tt8}), (\ref{eq:linear}) of the half-maximal 
theory, whereas the last part drops out in the projection to the half-maximal theory.
Specifically, splitting the ${\rm E}_{8(8)}$ generators according to
\bea
\left\{
L^{[{\cal MN}]}, Y^{{\cal A}}
\right\}\;,
\qquad
{\cal M}\in[\![1,16]\!]\;,\;\; {\cal A}\in[\![1,128]\!]\;,
\eea
into $\SO(8,8)$ and its orthogonal complement (transforming in the spinor representation ${\bf 128}_s$ of $\SO(8,8)$),
an embedding tensor of the maximal theory triggered by the first three terms in Eq.~\eqref{eq:maxhalf} takes the form
\begin{equation}
  \begin{cases}
  \Theta_{{\cal MN|PQ}} = \theta_{{\cal MNPQ}}+2\left(\eta_{{\cal M}[{\cal P}}\,\theta_{{\cal Q}]{\cal N}}-\eta_{{\cal N}[{\cal P}}\,\theta_{{\cal Q}]{\cal M}}\right)+2\,\eta_{{\cal M}[{\cal P}}\,\eta_{{\cal Q}]{\cal N}}\,\theta\;,\smallskip\\
\Theta_{{\cal A}|{\cal B}} =-\theta\,\eta_{{\cal AB}}+\frac1{96}\,\Gamma^{{\cal MNPQ}}_{{\cal AB}}\,\theta_{{\cal MNPQ}}\;,
  \end{cases}\label{eq:embmax}
\end{equation}
where $\Gamma^{{\cal MNPQ}}_{{\cal AB}}$ denotes the four-fold product 
of $\SO(8,8)$ $\Gamma$-matrices. Introducing covariant derivatives
\bea
D_{\mu} &=& \partial_{\mu} - g \,A_\mu{}^{{\cal MN}}\,\Theta_{{\cal MN|PQ}} \,L^{[{\cal PQ}]}
-g\,A_\mu{}^{{\cal A}}\, \Theta_{{\cal A}|{\cal B}}\,Y^{{\cal B}}
\;,
\label{eq:covD}
\eea
in the maximal theory, the gauge group is thus given by an extension of the gauge group
of the half-maximal theory by the additional generators $X_{{\cal A}} =  \Theta_{{\cal A}|{\cal B}}\,Y^{{\cal B}}$\,.
We may now address the following question: given an embedding tensor (\ref{eq:thetafull}) of the half-maximal
$\SO(8,8)$ theory, satisfying the quadratic constraints (\ref{eq:QC1}), (\ref{eq:QC2}),
does the associated embedding tensor (\ref{eq:embmax}) satisfy the quadratic constraints of the maximal
theory and thereby define a consistent gauging of the maximal theory?
By consistency of the truncation, the ${\cal N}=(8,0)$ vacuum of the half-maximal theory then turns into
a vacuum of some maximal gauged supergravity, breaking half of the supersymmetries spontaneously.

To answer this question, we recall that the quadratic constraints of the maximal theory transform in the
\bea
{\bf 3875} \oplus {\bf 147250}
\eea
under ${\rm E}_{8(8)}$. Breaking these under $\SO(8,8)$ and restricting to the representations that can actually 
appear in the symmetric tensor product of two half-maximal embedding tensors shows that in order to define a maximal ${\cal N}=16$ gauging, the components of the embedding tensor~(\ref{eq:embmax}) must satisfy additional constraints transforming as ${\bf 35}\oplus {\bf 6435}_c$, \textit{i.e.}\
\begin{equation}
{\scriptsize\yng(2)} ~\oplus~
{\scriptsize\yng(1,1,1,1,1,1,1,1)}_{\,\,\rm anti-sd} \;,
\label{eq:maxex}
\end{equation}
where the last term refers to the anti-selfdual contribution in the 8-fold antisymmetric tensor.
These additional conditions may be worked out explicitly
in analogy to Eq.~(\ref{eq:quadratic}) and take the following form
\begin{equation}
\begin{cases}
0 \ =\ 
48\,\theta\, \theta_{{\cal MN}} +
\theta_{{\cal M}}{}^{{\cal PQR}}\theta_{{\cal NPQR}}
-\frac1{16}\,\eta_{{\cal MN}}\,\theta^{{\cal PQRS}}\theta_{{\cal PQRS}}
\;,\\
0 \ =\ 
\Gamma^{{\cal KLMNPQRS}}_{\dot{{\cal A}}\dot{{\cal B}}}\,\theta_{{\cal KLMN}}\theta_{{\cal PQRS}}\;.
\end{cases}
\label{eq:addmax}
\end{equation}
Let us note that the first equation has the same representation content $\,{\scriptsize\yng(2)}\,$ as the constraint
\bea
\theta\, \theta_{{\cal MN}}+    \theta_{{\cal M}}{}^{{\cal K}} \,  \theta_{{\cal KN}}\, 
-\frac1{16}\,\eta_{{\cal MN}} \theta^{{\cal KL}} \,  \theta_{{\cal KL}}=0
\;,
\label{comb2s}
\eea
obtained from proper contraction of the constraints (\ref{eq:QC2}) of the ${\cal N}=8$ theory.
Furthermore, the choice of anti-selfduality (vs.\ selfduality) in Eq.~(\ref{eq:maxex}) is a pure convention here, 
depending on the embedding of ${\rm SO}(8,8)$ into ${\rm E}_{8(8)}$\,.

To summarize, an embedding tensor of the half-maximal theory
which in addition to the quadratic constraints (\ref{eq:QC1}), (\ref{eq:QC2}) of the half-maximal theory
satisfies the additional constraints~(\ref{eq:addmax}) defines a consistent maximal three-dimensional supergravity.
The half-maximal theory is recovered upon truncation (\ref{eq:E8SO88}).
Vacua of the half-maximal theory then give rise to vacua within the maximal theory.
It is a straightforward task to check the additional constraints~(\ref{eq:addmax}) for all the vacua we have identified in this paper.
For theories with a free parameter these constraints single out specific values of the parameter.
We collect the results of the possible embedding  into the maximal theory 
for every vacuum in the summarizing Tabs.~\ref{tab:summary1}--\ref{tab:summary3}.


\section{Summary and outlook}
\label{sec:conclusions}


In this paper, we have presented a classification of ${\cal N}=(8,0)$ AdS$_3$ vacua
in half-maximal $D=3$ supergravities.
Analyzing the consistency constraints on the embedding tensor,
we have determined the full set of possible gauge groups embedded in the $\SO(8,n)$
isometry group of ungauged $D=3$ supergravity, for $n\le8$.
There are four classes of such vacua according to the different superisometry groups 
OSp$(8|2,\mathbb{R})$, F(4), SU$(4|1,1)$, and OSp$(4^*|4)$, respectively.
For each of the vacua, we have determined the explicit embedding tensor,
the gauge group embedded into the $\SO(8,n)$
isometry group of ungauged $D=3$ supergravity,
and the physical mass spectrum, organized in terms of supermultiplets.
We summarize our results in Tabs.~\ref{tab:summary1}, \ref{tab:summary2}.
For all the vacua identified, we have furthermore determined if and under which conditions
on the free parameters, the half-maximal theories admit an embedding into a 
maximal (${\cal N}=16$) supergravity, with the gauge group enhanced by additional
generators according to Eqs.~\eqref{eq:embmax}, \eqref{eq:covD} above.

In Tab.~\ref{tab:summary3}, we collect our findings of
${\cal N}=(7,1)$ and ${\cal N}=(7,0)$ AdS$_3$ vacua. 
We have shown that the latter vacua are realized in half-maximal theories
that also admit fully supersymmetric ${\cal N}=(8,0)$ AdS$_3$ vacua
as different stationary points in their scalar potential, \textit{c.f.}~Figs.~\ref{fig:potential_n8}, \ref{fig:potential_n7}
above.
In particular, this indicates the existence of domain wall solutions interpolating between
${\cal N}=(8,0)$  and ${\cal N}=(7,0)$ AdS$_3$ vacua.
It would be very interesting to generalize our findings to a classification of
general fully supersymmetric ${\cal N}=(p,8-p)$ vacua of the half-maximal theories.
In this case, the classification will be organized by products of smaller supergroups,
which presumably leaves even more possibilities for the potential embeddings 
of their bosonic parts into $\SO(8,n)$.

Throughout, we have restricted the analysis to theories with $n\le8$ matter multiplets. In this range, our
classification is exhaustive. For general $n$, we expect the classification to straightforwardly
extend to theories with chiral embedding of the $R$-symmetry group into the first factor of the 
compact ${\rm SO}(8)\times{\rm SO}(n)$ invariance group.  Indeed, we have identified various 
families of theories labelled by integers $n_+$, $n_-$ which are defined for arbitrary (unbounded) values of these integers.
For diagonal embedding of the $R$-symmetry group on the other hand, one may expect new patterns to arise,
since with increasing $n$, also the number of possible distinct embeddings of ${\rm SO}(8)$
into ${\rm SO}(n)$ increases. In particular, there are maximal embeddings for arbitrarily high values of $n$.
For ${\cal N}=(4,4)$ vacua, many theories based on different such embedding patterns have been identified in Ref.~\cite{Nicolai:2003ux}.

An immediate question to address for the vacua identified in this paper is
the existence and the structure of their moduli spaces as submanifolds of
the scalar target space $\SO(8,n)/(\SO(8)\times\SO(n))$.
In the holographic context, this relates to the
exactly marginal operators of the putative holographically dual conformal field theories.

Probably the most outstanding issue about these vacua and theories concerns 
their possible higher-dimensional origin. More precisely, it would be very interesting to embed
the half-maximal $D=3$ supergravities identified in this paper as consistent truncations into
ten- or eleven-dimensional supergravities, such that in particular the AdS$_3$ vacua would uplift to 
full $D=11$ or IIB solutions, subject to the constraints from
higher-dimensional classifications and no-go results~\cite{Gran:2016zxk,Beck:2017wpm,Haupt:2018gap}.
For example, it would be interesting to see if the type IIA compactifications with exceptional supersymmetry 
of Ref.~\cite{Dibitetto:2018ftj} can be embedded into consistent truncations to $D=3$ supergravities
constructed in this paper. In particular, for the solution with F(4) supersymmetry,
Tab.~\ref{tab:summary2} offers several candidates with gauge group factors
$\SO(7)$ potentially realized as the isometry group of the round six sphere
$S^6=\SO(7)/\SO(6)$\,.
For the theories with free continuous parameter, it would be very interesting to identify its possible higher-dimensional origin.

A systematic approach for higher-dimensional uplifts builds on the reformulation
of the higher-dimensional supergravities as exceptional field theories based on the 
group $\SO(8,n)$ \cite{Hohm:2017wtr}. In this framework, consistent truncations
are described as generalized Scherk-Schwarz reductions, leaving the task of solving
the consistency equations for suitable Scherk-Schwarz twist matrices.
An apparent obstacle to a standard geometrical uplift of many of the theories collected
in Tabs.~\ref{tab:summary1}--\ref{tab:summary3} is the rank and the size of their gauge groups
which do not admit a geometric realization as the isometry group of a 7- or 8-dimensional 
internal manifold. 
It remains to be seen if the rich structure of three-dimensional supergravity 
hints at some more general reduction mechanisms specific to three-dimensional theories.
In this context, it may be advantageous to exploit the possibility of embedding the $D=3$ half-maximal
theory into maximal higher-dimensional supergravities via their formulation as an E$_{8(8)}$ 
exceptional field theory~\cite{Hohm:2014fxa} 
upon suitable generalization of the methods developed in Ref.~\cite{Malek:2017njj}.
Another interesting option to explore is the possible existence of Scherk-Schwarz twist matrices realising
these gauged supergravities while explicitly violating the section constraints, although the higher-dimensional
interpretation of such a construction remains somewhat mysterious.

Finally, it would be very interesting to extend the present analysis to include
AdS$_3$ solutions with less supersymmetry and relate to known solutions and structures such as Refs.~\cite{Kim:2005ez,Gauntlett:2006af,DHoker:2008lup,Colgain:2010wb,Lozano:2015bra,Couzens:2017way,Wulff:2017zbl,Eberhardt:2017uup}.

\paragraph{Acknowledgements}
We are grateful to 
G. Dibitetto,
Y. Herfray,
W. M\"uck,
K. Sfetsos, and
M. Trigiante
for helpful and inspiring discussions. NSD wishes to thank ENS de Lyon for hospitality during the course of this work. NSD is partially supported by 
the Scientific and Technological Research Council of Turkey (T\"ubitak) Grant No.116F137.

\begin{landscape}
\begin{table}
  \centering
  \captionsetup{width=1.5\textwidth}
  \begin{tabular}{ccccccc}
  $n\le8$ & Gauge group & $\G_{\mathrm{ext}}$ & Param. & Embedding max.& Spectrum & $\Theta_{\cal MN\vert PQ}$\\\hline
  \multicolumn{7}{c}{\multirow{2}{*}{\normalsize $\boldsymbol{\mathcal{G}_R=\OSphuit}$}} \\
  & & & & \\\hline
  $n_{+}+n_{-}$ & $\SO(8,n_{+})\times\SO(n_{-})$ & $\SO(n_{+})\times\SO(n_{-})$ &  & $(n_{+},n_{-})=(8,0),\;(0,8)$ & Tab.~\subref{tab:osp8multchiral} & Eq.~\eqref{eq:Tklmn=0OSp8}\\
  $n_{+}+4$ & $\SO(8,n_{+})\times\SO(4)$ & $\SO(n_{+})\times\SO(4)$ & $\xi$ & $n_{+}=4$ and $\xi=\pm2$ & Tab.~\subref{tab:osp8multchiral} & Eq.~\eqref{eq:OSp8SO3}\\
  $n_{+}+4$ & $\SO(8,n_{+})\times\SO(3)$ & $\SO(n_{+})\times\SO(3)$ &  & - & Tab.~\subref{tab:osp8multchiral} & Eq.~\eqref{eq:OSp8SO3} \\
  $n_{+}+6$ & $\SO(8,n_{+})\times\U(3)$ & $\SO(n_{+})\times\U(3)$ &  & $n_{+}=2$ & Tab.~\subref{tab:osp8multchiral} & Eq.~\eqref{eq:OSp8U3} \\
  $n_{+}+7$ & $\SO(8,n_{+})\times\G_{2}$ & $\SO(n_{+})\times\G_{2}$ &  & $n_{+}=1$ & Tab.~\subref{tab:osp8multchiral} & Eq.~\eqref{eq:OSp8G2} \\
  $8$ & $\SO(8)\times\SO(8-p)\times\SO(p)$ & $\SO(8-p)\times\SO(p)$ &  & - & Tab.~\subref{tab:osp8multchiral} & Eq.~\eqref{eq:OSp8SOp} \\
  $8$ & $\GL(8)$ & - &  &$\checkmark$ & Tab.~\subref{tab:osp8multdiag} & Eq.~\eqref{eq:OSp8GL8} \\\hline
\multicolumn{7}{c}{\multirow{2}{*}{\normalsize $\boldsymbol{\mathcal{G}_R=\SU(4|1,1)}$}} \\
  & & & & \\ \hline
  $n_{+}+n_{-}$ & $\SO(6,n_{+})\times\SO(2,n_{-})$ & $\SO(n_{+})\times\SO(n_{-})$ &  & $(n_{+},n_{-})=(2,6)$ & Tab.~\subref{tab:SU411multchiral(ii)a}& Eq.~\eqref{eq:SU411Tklmn=0} \\
  $n_{+}+2m$ & $\SO(6,n_{+})\times\U(m,1)$ & $\SO(n_{+})\times\U(m)$ &  & $(n_{+},m)=(4,2)$ & Tab.~\subref{tab:SU411multchiral(ii)b}& Eq.~\eqref{eq:SU411Um} \\
  $n_{+}+2$ & $\SO(6,n_{+})\times\SO(2,1)$ & $\SO(n_{+})$ &  & - & Tab.~\ref{tab:SU411multchiral(ii)2} & Eq.~\eqref{eq:SU411n2} \\
  $n_{+}+2$ & $\SO(6,n_{+})\times\SO(2,2)$ & $\SO(n_{+})\times\U(1)$ & $\xi$ & $n_{+}=6$ and $\xi=\pm2$ & Tab.~\ref{tab:SU411multchiral(ii)2} & Eq.~\eqref{eq:SU411n2} \\
  $6\leq n<8$ &  $\GL(6)\times\SO(2,n-6)$ & $\SO(n-6)$ &  & - & Tab.~\ref{tab:SU411diag(ii)} & Eq.~\eqref{eq:SU411GL6} \\
  $8$ &  $\GL(6)\times\SO(2,1)$ & $\U(1)$ &  & - & Tab.~\ref{tab:SU411diag(ii)8}& Eq.~\eqref{eq:SU411GL6} \\
  $6$ &  $\GL(6)\times\SO(2,2)$ & $\SO(2)$ & $\xi$ & $\xi=\pm3$ & Tab.~\ref{tab:SU411diag(ii)8}& Eq.~\eqref{eq:SU411GL6} \\
  $n$ & $\SO(6)\times\SO(2)\times\SO(n)$ & $\SO(n)$ & & - & Tab.~\ref{tab:su411multchiral(i)} & Eq.~\eqref{eq:SU411SOn} \\
  $4$ & $\SO(6)\times\SO(2)\times\SO(4)$ & $\SO(4)$ & $\xi$ & - & Tab.~\ref{tab:su411multchiral(i)} & Eq.~\eqref{eq:SU411SO3} \\
  $4$ & $\SO(6)\times\SO(2)\times\SO(3)$ & $\SO(3)$ & & - & Tab.~\ref{tab:su411multchiral(i)} & Eq.~\eqref{eq:SU411SO3} \\
  $6$ & $\SO(6)\times\SO(2)\times\U(3)$ & $\U(3)$ & & - & Tab.~\ref{tab:su411multchiral(i)} & Eq.~\eqref{eq:SU411U3} \\
  $7$ & $\SO(6)\times\SO(2)\times\G_{2}$ & $\G_{2}$ & & - & Tab.~\ref{tab:su411multchiral(i)} & Eq.~\eqref{eq:SU411G2} \\
  $8$ & $\SO(6)\times\SO(2)\times\SO(8-p)\times\SO(p)$ & $\SO(8-p)\times\SO(p)$ &  & $p=2$ & Tab.~\ref{tab:su411multchiral(i)} & Eq.~\eqref{eq:SU411SOp} \\
  $8$ & $\U(4,4)$ & $\U(4)$ &  & $\checkmark$ & Tab.~\ref{tab:su411multchiral(i)U4} & Eq.~\eqref{eq:SU411U4} \\
  $8$ & $\SL(2)\times\Sp(4,\mathbb{R})$ & $\U(1)$ & & $\checkmark$ & Tab.~\ref{tab:su411multchiral(i)SL2Sp4} & Eq.~\eqref{eq:SU411SL2Sp4} \\
  $2+n_{-}$ & $\U(4,1)\times\SO(n_{-})$ & $\SO(n_{-})\times\U(1)$ &  & - & Tab.~\ref{tab:su411multchiral(i)SO6U41} & Eq.~\eqref{eq:SU411SO6U41}
  \end{tabular}
  \caption{Theories with $n$ matter multiplets admitting an ${\cal N}=(8,0)$ AdS$_3$ vacuum which preserves a global
  ${\rm SL}(2,\mathbb{R})_L \times \mathcal{G}_R \times  \G_{\mathrm{ext}}$ symmetry.
  References to the explicit embedding tensors and the associated mass spectra are given.
  The fourth  column  indicates if the embedding tensor possesses a free parameter, the fifth column indicates, if possible, 
  the condition for the theory to admit an embedding into  maximal (${\cal N}=16$) $D=3$ supergravity.}
  \label{tab:summary1}
\end{table}
\end{landscape}

\begin{landscape}
\begin{table}
  \centering
  \captionsetup{width=1.5\textwidth}
  \begin{tabular}{ccccccc}
  $n\le8$ & Gauge group & $\G_{\mathrm{ext}}$ & Param. & Embedding max.& Spectrum & $\Theta_{\cal MN\vert PQ}$\\\hline
\multicolumn{7}{c}{\multirow{2}{*}{\normalsize $\boldsymbol{\mathcal{G}_R=\F(4)}$}} \\
  & & & & \\ \hline
  $n_{+}+n_{-}$ & $\SO(7,n_{+})\times\SO(1,n_{-})$ & $\SO(n_{+})\times\SO(n_{-})$ &  & $(n_{+},n_{-})=(1,7)$ & Tab.~\subref{tab:f4multchiral(ii)}& Eq.~\eqref{eq:Tklmn=0F4} \\
  $n_{+}+3$ & $\SO(7,n_{+})\times\SO(1,3)$ & $\SO(n_{+})\times\SO(3)$ &  $\xi$ & - & Tab.~\subref{tab:f4multchiral(ii)}& Eq.~\eqref{eq:F4SO31} \\
  $n\geq7$ &  $\GL(7)\times\SO(1,n-7)$ & $\SO(n-7)$ & & - & Tab.~\subref{tab:f4multdiag(ii)}& Eq.~\eqref{eq:F4GL7} \\
  $n$ & $\SO(7)\times\SO(n)$ & $\SO(n)$ &  & - & Tab.~\ref{tab:f4multchiral(i)} & Eq.~\eqref{eq:F4SOn} \\
  $4$ & $\SO(7)\times\SO(4)$ & $\SO(4)$ &$\xi$  & - & Tab.~\ref{tab:f4multchiral(i)} & Eq.~\eqref{eq:F4SO3} \\
  $4$ & $\SO(7)\times\SO(3)$ & $\SO(3)$ &   & - & Tab.~\ref{tab:f4multchiral(i)} & Eq.~\eqref{eq:F4SO3} \\
  $6$ & $\SO(7)\times\U(3)$ & $\U(3)$ &  & - & Tab.~\ref{tab:f4multchiral(i)} & Eq.~\eqref{eq:F4U3} \\
  $7$ & $\SO(7)\times\G_{2}$ & $\G_{2}$ &  & - & Tab.~\ref{tab:f4multchiral(i)} & Eq.~\eqref{eq:F4G2} \\
  $8$ & $\SO(7)\times\SO(8-p)\times\SO(p)$ & $\SO(8-p)\times\SO(p)$ &  & $p=1$ & Tab.~\ref{tab:f4multchiral(i)} & Eq.~\eqref{eq:F4SOp} \\\hline
\multicolumn{7}{c}{\multirow{2}{*}{\normalsize $\boldsymbol{\mathcal{G}_R=\OSpquatre}$}} \\
  & & & & \\ \hline
  $n_{+}+n_{-}$ & $\SO(5,n_{+})\times\SO(3,n_{-})$ & $\SO(n_{+})\times\SO(n_{-})$ &  & $(n_{+},n_{-})=(3,5)$ & Tab.~\ref{tab:OSp44multchiral(ii)}& Eq.~\eqref{eq:OSp44Tklmn=0} \\
  $n\geq4$ & $\SO(5,n-4)\times\G_{2(2)}$ & $\SO(n-4)\times\SO(3)$ &  & $n=8$ & Tab.~\subref{tab:OSp44multdiag(ii)G22}& Eq.~\eqref{eq:OSp44G22} \\
  $n\geq5$ &  $\GL(5)\times\SO(3,n-5)$ & $\SO(n-5)$& & - & Tab.~\subref{tab:OSp44multdiag(ii)GL5}& Eq.~\eqref{eq:OSp44GL5} \\
  $n\geq3$ &  $\GL(3)\times\SO(5,n-3)$ & $\SO(n-3)$& & $n=8$ & Tab.~\subref{tab:OSp44multdiag(ii)GL3}& Eq.~\eqref{eq:OSp44GL3} \\
  $8$ &  $\GL(5)\times\GL(3)$ & - &  & - & Tab.~\subref{tab:OSp44multdiag(ii)GL5GL3} & Eq.~\eqref{eq:OSp44GL5GL3} \\
  $n$ & $\SO(5)\times\SO(3)\times\SO(n)$ & $\SO(n)$ & & - & Tab.~\ref{tab:osp44multchiral(i)} & Eq.~\eqref{eq:OSp44SOn} \\
  $4$ & $\SO(5)\times\SO(3)\times\SO(4)$ & $\SO(4)$ &$\xi$  & - & Tab.~\ref{tab:osp44multchiral(i)} & Eq.~\eqref{eq:OSp44SO3} \\
  $4$ & $\SO(5)\times\SO(3)\times\SO(3)$ & $\SO(3)$ &  & - & Tab.~\ref{tab:osp44multchiral(i)} & Eq.~\eqref{eq:OSp44SO3} \\
  $6$ & $\SO(5)\times\SO(3)\times\U(3)$ & $\U(3)$ & & - & Tab.~\ref{tab:osp44multchiral(i)} & Eq.~\eqref{eq:OSp44U3} \\
  $7$ & $\SO(5)\times\SO(3)\times\G_{2}$ & $\G_{2}$ & & - & Tab.~\ref{tab:osp44multchiral(i)} & Eq.~\eqref{eq:OSp44G2} \\
  $8$ & $\SO(5)\times\SO(3)\times\SO(8-p)\times\SO(p)$ & $\SO(8-p)\times\SO(p)$ &  & $p=3$ & Tab.~\ref{tab:osp44multchiral(i)} & Eq.~\eqref{eq:OSp44SOp} \\
  $8$ & $\Sp(2,2)\times\SO(3)$ & $\USp(4)$ &  & $\checkmark$ & Tab.~\ref{tab:osp44multchiral(i)USp4} & Eq.~\eqref{eq:OSp44USp44}
  \end{tabular}
  \caption{Theories with $n$ matter multiplets admitting an ${\cal N}=(8,0)$ AdS$_3$ vacuum which preserves a global
  ${\rm SL}(2,\mathbb{R})_L \times \mathcal{G}_R \times  \G_{\mathrm{ext}}$ symmetry.
  References to the explicit embedding tensors and the associated mass spectra are given.
  The fourth  column  indicates if the embedding tensor possesses a free parameter, the fifth column indicates, if possible, 
  the condition for the theory to admit an embedding into  maximal (${\cal N}=16$) $D=3$ supergravity.}
  \label{tab:summary2}
\end{table}
\end{landscape}

\begin{landscape}
\begin{table}
  \centering
  \captionsetup{width=1.1\textwidth}
  \begin{tabular}{cccccccc}
  $\cal N$ & $n\le8$ & Gauge group & $\G_{\mathrm{ext}}$ & Param. & Embedding max. & Spectrum & $\Theta_{\cal MN\vert PQ}$\\\hline
\multicolumn{8}{c}{\multirow{2}{*}{$\boldsymbol{\mathcal{G}_R=\OSpsept}$}} \\
  & & & & \\ \hline
  & & & & \\
  $(7,0)$& $8$ &  $\SO(7)\times\SO(8)$ & - & & - & Tab.~\ref{tab:OSp7mult}& Eq.~\eqref{eq:OSp7SO8SO7}\\\
  & & & & \\\hline
\multicolumn{8}{c}{\multirow{2}{*}{$\boldsymbol{\mathcal{G}_R= \G(3)}$}} \\
  & & & & \\ \hline
  & & & & \\
  $(7,1)$& $n$ & $\SO(n,1)\times\G_{2}$ & $\SO(n)$ & & $n=8$ & Tab.~\ref{tab:G3mult71} & Eq.~\eqref{eq:G3SOn1G2} \\
  $(7,0)$& $7$ & $\SO(7)\times\G_{2}$ & - & & - & Tab.~\ref{tab:G3mult70} & Eq.~\eqref{eq:G3SO7G2}
\end{tabular}
  \caption{Theories with $n$ matter multiplets admitting an ${\cal N}=(7,q)$ AdS$_3$ vacuum which preserves a global
  ${\rm SL}(2,\mathbb{R})_L \times \mathcal{G}_R \times  \G_{\mathrm{ext}}$ symmetry.
  References to the explicit embedding tensors and the associated mass spectra are given.
  The fifth  column  indicates if the embedding tensor possesses a free parameter, the sixth column indicates, if possible, 
  the condition for the theory to admit an embedding into  maximal (${\cal N}=16$) $D=3$ supergravity.}
  \label{tab:summary3}
\end{table}
\end{landscape}

\section*{Appendix}

\begin{appendix}
\section{Notations and supersymmetry transformation}
\label{sec:appendix_notations}
We list here the conventions for the gauging and the supersymmetry transformations in the Lagrangian~\eqref{L}, following Refs.~\cite{Nicolai:2001ac,Nicolai:2001sv}. The coset representative ${\cal V}(\phi)$ defined in Eq.~\eqref{eq:cosetrepresentative} transforms as
\begin{equation}
{\cal V}(\phi')=g\,{\cal V}(\phi)\,h^{-1}(\phi), 
\end{equation}
with $g\in\SO(8,n)$ and $h(\phi)\in\SO(8)\times\SO(n)$. For the ungauged theory, its coupling to fermions is 
described in terms of the Cartan-Maurer form
\begin{equation}
{\cal V}^{-1}\partial_{\mu}{\cal V} = \frac{1}{2}\,Q_{\mu}^{IJ}L^{IJ}+\frac{1}{2}\,Q_{\mu}^{rs}L^{rs} +P_{\mu}^{Ir}L^{Ir}.
\end{equation}
Once the theory is gauged, the covariantization gives
\begin{equation}
{\cal V}^{-1}{\cal D}_{\mu}{\cal V} = {\cal V}^{-1}\partial_{\mu}{\cal V}+ A_{\mu}^{\cal MN}\,{\cal V}^{-1}\,\Theta_{\cal MN\vert PQ}\,L^{\cal MN}\,{\cal V} =\frac{1}{2}\,{\cal Q}_{\mu}^{IJ}L^{IJ}+\frac{1}{2}\,{\cal Q}_{\mu}^{rs}L^{rs} +{\cal P}_{\mu}^{Ir}L^{Ir}.
\end{equation}
The full covariant derivatives of the fermions then read
\begin{equation}
    \begin{cases}
    {\cal D}_{\mu}\psi^{A}_{\nu} = \partial_{\mu}\psi_{\nu}^{A} + \dfrac{1}{4}\,{\omega_{\mu}}^{\alpha\beta}\gamma_{\alpha\beta}\psi_{\nu}^{A}+\dfrac{1}{4}{\cal Q}_{\mu}^{IJ}\Gamma_{AB}^{IJ}\,\psi_{\nu}^{B}, \\
    {\cal D}_{\mu}\chi^{\dot Ar} = \partial_{\mu}\chi^{\dot Ar} + \dfrac{1}{4}\,{\omega_{\mu}}^{\alpha\beta}\gamma_{\alpha\beta}\chi^{\dot Ar}+\dfrac{1}{4}{\cal Q}_{\mu}^{IJ}\Gamma_{\dot A\dot B}^{IJ}\,\chi^{\dot Br} + {\cal Q}_{\mu}^{rs}\chi^{\dot As},
    \end{cases}
\end{equation}
with the spin-connection ${\omega_{\mu}}^{\alpha\beta}$, the three-dimensional $\gamma$ matrices $\gamma^{\alpha}$ in flat spacetime and $\gamma^{\alpha\beta} = \gamma^{[\alpha}\gamma^{\beta]}$. Finally, the full supersymmetry variations are
\begin{equation}
    \begin{cases}
    {\cal V}^{-1}\delta{\cal V} = L^{Ir}\,\bar\varepsilon^{A} \,\Gamma_{A\dot A}^{I}\chi^{\dot Ar}, \\
    \delta {e_{\mu}}^{\alpha} = i\,\bar\varepsilon^{A}\,\gamma^{\alpha}\,\psi_{\mu}^{A}, \\
    \delta {A_{\mu}}^{\cal MN} = -\dfrac{1}{2}{{\cal V}^{\cal MN}}_{IJ}\,\bar\varepsilon^{A}\Gamma_{AB}^{IJ}\,\psi_{\mu}^{B}+i\,{{\cal V}^{\cal MN}}_{Ir}\,\bar\varepsilon^{A}\Gamma_{A\dot A}^{I}\,\gamma_{\mu}\,\chi^{\dot A r}, \\
    \delta\chi^{\dot A r} = \dfrac{i}{2}\,\Gamma_{A\dot A}\,\gamma^{\mu}\,\varepsilon^{A}\,{\cal P}_{\mu}^{Ir}+g\,A_{2}^{A\dot A r}\,\varepsilon^{A}, \\
    \delta\psi_{\mu}^{A} = {\cal D}_{\mu}\varepsilon^{A}+i\,g\,A_{1}^{AB}\,\gamma_{\mu}\,\varepsilon^{B},
    \end{cases}
\end{equation}
with supersymmetry parameter $\varepsilon^{A}$ and $\gamma^{\mu} = {e^{\mu}}_{\alpha}\gamma^{\alpha}$.


\section{Second quadratic constraint}
\label{app:QC2}


We list here the full set of independent equations contained in the second quadratic constraint \eqref{eq:QC2} with parametrization~\eqref{eq:thetafullparam}
for the various values of the free indices.

\paragraph*{$\boldsymbol{({\cal MNPQ}) = (IJKL)}$}
\begin{subnumcases}{\label{eq:QC2IJKL}}
\Lambda_{MN}\left[\theta_{MNP[I}\theta_{JKL]P}+4\,\theta_{M[I}\theta_{JKL]N}-2\,\kappa\,\delta_{M[I}\theta_{JKL]N}\right] = 0, \label{eq:QC2IJKLa} \\
\theta_{uvP[I}\theta_{JKL]P} = 0. \label{eq:QC2IJKLb} 
\end{subnumcases}

\paragraph*{$\boldsymbol{({\cal MNPQ}) = (IJKr)}$}
\begin{subnumcases}{\label{eq:QC2IJKr}}
3\,\theta_{uvs[I}\theta_{JK]rs}-\theta_{uvrL}\theta_{IJKL} = 0, \label{eq:QC2IJKra} \\
-3\,\theta_{usM[I}\theta_{JK]rs}-3\,\delta_{M[I}\theta_{JK]rs}\theta_{su}+ 3\,\theta_{M[I}\theta_{JK]ru}-3\,\kappa\,\delta_{M[I}\theta_{JK]ru}  \nonumber\\
+ \theta_{urML}\theta_{IJKL}+\left(\theta_{ur}+\kappa\,\delta_{ur}\right)\,\theta_{IJKM}-\delta_{ur}\,\theta_{ML}\theta_{IJKL} = 0. \label{eq:QC2IJKrb} 
\end{subnumcases}

\paragraph*{$\boldsymbol{({\cal MNPQ}) = (IJrs)}$}
\begin{subnumcases}{\label{eq:QC2IJrs}}
\Lambda_{uv}\big[\theta_{uvp[I}\theta_{J]prs} - \theta_{uvL[I}\theta_{J]Lrs} - \theta_{uvp[r}\theta_{s]pIJ} \nonumber \\
+2\,\theta_{up}\theta_{IJp[r}\delta_{s]v} + 2\,\theta_{IJu[r}\theta_{s]v} - 2\,\kappa\,\theta_{IJu[r}\delta_{s]v}\big] = 0, \label{eq:QC2IJrsa} \\
 \Lambda_{MN}\big[\theta_{MNL[I}\theta_{J]Lrs}+\theta_{MNp[r}\theta_{s]pIJ}+4\,\theta_{M[I}\theta_{J]Nrs}-2\kappa\,\delta_{M[I}\theta_{J]Nrs}\big] = 0, \label{eq:QC2IJrsb} \\
 \theta_{Mup[I}\theta_{J]prs}-\theta_{Mup[r}\theta_{s]pIJ}+\delta_{M[I}\theta_{J]rsp}\theta_{pu}+\kappa\,\delta_{M[I}\theta_{J]rsu}-\theta_{M[I}\theta_{J]rsu} = 0. \label{eq:QC2IJrsc}
\end{subnumcases}

\paragraph*{$\boldsymbol{({\cal MNPQ}) = (Ipqr)}$}
\begin{subnumcases}{\label{eq:QC2Ipqr}}
\Lambda_{uv}\big[\theta_{Iuvs}\theta_{spqr}-\theta_{uvIL}\theta_{Lpqr}+3\,\theta_{Is[pq}\theta_{r]uvs} -3\,\theta_{IL[pq}\theta_{r]uvL} \nonumber \\
 -6\,\theta_{vs}\theta_{Is[pq}\delta_{r]u}+6\,\theta_{Iv[pq}\theta_{r]u}+6\,\kappa\,\theta_{Iv[pq}\delta_{r]u}\big] = 0, \label{eq:QC2Ipqra} \\
 \Lambda_{uv}\big[-\theta_{MNIL}\theta_{Lpqr}+3\,\theta_{Is[pq}\theta_{r]sMN}+4\,\theta_{MI}\theta_{Npqr}-2\kappa\,\delta_{MI}\theta_{Npqr}\big] = 0, \label{eq:QC2Ipqrb} \\
 \theta_{MIus}\theta_{spqr}-3\,\theta_{Is[pq}\theta_{r]usM}-3\,\theta_{IL[pq}\theta_{r]uML} +\delta_{MI}\theta_{us}\theta_{spqr}+\kappa\,\delta_{MI}\theta_{upqr}\nonumber \\
 -\theta_{MI}\theta_{upqr}-3\,\theta_{ML}\theta_{IL[pq}\delta_{r]u}+3\,\kappa\,\theta_{IM[pq}\delta_{r]u}+3\,\theta_{IM[pq}\theta_{r]u}=0. \label{eq:QC2Ipqrc}
\end{subnumcases}

\paragraph*{$\boldsymbol{({\cal MNPQ}) = (pqrs)}$}
\begin{subnumcases}{\label{eq:QC2pqrs}}
\Lambda_{uv}\big[\theta_{uvl[p}\theta_{qrs]l}-\theta_{Luv[p}\theta_{qrs]L}-2\,\delta_{u[p}\theta_{qrs]l}\theta_{lv}-2\,\theta_{u[p}\theta_{qrs]v}-2\,\delta_{u[p}\theta_{qrs]v}\big]=0, \label{eq:QC2pqrsa} \\
-\theta_{MLu[p}\theta_{qrs]L}+\theta_{Mlu[p}\theta_{qrs]l}+\delta_{u[p}\theta_{qrs]L}\theta_{ML}-\kappa\,\delta_{u[p}\theta_{qrs]M}-2\,\theta_{u[p}\theta_{qrs]M}=0, \label{eq:QC2pqrsb} \\
\theta_{MNl[p}\theta_{qrs]l}=0. \label{eq:QC2pqrsc}
\end{subnumcases}

\end{appendix}


\providecommand{\href}[2]{#2}\begingroup\raggedright\endgroup

\end{document}